\documentclass[aps,twocolumn,superscriptaddress,prd,floatfix,citeautoscript,preprintnumbers,nofootinbib]{revtex4-1}
\pdfoutput=1
\usepackage{color}
\usepackage{graphicx}
\usepackage{amssymb}
\usepackage{amsmath}
\usepackage{epstopdf}
\usepackage[normalem]{ulem}
\usepackage{comment}
\usepackage{breakcites}
\usepackage{subfigure,epsfig}
\usepackage[breaklinks=true]{hyperref}
\usepackage{titlesec}

\titleformat*{\section}{\large\bfseries}
\titleformat*{\subsection}{\large\bfseries}
\titleformat*{\subsubsection}{\normalsize\bfseries}
\usepackage{bbold}

\newcommand{\beq}{\begin{equation}}
\newcommand{\eeq}{\end{equation}}
\newcommand{\bal}{\begin{align}}
\newcommand{\eal}{\end{align}}

\newcommand{\ket}[1]{\mbox{$ | #1 \rangle $}}
\newcommand{\bra}[1]{\mbox{$ \langle #1 | $}}

\newcommand{\etal}{\emph{et al.}}

\begin{document}

\title{Review on novel methods for lattice gauge theories}
\begin{abstract}
Formulating gauge theories on a lattice offers a genuinely non-perturbative
way of studying quantum field theories, and has led to impressive achievements.
In particular, it significantly deepened our understanding of quantum chromodynamics.
Yet, some very relevant problems remain inherently challenging, such as 
real time evolution, or the presence of a chemical potential, cases in which
Monte Carlo simulations are hindered by a sign problem.

In the last few years, a number of possible alternatives have been put forward, based on quantum
information ideas, which could potentially open the access to areas of research that have so far
eluded more standard methods.
They include  tensor network calculations, quantum simulations with different physical platforms and quantum computations,
and constitute nowadays a vibrant research area.
Experts from different fields, including experimental and theoretical high energy physics,
condensed matter, and quantum information, are turning their attention to these
interdisciplinary possibilities, and driving the progress of the field. 
The aim of this article is to review the status and perspectives of these new avenues for the exploration of lattice gauge
theories.

\end{abstract}
\author{Mari Carmen Ba\~nuls}
\affiliation{ Max-Planck Institut, f\"{u}r Quantenoptik,
    Garching 85748, Germany}
\affiliation{Munich Center for Quantum Science and Technology (MCQST), Schellingstr. 4, Munich 80799, Germany}
\author{Krzysztof Cichy}
\affiliation{Faculty of Physics, Adam Mickiewicz University, Uniwersytetu Pozna\'nskiego 2, 61-614 Pozna\'{n}, Poland}
% Activate to display a given date or no date
%\date{\today}							

\maketitle

\tableofcontents

%\section{}
%\subsection{}

\section{Introduction}
\label{sec:intro}
\subsection{Lattice gauge theories:\\ achievements and limitations}
\label{sec:generalintro}

Gauge theories, i.e.\ field theories invariant under a particular local (gauge) symmetry transformation, are of fundamental importance in theoretical physics, 
where they appear in very different fields.
They underlie the Standard Model of elementary particle physics, being a non-Abelian gauge theory based on the gauge group $U(1)\times SU(2)\times SU(3)$, the former two groups corresponding to the electroweak theory and the latter to quantum chromodynamics (QCD).
The fourth fundamental interaction in nature, gravity, can also be formulated as a gauge theory.
In condensed matter physics, gauge theories emerge as effective descriptions of strongly correlated phenomena, such as superconductivity and
the fractional quantum Hall effect.

One of the most significant strategies for studying gauge theories is to formulate them on the lattice.
This provides a mathematical definition of the theory, by providing a momentum cutoff, both in the ultraviolet (the lattice spacing) and in the infrared (the lattice size).
Formally, such a regulated theory becomes a statistical mechanical system, amenable to established methods of statistical physics (see, e.g., Ref.~\cite{Kogut1983}).
The lattice approach plays a particularly important role in QCD, believed to be the correct theory of the strong interaction.
This interaction binds quarks together to form protons and neutrons, the basic building blocks of matter.

It is difficult to overstate the importance of lattice QCD (LQCD) amongst lattice gauge theories (LGT). Its fundamental character
has motivated an intense research effort during decades, which, in turn, has led to crucial theoretical and algorithmic developments
whose implications extend to other lattice field theories.
The main features and limitations of LQCD, which we discuss next, thus set the background
on which new LGT techniques are to be developed and tested.

QCD is a theory that features rather large dependence of its coupling on the energy scale.
On the one hand, there is the high-energy regime with relatively small coupling, allowing the use of perturbative methods.
On the other hand, at low energies, the strong coupling becomes indeed strong and perturbation theory is bound to fail.
In this non-perturbative regime, a quantitative description \emph{from first principles} is only possible by formulating the theory in a discretized form, on a spacetime lattice, and performing numerical computations.
Initially proposed by Kenneth G.\ Wilson in 1974~\cite{Wilson:1974sk}, the LQCD approach has succeeded in answering many questions about the physics of the strong interaction.
This includes calculations of fundamental properties of QCD (such as quark masses and the running of the coupling), masses of QCD bound states (such as protons, neutrons, pions and other mesons and baryons), structure of hadrons (e.g.\ how quarks and gluons interact with one another inside the proton or the proton's spin content), flavour physics (leading to constraints on the CKM matrix elements), as well as non-zero temperature properties.\footnote{For an overview of the current challenges and perspectives of QCD, see, e.g., Ref.~\cite{Brambilla:2014jmp}.}

The lattice formulation of QCD is usually based on the path integral quantization approach.
The QCD path integral, formally infinite-dimensional, is given proper mathematical meaning (regularized) by discretization.
Yet, the obtained object is still highly multi-dimensional, prohibiting analytical solution.
Instead, the multi-dimensional integral can be computed by employing Monte Carlo (MC) simulations.
The prerequisite for performing such simulations is the formulation of LQCD in Euclidean spacetime, since Minkowski spacetime path integral would include an ill-behaved exponential of the action ($S$) multiplied by the imaginary unit, $\exp(iS)$.
Instead, the Euclidean path integral includes the factor $\exp(-S)$, formally a Boltzmann factor that, as mentioned above, brings the task into a problem of investigating a statistical mechanical system, amenable to, e.g., MC simulations.
The Boltzmann factor provides a probability measure for sampling the path integral.
Obviously, the numerical solution is still highly non-trivial.
The physics of QCD dictates the sizes of lattices that need to be used to connect a lattice calculation to the natural world.
Even though numerics necessarily needs a finite lattice spacing and finite volume, the effect thereof can be extrapolated away by performing simulations with several lattice spacings, volumes  and other relevant characteristics. % of the simulation.
Nevertheless, the lattice spacings and volumes can not be arbitrary.
To maintain control over non-physical lattice effects, all the parameters have to be chosen such that a reliable extrapolation is possible.
In practice, this leads to discretized path integrals with dimensions between millions and billions, thus requiring huge computational resources, provided by the world's most powerful supercomputers.

Despite the amazing successes of LQCD, shedding light on various aspects of the strong interaction, there are still areas where LQCD cannot provide satisfactory answers.
As already mentioned, LQCD is formulated in Euclidean spacetime.
For many physical quantities, this is not a restriction, as the analytical continuation back to Minkowski spacetime poses no problem.
However, there are classes of observables that cannot be simply extracted from a Euclidean simulation.
One prominent example are quantities that depend on real time.
It is not possible to connect Euclidean results to real-time properties.
The latter can only be accessed with a simulation in Minkowski spacetime, where the temporal direction has a different signature.
An important subclass of this kind are also light-cone correlations, in terms of which partonic properties are formulated and expressed, e.g.\ as parton distribution functions, see Ref.~\cite{Cichy:2018mum} for an overview of methods how the issue can be overcome.
The other highly important class of problems not amenable to Euclidean MC simulations is 
the presence of a non-zero chemical potential (yielding non-zero baryon density in QCD).
In this case, the Boltzmann factor of the Euclidean path integral 
becomes complex and can no longer be interpreted as a probability measure, undermining the whole principle of MC simulations.
Such situation, when the probability measure becomes complex (or negative) and can not be used to sample a probability distribution, is commonly referred to as the sign problem.
It is well-known in quantum many-body physics and is often associated with the presence of fermions.
However, the QCD sign problem is not of fermionic origin, but it is more fundamental, directly related to the signature of spacetime.
There are also cases where it is of topological origin, as in QCD with the $CP$-violating $\theta$-term (thus, it would appear also in LQCD with Minkowski signature).
In the context of LQCD, the sign problem is also often termed the complex action problem.

The sign problem is NP-hard, a statement proved by Troyer and Wiese for the case of a classical three-dimensional Ising spin glass~\cite{Troyer2005}.
This means that if it can be solved in polynomial time, so could any possible NP problem (i.e.\ one verifiable in polynomial time) and thus NP=P.
The NP=P problem has not been solved despite many intense efforts and a million dollar financial incentive from the Clay Mathematics Institute.
Hence, it is widely believed that NP$\neq$P and it is extremely unlikely that the sign problem has a generic solution.
This does not, however, mean that the physics hidden behind the sign problem is inaccessible.
The NP-hardness prevents a generic solution, but ways to alleviate the problem may still exist, by using some physical property of the system.
One of the strategies described in this review, tensor network (TN) methods, avoid the sign problem
altogether by not relying on MC simulations and working directly in Hilbert space of possible states of the system.
Obviously, the Hilbert space grows exponentially fast with the system size.
The likely exponentially-hard sign problem may seem to be traded for another exponentially-hard problem.
However, the essential ingredient of TN is a parametrization of a subset of Hilbert space that corresponds to the entanglement properties of the system under study. 
In many cases, this parametrization allows for an efficient solution of the problem.
We discuss the details of TN methods, and the recent progress in applying them to LGT, in the next section.

Even without a sign problem, the computational complexity of real time simulations is expected to scale exponentially in terms of the
number of particles, the desired precision and the time length.
A possible way around these limitations is to follow Feynman's idea, formulated in the early 1980s~\cite{Feynman1982}.
Feynman believed that the only really appropriate way to study quantum-mechanical systems is to employ quantum mechanics directly.
This can ensue by engineering a physical system that, as closely as possible, mimics the dynamics of the 
considered theory, but is experimentally realizable, controllable and measurable.
Alternatively, one can also construct a more universal device, a quantum computer, that can be programmed to simulate the desired theory.
Different from a classical computer, a quantum one operates according to the laws of quantum mechanics, 
by applying (unitary) quantum operations on quantum bits (or qubits) that can be in any superposition state.
Born in pursue of Feynman's vision, the field of quantum simulation has made impressive technical and conceptual advances
in the last decade. Its prospective application to LGT is actively discussed, with some proof of principle experiments having already taken place.
We review this research direction, as well as the quantum computation prospects for the study of lattice gauge models in Sec.~\ref{sec:quantum}.

\subsection{Overcoming the sign problem in LQCD}
\label{sec:signintro}

Before we embark on this discussion and review the quantum inspired approaches to lattice gauge theories,
we first discuss other ways of overcoming the sign problem in LQCD, concentrating on the sign problem triggered by the chemical potential.
They can be divided into two classes.
The first one encompasses methods that can be successful in the case of a mild sign problem, such as QCD at a small chemical potential.
Unfortunately, the interesting physics of non-zero chemical potential lies in the regime where the sign problem becomes severe.
The second class of methods shortly discussed here have, in principle, the capacity to access also this more difficult regime, in particular the region of the conjectured tricritical point in the QCD phase diagram.
For extensive reviews of the different approaches, see, e.g., Refs.~\cite{Schmidt:2006us,Philipsen:2008gf,Ejiri:2008nv,Karsch:2009zz,Philipsen:2010gj,deForcrand:2010ys,Gupta:2011ma,Philipsen:2012nu,Levkova:2012jd,Petreczky:2013qj,Aarts:2013naa,Sexty:2014dxa,Aarts:2015kea,Aarts:2015tyj,Gattringer:2016kco,Ding:2017giu,Ratti:2018ksb,Sharma:2019wiv,Berger:2019odf}.

\vspace*{1mm}
\noindent\textbf{Taylor expansion.} Historically, the first method widely used to simulate QCD at non-zero chemical potential ($\mu$) was the Taylor expansion.
Already in 1988, it was applied for the computation of the quark number susceptibility~\cite{Gottlieb:1988cq}, related to the derivative of the partition function with respect to the chemical potential.
Any observable at $\mu\neq0$ can be formally expressed as a Taylor expansion in $\mu$.
The even ($2k$) coefficients of the expansion are given by the derivatives of the observable and of the fermionic determinant with respect to $\mu$, all taken at $\mu=0$ and, thus, avoiding the sign problem (all odd coefficients vanish by symmetry).
Several thermodynamic observables were studied this way, but it is clear that the Taylor series does not converge for large chemical potentials.
In principle, one can increase the expansion order, but this encounters exponentially increasing statistical error~\cite{deForcrand:2010ys}, rendering it impractical beyond $k\approx4-5$.
Moreover, derivatives defining the expansion are discontinuous at phase transitions, thus not allowing for investigations of the latter.
Hence, the Taylor series method allowed only insight into the small-$\mu/T$ ($T$ -- temperature) area of the QCD phase diagram.

\vspace*{1mm}
\noindent\textbf{Reweighting.} Another classical technique of accessing non-zero chemical potential observables is reweighting.
In principle, results for certain simulation parameters, including $\mu\neq0$, can be obtained from a simulation with other parameters, e.g.\ $\mu=0$.
This proceeds via the computation of the ratio of fermionic determinants for both cases, evaluated in the $\mu=0$ theory.
It was first proposed in 1988~\cite{Ferrenberg:1988yz}, with an application to the two-dimensional Ising and 8-state Potts models.
In 1997~\cite{Barbour:1997bh}, it was used for QCD and with further improvements, such as reweighting both in $\mu$ and $T$~\cite{Fodor:2001au} or expanding the reweighting factor in terms of $\mu$~\cite{Allton:2002zi}, the reweighting method led to a lot of insight into the physics of QCD at finite baryon density.
However, again, this was restricted to relatively small chemical potentials.
The reason in this case is that the huge configuration space in MC simulations is probed by means of importance sampling, with only order of hundreds or, at best, thousands gauge field configurations peaked in configuration space for the given simulation parameters.
In turn, the peak of a distribution with different parameters is located somewhere else and there is sufficient overlap between the simulated and desired distributions only if the parameters are not too far from each other.
Thus, despite the fact that, in principle, reweighting is exact, the exponentially small overlap between large-$\mu$ and $\mu=0$ distributions prohibits the access to the former.

\vspace*{1mm}
\noindent\textbf{Analytic continuation.} The third method that we discuss in the context of early ways of alleviating the sign problem is analytic continuation~\cite{Alford:1998sd,Lombardo:1999cz,deForcrand:2002hgr,DElia:2002tig}. It utilizes the fact that the problem does not appear for purely imaginary chemical potentials.
Thus, one can simulate at a few imaginary values of $\mu$ and fit the data with a chosen ansatz (e.g.\ polynomial or in terms of rational functions).
Then, the ansatz can be analytically continued to real chemical potentials.
However, the method is limited to rather small values of $\mu$ by unphysical Roberge-Weiss phase transitions~\cite{Roberge:1986mm}, due to the periodicity of the partition function in the imaginary $\mu$.

\vspace*{1mm}
\noindent\textbf{Canonical ensemble simulations.}
A different strategy to simulate QCD with non-zero baryon density is to work in the canonical ensemble, i.e.\ fix the baryon number~\cite{Miller:1986cs,Engels:1999tz,Liu:2002qr,deForcrand:2006ec}, instead of the chemical potential.
This old approach has been revisited a few years ago~\cite{Danzer:2008xs,Li:2010dya,Li:2011ee,Bornyakov:2018cjx,Bornyakov:2019kni}.
The grand canonical partition function, which is the object of ultimate interest, is related to the canonical partition function via a fugacity\footnote{The fugacity is defined as $\exp(\mu/T)$.} expansion, where the coefficients of the powers of fugacity are the canonical partition functions, $Z_n$.
The latter can be obtained from the inverse Fourier transforms of the grand canonical partition function corresponding to non-zero imaginary chemical potentials, discussed in the previous paragraph.
This procedure is exact, however, on the lattice, some assumptions need to be made that restrict the available range of chemical potentials.
Usually one proceeds by employing an ansatz for the functional dependence on the imaginary chemical potential and/or one uses analytic continuation to real chemical potential values.

\vspace*{1mm}
\noindent\textbf{Effective theory.}
Another approach that can allow simulations at finite density is to derive an effective theory.
Two examples of such theories are an effective 3D model of QCD with heavy quarks combined with strong coupling and hopping parameter expansions~\cite{Fromm:2011qi,Fromm:2012eb,Langelage:2014vpa,Glesaaen:2015vtp} and hadron worldlines reformulation of the partition function for chiral quarks using strong coupling expansion~\cite{Wolff:1984we,Karsch:1988zx,deForcrand:2014tha}.
In both cases, the sign problem at finite density becomes mild, allowing for reliable reweighting of $\mu=0$ simulations, or is absent altogether when the partition function is suitably rearranged, allowing for effective simulation with worm algorithms.
Thus, the phase diagram can be explored in the limiting regimes of heavy quarks close to the continuum or light quarks on coarse lattices.
However, it is not clear whether direct contact can be made to the physical regime with continuum light quarks.
For more details on this approach, see the review of Ref.~\cite{Philipsen:2016wjt} and references therein.

\vspace*{1mm}
\noindent\textbf{Complex Langevin.}
An intensely investigated method of avoiding the sign problem at non-zero chemical potential is to use the approach of stochastic quantization, also termed complex Langevin (CL) simulations for complex actions.
It was invented already in 1983 independently by Parisi~\cite{Parisi:1984cs} and Klauder~\cite{Klauder:1983nn}, see also Ref.~\cite{Damgaard:1987rr} for an early extensive review.
It is not based on Monte Carlo simulations, but rather on solving for the evolution of the system in a fictitious Langevin time\footnote{Thus, it was investigated also as an alternative to MC simulations for LGT, not necessarily as means to solve the sign problem~\cite{Batrouni:1985jn}.}.
For complex actions, like finite-density QCD, one needs an analytic continuation of the real Langevin process to a complexified manifold.
One sets up a stochastic process on a complexification of configuration space, i.e.\ with gauge links belonging to the special linear group $SL(N,\mathbb{C})$, instead of $SU(3)$.
After evolving long enough with CL equations, observables satisfying certain criteria should be distributed according to the QCD action. 
The method is guaranteed to give correct results for systems with real-valued actions, while for complex actions, problems with convergence to the right limit may appear.
Indeed, failure of the algorithm to converge to the right answer have been seen early~\cite{Ambjorn:1985iw,Ambjorn:1986fz} and the reasons for this were understood theoretically.
Consequently, criteria for correctness of the results have been formulated and examined~\cite{Aarts:2009uq,Aarts:2010aq,Aarts:2011ax,Aarts:2012ft,Aarts:2013uza,Nishimura:2015pba,Nagata:2015uga,Hayata:2015lzj,Wosiek:2015bqg,Salcedo:2016kyy,Nagata:2016alq,Nagata:2016vkn,Seiler:2017vwj,Nagata:2018net,Scherzer:2018hid,Kogut:2019qmi}.
CL was also validated against other techniques, such as reweighting~\cite{Fodor:2015doa}.
In addition to usage of CL as remedy for the sign problem related to non-zero chemical potential (see, e.g.,~\cite{Aarts:2008rr,Aarts:2008wh,Aarts:2009hn,Aarts:2009dg} for early attempts), the method was tested for real-time simulations, see, e.g.,~\cite{Berges:2005yt,Berges:2006xc,Berges:2007nr}.
The problems, along with general features of CL simulations, have been investigated, e.g., in the framework of random matrix theory~\cite{Aarts:2010gr,Mollgaard:2013qra,Bloch:2015coa,Nagata:2016alq,Schmalzbauer:2016pbg,Bloch:2017sex}
\footnote{
Random matrix theory, as well as chiral perturbation theory, was also successfully applied to QCD and its toy models at non-zero density (e.g.\ two-colour QCD or QCD in lower dimensions), predicting analytically properties like the distribution of low-lying eigenvalues of the Dirac operator and testing various issues related to the sign problem and to the physics of finite-density QCD, see, e.g., Refs.\ \cite{Stephanov:1996ki,Halasz:1997he,Halasz:1998qr,Halasz:1999gc,Kogut:2000ek,Akemann:2001bf,Akemann:2003wg,Osborn:2004rf,Akemann:2004dr,Osborn:2005ss,Bloch:2006cd,Splittorff:2006fu,Lombardo:2009aw,Akemann:2010tv,Bloch:2011jx,Splittorff:2014zca,Janssen:2015lda,Akemann:2016keq}.}.
Significant progress has been achieved employing the method of gauge cooling~\cite{Seiler:2012wz}, see, e.g., Refs.~\cite{Aarts:2013uxa,Sexty:2013ica,Makino:2015ooa,Nagata:2015uga,Nagata:2016alq,Aarts:2016qrv,Aarts:2017vrv,Sinclair:2017zhn,Cai:2019vmt}.
The most recent developments are the deformation technique~\cite{Ito:2016efb,Nagata:2018mkb} and the method of dynamical stabilisation~\cite{Aarts:2016qhx,Attanasio:2018rtq}.
For extensive reviews of the progress of the CL approach, we refer to Refs.~\cite{Aarts:2013lcm,Sexty:2014zya,Aarts:2014fsa,Sexty:2014dxa,Seiler:2017wvd,Berger:2019odf}.

\vspace*{1mm}
\noindent\textbf{Dual variables.}
Full reformulation of the partition function in terms of ``dual'' variables was also extensively used.
In this case, one looks for an exact rewriting of the partition function such that no expansion is needed.
The dual variables, such as worldlines, fermion bags, loops, dimers, plaquette occupation numbers, can have manifestly real and positive weights and MC simulations or other approaches are then possible.
The strategy is particularly well-suited in bosonic theories and Abelian gauge theories, such as the $\phi^4$ theory~\cite{Giuliani:2017qeo}, the Abelian gauge-Higgs model~\cite{Mercado:2013ola} or the Schwinger model~\cite{Gattringer:2015nea,Goschl:2017kml}.
The fermion bag approach was used, e.g., to solve massless fermionic models such as the two- and three-dimensional Thirring models~\cite{Chandrasekharan:2009wc,Chandrasekharan:2011mn,Ayyar:2017xmi} or models belonging to the Ising Gross-Neveu universality class~\cite{Huffman:2017swn}. 
The difficulty lies in including the non-Abelian interactions and in such cases so far the dual formulation can be a basis for an effective model, such as an effective Polyakov loop model~\cite{Gattringer:2011gq} or the above mentioned 3D theory with heavy quarks~\cite{Fromm:2011qi,Fromm:2012eb,Langelage:2014vpa,Glesaaen:2015vtp}.
For reviews of this thread, see, e.g., Refs.~\cite{Chandrasekharan:2008gp,Chandrasekharan:2013rpa,Gattringer:2014nxa,Gattringer:2016kco} and references therein.

\vspace*{1mm}
\noindent\textbf{Partition function modification.}
Modification of a partition function underlies also a method proposed by Doi and Tsutsui~\cite{Tsutsui:2015tua,Doi:2017gmk}.
The criterion for modification is such that the modified model does not have a sign problem and the desired observables of the original model (with a sign problem) can be related to the ones in the modified model through an identity.
This proceeds via a reweighting factor in Ref.~\cite{Tsutsui:2015tua}, which can suffer from similar constraints as the original reweighting technique, but in Ref.~\cite{Doi:2017gmk} (``multi-modification method''), this factor is avoided.
The method was applied to a Gaussian model and its analytical solution was reproduced, while the CL simulations failed to converge to the right result.
It remains to be shown that the modification approach can work also in more complicated models and ultimately in finite-density QCD.

\vspace*{1mm}
\noindent\textbf{Density of states.}
The next approach that we mention is the ``density of states'' (DOS) technique.
It belongs to the class of non-Markovian random walk methods, whose best known representatives are the multicanonical algorithm of Berg and Neuhaus~\cite{Berg:1992qua} (1992) and the Wang-Landau algorithm~\cite{Wang:2000fzi} (2000).
However, already in 1988 a similar method was proposed in the context of QCD by Gocksch~\cite{Gocksch:1988iz}.
A modification of the Wang-Landau algorithm gave rise to the recently introduced Linear Logarithmic Relaxation (LLR) method of Langfeld, Lucini and Rago~\cite{Langfeld:2012ah,Langfeld:2015fua}.
Knowing the DOS of a system, the partition function is recovered upon a one-dimensional integration thereof with the Gibbs factor.
However, determining the DOS is, obviously, highly non-trivial.
In a simple histogram approach, the simulation mostly probes irrelevant configurations with energy close to zero and the relevant ones are suppressed, yielding a large uncertainty.
The LLR technique aims at calculating, instead of the DOS directly with a histogram, its slope with respect to the energy.
Langfeld et al.\ argue that this slope can be obtained with roughly the same statistical precision for all regions of the energy, using MC simulations with an external parameter and restricted to an interval $\delta E$ around the desired energy $E$ by an introduction of a so-called window function.
The slopes can be obtained by solving a stochastic non-linear equation up to discretization errors of $\mathcal{O}(\delta E^2)$.
An alternative way to get these slopes, the functional fit approach (FFA)~\cite{Gattringer:2015lra,Giuliani:2016tlu,Giuliani:2017fss}, was also proposed and is being pursued.
Several tests of the method were performed and they are reported in Refs.~\cite{Fodor:2007vv,Bazavov:2012ex,Langfeld:2013xbf,Langfeld:2014nta,Garron:2016noc,Garron:2017fta,Lucini:2019abc} and reviewed in Refs.~\cite{Gattringer:2016kco,Langfeld:2016kty,Lucini:2019abc}.

\vspace*{1mm}
\noindent\textbf{Lefschetz thimbles.}
A relatively new approach to the sign problem is to regularize the theory on a Lefschetz thimble.
It was first proposed by Witten in 2010 \cite{Witten:2010cx,Witten:2010zr} and in the context of QCD by Cristoforetti, Di Renzo and Scorzato in 2012~\cite{Cristoforetti:2012su}. Its idea is based on the observation that the sign problem may be mildened or eliminated by choosing a different domain of integration within a complexified extension of the path integral.
This is conceptually similar to the method of saddle-point integration to compute one-dimensional oscillatory integrals, where the integration path is deformed to follow the steepest descent of the real part and keep the phase stationary.
Lefschetz thimbles are generalizations of the path of steepest descent for multi-dimensional functions and each stationary point of the function has an associated Lefschetz thimble.
The path integral can then be obtained as a sum over integrals over thimbles.
However, Cristoforetti et al.\ showed that for a wide class of theories, regularization on a single thimble is enough to obtain a theory with the same degrees of freedom and the same perturbative expansion and naive continuum limit, which, by universality, is enough to extract the desired physics.
The authors considered in detail the cases of a scalar field theory and of QCD with a chemical potential and proposed a simulation algorithm, while first results of its application were given in Ref.~\cite{Cristoforetti:2013wha} for the (3+1)-dimensional relativistic Bose gas, where the sign problem is severe in traditional MC.
Several algorithms for simulations on thimbles were proposed: based on Metropolis sampling~\cite{Mukherjee:2013aga} (with an application to a one-plaquette model) and~\cite{Alexandru:2015xva} (applied to (0+1)-dimensional finite density Thirring model), Hybrid Monte Carlo~\cite{Fujii:2013sra} (application to the finite density scalar field theory).
As argued theoretically, the thimble approach leads to a new, but mild sign problem and it was found in Ref.~\cite{Fujii:2013sra} that the residual phase factors averaged to values never smaller than 0.99, i.e.\ fully tractable by reweighting.
Explicit comparison with the CL approach gave fully compatible results.
Similar comparison was also performed in Ref.~\cite{Aarts:2013fpa} for a simple quartic model.
Regularization of quantum field theories on Lefschetz thimbles, in the framework of resurgence theory, was also considered in Ref.~\cite{Basar:2013eka,Cherman:2014ofa}.
Further theoretical and practical developments, as well as tests in various setups were reported in Refs.~\cite{Cristoforetti:2014gsa,Mukherjee:2014hsa,Aarts:2014nxa,Tanizaki:2014tua,Kanazawa:2014qma,Behtash:2015kna,DiRenzo:2015foa,Fukushima:2015qza,Tanizaki:2015rda,Fujii:2015bua,Fujii:2015vha,Tanizaki:2016xcu,Fukuma:2017fjq,Mori:2017zyl,Tanizaki:2017yow,DiRenzo:2017igr,Ulybyshev:2017hbs,Bluecher:2018sgj,Bloch:2018yhu,Ulybyshev:2019hfm,Ulybyshev:2019fte}.
It was also pointed out that there are cases were important contributions come from more than one thimble and a generalized Lefschetz thimbles method of handling such cases, termed the holomorphic gradient flow, was proposed and tested~\cite{Alexandru:2015xva,Alexandru:2015sua,Alexandru:2016ejd,Alexandru:2017oyw,Nishimura:2017vav,Alexandru:2017czx,Alexandru:2018ngw}.
Another related method, consisting in performing MC simulations on sign-optimized manifolds was also recently introduced~\cite{Alexandru:2018fqp,Alexandru:2018ddf}.
The most comprehensive reviews of the Lefschetz thimbles approach were given in Refs.~\cite{Scorzato:2015qts,Bedaque:2017epw}.
Lefschetz thimbles were also used to improve the CL dynamics~\cite{Tsutsui:2015tua}, yielding correct results in a case where the CL failed to convergence to the right result

\vspace*{1mm}
\noindent\textbf{Path optimization.}
Somewhat related to the Lefschetz thimbles method is the path optimization approach~\cite{Mori:2017pne}.
One introduces a so-called cost function, related to the seriousness of the sign problem.
By minimizing this function, one constructs an integration path where the sign problem is mildened.
The authors of Ref.~\cite{Mori:2017pne} demonstrated this technique in a toy one-dimensional integral with a severe sign problem and showed that the latter is solved, while the presence of singular points can make the Lefschetz thimbles method fail.
In subsequent publications, neural networks were employed as a tool to solve the optimization problem of minimizing the cost function in a finite-density scalar field theory~\cite{Mori:2017nwj}, the Polyakov loop extended Nambu-Jona-Lasinio model~\cite{Kashiwa:2018vxr,Kashiwa:2019lkv} and (0+1)-dimensional QCD~\cite{Ohnishi:2018jjw,Mori:2019tux}.
Ref.~\cite{Ohnishi:2018jjw} provides also a review of the path optimization technique.
Another proposal to optimize the integration path was introduced in Ref.~\cite{Bursa:2018ykf}.
In this case, the optimization does not rely on a cost function, but on a list of criteria, e.g.\ having a simple functional form, allowing for approximations in the functional form such that a mild sign problem is tolerated at the advantage of simplicity, locality, preservation of symmetry properties.
Since optimizing a fitting ansatz that describes the integration contour may be a daunting task on large lattices, the authors proposed to tune the ansatz parameters on small lattices.
An application to a one-dimensional Bose gas was also presented.
Obviously, it remains to be shown whether path optimization methods can succeed in more complicated theories, in particular in (3+1)-dimensional QCD.

\section{Tensor Networks}
\label{sec:TN}
\subsection{Tensor Networks: a new tool for classical computations}
\label{subsec:TNintro}
\noindent\textbf{What are Tensor Networks?}
Tensor network states (TNS) constitute a relatively new field of study, 
which has already left a mark in the condensed matter research.
The interest in this research area has been boosted by
 the success of the density matrix renormalization group (DMRG) algorithm~\cite{White1992,Schollwoeck2005},
for the description of one-dimensional strongly correlated systems,
which in turn can be best understood and extended in the context of tensor networks.
The theory of TNS is currently an active research topic, at the intersection of different fields, 
among them quantum information and applied mathematics, 
and strives to characterize the physical and mathematical properties of TNS.
From an applied perspective, TNS provide numerical tools to 
tackle strongly correlated systems in a non-perturbative way,
including some of the most challenging scenarios for other methods,
such as fermionic systems and (some) out-of-equilibrium dynamics.
Since the invention of DMRG,  TNS
have been successfully applied to many different problems in 
fields as diverse as condensed 
matter physics, quantum chemistry, quantum optics or classical
stochastic models.
During the last years, there has been a systematic effort to investigate the
applicability of such techniques to quantum field theories,
and in particular to lattice gauge theories.

\begin{figure}[th!]
\centering
\includegraphics[width=.75\columnwidth]{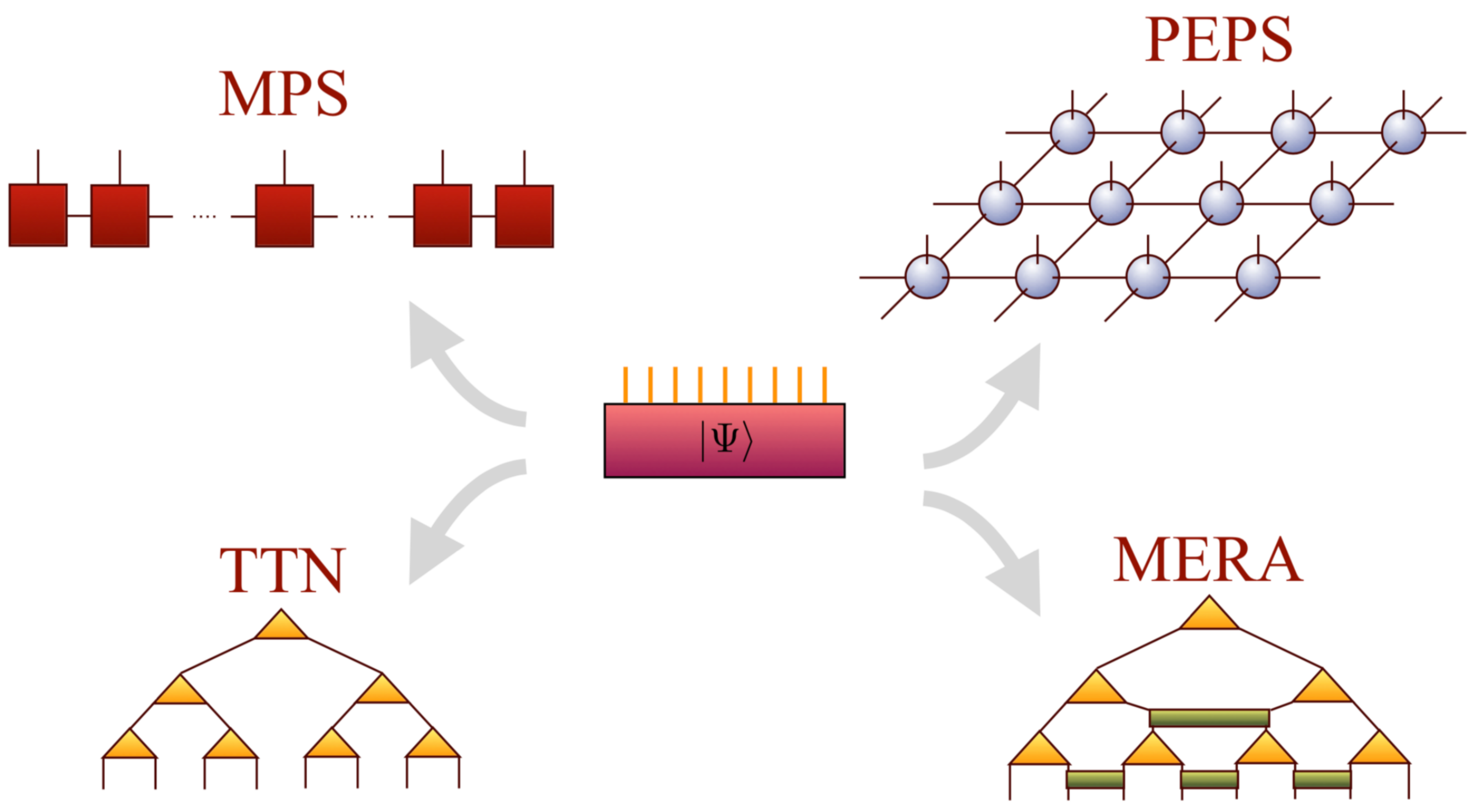}
\caption{Pictorial representation of an arbitrary state as a tensor (center) and the most significant TNS families.}
\label{fig:tns}
\end{figure}

The concept of tensor networks (TN) has emerged in various occasions and contexts in the scientific literature.
Here we focus on its introduction and 
applications in the field of quantum many body systems.
In this realm, and generally speaking, the name \emph{tensor network states} (or tensor product states) is used to designate families of ansatzes which can 
efficiently parametrize the state of such quantum systems
and fundamentally encode particular patterns of entanglement~\cite{Cirac2009rg,Verstraete2008,Schollwoeck2011,Orus2014a,Silvi2017tns}.
Alternatively, a TN (without open indices) may be used to represent
 the partition function of a certain (classical or quantum) model.
In this case, the numerical contraction of the network using 
different algorithms results in an approximation to the desired value~\cite{Ran2017tncontr}.

% name some of the progress: phase classification, refs. Maybe a bit of history in a paragraph 
% all citations need to be here
\vspace*{2mm}
\noindent\textbf{An efficient parametrization of physical states based on entanglement.}
The complexity of describing exactly the possible states of a quantum many body system scales exponentially with the 
number of its constituents. 
Indeed, for a system of $N$ sites, each of them with a Hilbert space of finite dimension $d$,
specifying an arbitrary global state 
in a tensor product basis requires $d^N$ complex coefficients.

But states of physical interest are not arbitrary ones. On the contrary, 
they include 
ground states and (low-energy) excitations, as well as 
thermal equilibrium states of models that usually involve a certain locality. 
The description of these physical states
is often much less complex than that of generic ones, 
%the full Hilbert space of the system, 
and the search for efficient parametrizations
of this  \emph{relevant corner of the Hilbert space} is a meaningful quest.
A significant property of many states in this interesting corner turns out to be  
their low entanglement.
More specifically, we can consider the entanglement entropy with respect to a 
bipartition that divides the system in two regions.
While for a random state this quantity grows extensively with the number of sites (or volume) of the smallest
region, the interesting states often satisfy an \emph{area law}. 
Namely, 
the corresponding entropy of entanglement
scales only with the size of the boundary between both regions (see Ref.~\cite{Eisert2010} for a review).\footnote{In the case of thermal ensembles, correlations take the place of entanglement in that argument.}
This property has been rigorously demonstrated only for ground states of one-dimensional local gapped Hamiltonians~\cite{Hastings2007}.
In the case of critical one-dimensional systems, which can be described by conformal field theories, 
the area law is slightly violated, with the entanglement entropy scaling with the logarithm of the size of the subregion~\cite{Calabrese2004}.
For ground states of higher-dimensional systems, the few rigorous results~\cite{Plenio2005area}, and the numerical evidence
suggest that the area law would hold for systems with finite correlation lengths, while 
for critical systems 
small (logarithmic) corrections may (but do not need to) appear~\cite{Wolf06fermions}.
A rigorous result holds as well for 
thermal equilibrium states of local Hamiltonians in any number of spatial dimensions, 
where an area law can be proven for the quantum mutual information, a measure of the total correlations in 
the mixed state~\cite{Wolf08arealaw}.
\footnote{In this case, the bound of the entropy has a prefactor proportional to the inverse temperature, therefore the area law holds 
 for any non-zero temperature.}

\vspace*{2mm} 
\noindent\textbf{TNS reproduce the entanglement scaling in physical states.}
Tensor network families precisely consist of states with well-defined entanglement scaling.
The matrix product state (MPS) ansatz 
fulfills the area law in one spatial dimension by construction, and can be used to efficiently approximate
states with that property.
And it is possible to find an MPS approximation to critical ground states
with resources that scale only polynomially with the system size.
On the other hand, the multiscale entanglement renormalization ansatz (MERA)~\cite{Vidal2007,Evenbly2009}
in one dimension
can exactly accommodate the logarithmic scaling of entanglement that occurs in critical systems~\cite{Pfeifer2009,Giovanetti2009mera},
and the same (average) scaling can be achieved by tree tensor network states~\cite{Shi2006,Tagliacozzo2009,Silvi2010tree}.
Efficient numerical algorithms exist that allow one to use these ansatzes as variational families
in order to solve a wide variety of problems.
For higher dimensions, the MPS is naturally generalized by the
projected entangled pair states (PEPS)~\cite{Verstraete2004b},
which satisfy an area law by construction, too, 
and also contain the MERA in two dimensions~\cite{Barthel2010mera}.
A generalization of the latter, called branching-MERA~\cite{Evenbly2014branching}, can instead include logarithmic 
corrections to the area law in more than one spatial dimension.

In order to visualize tensor networks, it is common to employ a graphical representation,
in which general tensors are depicted as objects with as many legs as indices, 
and a common leg between two tensors indicates
 the contraction of the corresponding index. 
For instance, a general state in the $d^N$-dimensional Hilbert space of the $N$-site system
mentioned above
can be thought of as an $N$-rank tensor, i.e.\ an array with $N$ dimensions,
each of them corresponding to the physical index of one site in the system, and
is represented graphically by a solid box with $N$ legs (see Fig.~\ref{fig:tns}).
The states in TNS families correspond to contractions of different tensor networks, 
such that the collective states are constructed from local smaller tensors. 
This structure makes them suitable candidates for numerical algorithms 
that try to simulate quantum many-body problems.

% say that so far the numerics (which we review) has been focused on 1D stms with MPS thus the rest of the section

% the MPS ansatz

\subsubsection{Matrix Product States}
\label{ssec:mps}

The MPS family is the paradigmatic tensor network, best understood theoretically, and most frequently employed in numerical calculations.
The ansatz underlies the unparalleled success of the DMRG algorithm 
~\cite{White1992,Kluemper1993,Oestlund1995,Schollwoeck2005,Vidal2003,Verstraete2004,McCulloch2007,McCulloch2008idmrg}
in the solution of one-dimensional strongly correlated lattice problems 
(see~\cite{Schollwoeck2011} for a comprehensive review).

For a quantum system composed of $N$ sites, each of them with a Hilbert space of finite dimension $d$ (termed physical dimension), an MPS is a state of the form
\beq
|\Psi\rangle=\sum_{i_1,\ldots i_N=1}^d \mathrm{tr}\left ( A_1^{i_1} A_2^{i_2}\ldots A_N^{i_N} \right ) |i_1 i_2 \ldots i_N\rangle,
\label{eq:mps}
\eeq
being $\{\ket{i_k}\}_{i=1,\ldots d}$ a basis for the $k$-th site.
Each $A_k^{i_k}$ is a $D\times D$ matrix, so that the total number of parameters scales as $N d D^2$. 
In the case of open boundary conditions, the first and last matrices, $A_1^{i_1}$ and $A_N^{i_N}$, can be 
reduced to $D$-dimensional vectors. 
With all the tensors $A_k=A$ identical, the ansatz (called in this case uniform MPS or uMPS) can be used to describe a system in the
thermodynamic limit.\footnote{This translationally invariant ansatz can naturally be generalized to consist of a repeated unit cell with a finite number of distinct tensors.} 

The parameter $D$, called the bond dimension, determines 
the number of variational parameters, as well as 
the maximum amount of entanglement that the ansatz can describe.
For a partition of the system in two halves, the entanglement entropy is 
upper bounded by $S(\rho_{N/2})\leq \log D$  in the case of open boundary conditions (or twice this number for
periodic ones),  explicitly manifesting the entanglement area law for MPS.
For $D=1$, the MPS reduce to product states. 
The tensors for any larger $D$ include the smaller ones as particular cases, so 
that MPS establish a hierarchy of entanglement. 
Moreover, they are a complete family, and any state in the Hilbert space can be exactly written as 
an MPS with bond dimension $D \leq d^N$.

% properties, symmetries

For a state in an MPS form, most of the physical properties can be easily accessed, since expectation values 
of local observables or products of them 
(and actually of any operator with a matrix product structure)
 can be computed very efficiently, with a computational
cost that scales only linearly in the system size.

The formal properties of MPS as a family have been extensively studied for uniform and 
non-uniform MPS
~\cite{Fannes1992,PerezGarcia2007,Haegeman2014geometry},
and a good theoretical understanding of MPS is now available.
Especially relevant is the role of symmetries of the local tensors.
States with well-defined behaviour under global symmetries can be parametrized
in terms of symmetric tensors, namely tensors with well-defined transformation
properties under symmetry operations that act on their physical and virtual indices~\cite{PerezGarcia2008sym,Singh2010,PerezGarcia2010njp}.
This has crucial theoretical implications, 
allowing, for instance, the classification of symmetry-protected topological 
phases in one spatial dimension~\cite{Chen2011spt,Schuch2011}.
The concept, which is not restricted to MPS, can be extended to construct 
gauge-invariant TNS~\cite{Tagliacozzo2014,Haegeman2015,Zohar2015c,Zohar2015b,Kull2017gauge}.
From a practical perspective, symmetric tensors can be used 
in the numerical algorithms described in the following paragraphs, in order to 
reduce the computational cost in cases where the target states are symmetric (for a recent review, see~Ref.\ \cite{Hubig2018sym}).

% algorithms

\subsubsection{Main numerical algorithms}
\label{ssec:algorithms}

During the past decades,
multiple numerical algorithms have been developed that 
exploit the TNS potential in very different problems.
For the sake of completeness, we review here the fundamental
aspects of the main techniques employed in the context 
of lattice gauge theories,
and we refer the reader to the 
 literature on the subject
for detailed technical descriptions of the algorithms (see, e.g., Refs.\ \cite{Verstraete2008,Schollwoeck2011,Lubasch2014peps,Silvi2017tns,Haegeman2017rev,Montangero2018book,Hubig2018ipeps,Vanderstraeten2018tangent}). 
Most of the algorithms we describe here are easy to implement,
and there exist several libraries available to the public that currently provide most of the functionality~\cite{alps,itensor,openmps,block}.

\vspace*{2mm}
\noindent\textbf{Variational search for ground (and excited) states}.
Approximating the ground state of quantum many body problems is one of the fundamental applications
of TNS in general.
Using MPS as a variational family, the goal is to minimize the energy over MPS with fixed bond dimension.
For a finite system, a successful strategy 
 is to sequentially optimize each of the tensors in the ansatz while 
keeping the remaining ones fixed. 
With this constraint, the problem reduces to a sequence
of local optimizations of the Ritz form, whose solution can be found
using conventional eigenvalue solvers.
The procedure is iterated for all the tensors, sweeping over the whole chain 
from left to right and backwards, until convergence.
Up to some technical details~\cite{McCulloch2007}, this is the
fundamental strategy of the finite DMRG method
~\cite{White1992,Schollwoeck2005,Vidal2003,Verstraete2004}.
Since the energy of the ansatz can only decrease after each iteration, 
the algorithm is guaranteed to converge (although it could do it to
a local minimum). %~\cite{White1993local,White2005,Hubig2015single}
Typically, the solution for a given bond dimension is used to initialize the ansatz for a larger $D$ and 
the calculation is repeated until enough precision is achieved (or 
the available computational resources are exhausted).
The method can be easily extended to compute also low-energy excited states,
 by simply searching for the lowest-energy state that is 
orthogonal to previously computed levels~\cite{Verstraete2008,Schollwoeck2011}.

A crucial property of MPS calculations is their computational efficiency.
The computational cost of the variational algorithm described above scales as ${O}(D^3)$ in the
case of an MPS with open (free) boundary conditions. For periodic MPS, the cost scales as 
${O}(D^5)$, although approximate versions of the algorithm exists that still scale
with the cube of the bond dimension~\cite{Pippan2010periodic,Pirvu2011pbc} (with an additional overhead).
However, for most problems, it turns out to be advantageous to work with open
boundary conditions and apply suitable finite size scaling 
analysis to extract the bulk properties.
The efficiency of these algorithms makes it possible to optimize MPS with very large bond dimension, 
and has helped DMRG-like methods becoming one of the most powerful tools
for the numerical investigation of condensed matter problems.
For instance, MPS have been successfully applied to describe states that violate the area law, including
critical systems, long range interactions and even two-dimensional 
problems~\cite{Yan2011spinliquid,Depenbrock2012spinliquid,Leblanc2015hubbard,Saadatmand2018cyl}.

Alternatively, uniform MPS can be used to directly target systems in the  thermodynamic limit, avoiding finite size effects. 
The most popular methods to find ground state approximations with
translationally invariant MPS involve imaginary time evolution as described next, 
but it is also possible to optimize the uMPS ansatz variationally~\cite{Zauner2018vuMPS}.
Excitations over a uMPS ground state, specified by a single tensor, can be parametrized by one (or few) different tensors
that are optimized variationally~\cite{Haegeman2011,Haegeman2013post}.
By taking appropriate superpositions over all translations of such construction, 
it is possible to construct states of well-defined momentum.

\vspace*{2mm}
\noindent\textbf{Time evolution algorithms}.
One of the most challenging applications of MPS is the simulation of 
dynamical scenarios.
Standard TNS techniques can be used to evolve an initial MPS state
under a certain Hamiltonian.
The most traditional algorithms~\cite{Vidal2004,Verstraete2004a,GarciaRipoll2006}
 use a Suzuki-Trotter approximation for the time evolution operator.
 The total time is discretized into a number of small steps,
 and the evolution operator for each of them is approximated by a product of terms which can be 
 applied within the MPS framework. 
 In general, each such application may transform the initial state into an MPS with larger bond dimension,
 so the algorithms involve a \emph{truncation} step in which the resulting state is approximated
 with a maximum bond dimension.
 Errors originate from this truncation and from the size of the Trotter step~\cite{Schollwoeck2006}.
The basic algorithms require a certain locality in the Hamiltonian, in order to approximate
the evolution operator in an efficient manner, but alternative algorithms have also been proposed that
can deal with long-range Hamiltonians~\cite{Haegeman2011,Zaletel2015long}.
%One of them is based on a time dependent variational principle~\cite{Haegeman2011}
%which performs the evolution using the concept of the tangent space to the manifold of MPS.

Typically, when evolving systems far from equilibrium, the truncation error becomes 
the limiting factor quite fast, a feature that is well understood in terms of the entanglement 
growth in the physical state.
If the entanglement of the state grows linearly in time,
as may happen in global quenches~\cite{Calabrese2005}, an exponentially growing bond dimension
will be required to maintain constant precision, which rapidly exhausts the 
available computational resources~ ~\cite{Osborne2006,Schuch2008a}.
On the other hand, these algorithms provide robust results for a variety of 
scenarios, such as close to adiabatic evolutions, or local quenches~\cite{Paeckel2019tevolRev}.
Exploring the power and limitations of TNS for out-of-equilibrium dynamics is 
currently an open research question~\cite{Dubail2017mpo,Leviatan2017,White2018therm,Surace2019}.

All the evolution algorithms mentioned above can be used to evolve an initial state in imaginary time.
Imaginary time evolution results, in the limit of infinite time, in an approximate projection onto the 
ground state of the Hamiltonian, since
\begin{equation}
 \lim_{\tau\to\infty} e^{-\tau H} \ket{\Phi_0} \propto \bra{E_0} \Phi_0\rangle \ket{E_0},
\end{equation}
assuming the initial state had non-vanishing overlap $\bra{E_0} \Phi_0\rangle$ with said ground state.
This is the most widely used strategy for finding infinite MPS ground 
states~\cite{Vidal2007infinite,Orus2008itebd,Haegeman2011}.

\vspace*{2mm}
\noindent\textbf{Thermal equilibrium states}.
Tensor networks can be used to describe not only pure states, but also operators.
In particular, a matrix product operator (MPO) is an operator which, in a tensor product basis,
has coefficients with an MPS structure.
The MPO description turns out to be exact for local Hamiltonians, and can be used to approximate 
long-range ones, as well as evolution operators.
This provides very efficient ways to deal
with the variational and evolution algorithms~\cite{McCulloch2007,Pirvu2010a}.

The MPO can be used as an ansatz for mixed states, 
such as equilibrium states at finite temperature or
the states of open quantum systems, and is then sometimes called matrix product density operator (MPDO)~\cite{Verstraete2004a,Zwolak2004}.
In the case of thermal states, it has actually been demonstrated that thermal states of local Hamiltonians
can be efficiently approximated by MPO or their generalizations in higher dimensions~\cite{Hastings2006,Molnar2015}.

In order for an MPO ansatz to represent a valid mixed state, it has to satisfy  the physical constraints of normalization, 
hermiticity and positivity. Being a global property of the operator, positivity in particular is not easy to impose at the level of the
local tensors.
It is possible however to consider a restricted form of MPO, called the \emph{purification} ansatz, which guarantees positivity by imposing
a positive structure to each of the tensors~\cite{Verstraete2004a}.
The construction corresponds to an MPS form for a purification of the mixed state~\cite{Nielsen2004}, 
i.e.\ a pure state of a larger system, which results in the desired mixed state when tracing out the extra degrees of freedom.
The relation between the efficiency of the generic MPO and the purification descriptions
is not trivial~\cite{delasCuevas2013,Kliesch2014mpdo,delasCuevas2016}.
In practice, a purification MPO can be used in the algorithms,
although involving, in general, a higher computational cost~\cite{Lubasch2014peps,Werner2016puri}.

In the particular case of thermal equilibrium states, the purification mentioned above 
is equivalent to a thermofield doubling, which can be constructed by 
constructing a maximally entangled state between the physical system and 
an ancillary copy, and then applying the 
exponential  operator $e^{-{\beta H/2}}$ to the physical indices,
where $\beta$ is the  inverse temperature.
The application of such exponential is formally identical to the imaginary time evolution
and can thus be performed by the standard TNS algorithms described above.
This yields the most popular TN algorithm for approximating thermal states of local Hamiltonians, 
although alternative algorithms have been developed to compute thermal states or their properties
~\cite{White2009metts,Chen2017series,August2018mpo}.

%\paragraph{MPS with explicit symmetry}

%\cc{MC}{Should add a paragraph about gauge symmetric MPS.  Here?}

\subsubsection{Other TNS families: TTN and MERA}

Other families of TN have been defined that have a different entanglement content.
In particular, the multiscale entanglement renormalization ansatz (MERA)~\cite{Vidal2007,Evenbly2009},
and its subfamily,  tree tensor networks (TTN)~\cite{Shi2006,Silvi2010tree},
were introduced with the motivation to describe one-dimensional critical systems, 
which violate the area law logarithmically.
For these families, the TN has a hierarchical structure, and realizes some kind of real-space coarse graining.
By restricting the constituting tensors to be isometric or unitary, 
local expectation values can be efficiently contracted for both ansatzes, which allows their use
in numerical algorithms.
The cost of the contraction, which depends on the particular coarse-graining chosen, 
scales however with a higher power than in the MPS case.
%~\cite{Gerster2014ttn} % unconstrained TTN can be more efficient than PBC

In one spatial dimension, TTN and MERA can violate the area law logarithmically and 
give rise to critical correlations, which decay as a power law (in the case of TTN,
this holds for the average over sites~\cite{Silvi2010tree}).
They can then be used to describe critical points.
In two dimensions, however, it was proven that MERA are subfamily 
of PEPS~\cite{Barthel2010mera}, and as a consequence, cannot violate the area law.\footnote{Nevertheless,
other tensor network structures can be defined with more entanglement~\cite{Evenbly2014branching}.}
Nevertheless, because of the isometric properties, contractions can be done efficiently,
and optimization algorithms
have been applied to two-dimensional problems~\cite{Tagliacozzo2009,Evenbly2009mera2d}.

\subsubsection{Higher dimensions: PEPS}

The natural generalization of MPS to higher dimensions are projected entangled pair states (PEPS)~\cite{Verstraete2004b},
which also satisfy the area law by construction but, different from their one-dimensional counterparts, 
can account for power-law decaying correlations even with a very small bond dimension~\cite{Verstraete2006a}.

Most of the algorithms described in the previous paragraphs can be generalized to work with PEPS, 
both for finite and for infinite systems, but with some important modifications.
Since, in general, it is not possible to contract a PEPS efficiently,
the norms and expectation values involved in the algorithms need to be approximated.
Moreover, the computational cost of the corresponding calculations is considerably larger 
than in the one-dimensional case (e.g.\ the $O(D^3)$ scaling of MPS is substituted by $O(D^{10})$ for 
PEPS on a square 2D lattice).
From the numerical point of view, the algorithms that work with the PEPS ansatz~\cite{Verstraete2008,Lubasch2014peps,Hubig2018ipeps}
are, thus, more challenging than their MPS counterparts, and the results that can be found in the literature are still limited.
Nevertheless, the active research in the subject and the invention and development
of improved techniques~\cite{Corboz2016,Corboz2016a,Vanderstraeten2016a,Czarnik2019time}
has allowed the method to provide competitive results, mainly in 
the context of condensed matter problems (e.g.~\cite{Rader2018prx,Corboz2018prx}).

\subsubsection{Tensor Renormalization}

In the previous sections, we discussed how TNS can be used as ansatzes for quantum many-body states. But tensor networks can also describe observables. For instance, the partition function of a quantum model in 1+1  or a statistical model in 2 dimensions can often be (exactly or via some approximation) expressed as a tensor network with no open indices, such that the complete contraction returns a scalar. In general~\cite{Schuch2007complexity}, contracting an arbitrary TN is a computationally intractable problem (in the \#P  complexity class), but, as it happens for states, the physically interesting instances often admit efficient approximations.

The tensor renormalization group (TRG) algorithm, introduced in 2006 by Levin and Nave~\cite{Levin2007} is the basis 
of a whole family of algorithms that approximate these contractions by performing a real-space coarse graining of the network.
In the original TRG algorithm, local tensors are truncated using SVD, and exact contractions of various terms (e.g.\ around a plaquette) yield the renormalized tensors that conform the coarse-grained TN.
This procedure is iterated until the tensors converge, at which point the partition function in the thermodynamic limit is obtained as the trace of the converged tensor.
Different variations of the algorithm have been proposed, to take into account the effect of the whole network when deciding the truncation (second order renormalization group or SRG~\cite{Xie2009}),
or to extend the applicability to TN in any number of dimensions, using the higher order SVD (higher order tensor renormalization group, or HOTRG~\cite{Xie2012}). \footnote{The computational cost, nevertheless, scales differently depending on the dimensionality of the system}
Further extensions of the method include its application to fermionic systems, using Grassmann variables~\cite{Gu2010gtrg}, and to non-compact bosonic fields~ \cite{Campos2019boson},
and excellent numerical results have been reported for a variety of systems.

The best-known limitation of TRG and its variations mentioned above is the inability to disentangle some types of short-time correlations, exemplified by the corner double line (CDL) tensors.
To avoid this difficulty, alternative procedures have been proposed that try to remove such short range correlations at each iteration, via the introduction of disentanglers, in the spirit of MERA (as in the tensor network renormalization, TNR, introduced by Evenbly and Vidal~\cite{Evenbly2015b}), or the optimization of the tensors decomposition (as in the loop-TRG by Yang \etal~\cite{yang2017loop}). 

\subsection{A successful approach for 1+1 dimensional LGT}
\label{subsec:TNresults}
The feasibility of TNS techniques for the study of LGT in one spatial dimension has been 
demonstrated beyond doubt by a number of studies that appeared in the last years.
They employed all the assortment of TNS algorithms and focused on models with different features.
We report here on this body of evidence, which is crucial to  justify the further effort required for higher-dimensional studies.

Since one of the most important aims of the whole programme is to eventually apply the TN formalism to QCD, we begin our discussion with the Abelian Schwinger model, the simplest lattice gauge theory with matter and the model that received most attention from the point of view of TN applications.
Then, we move on to the simplest non-Abelian theory, based on the $SU(2)$ gauge group.
Further, we review the progress in other lattice field theories, such as the scalar field theory, the $Z_2$ lattice gauge theory, $O(N)$ sigma models, the Thirring model and Abelian and non-Abelian Higgs models.

\subsubsection{The Schwinger model}
\label{sec:schwinger}
The Schwinger model \cite{Schwinger:1962tp} is one of the most widely used toy models of QCD, in particular for testing new lattice techniques or algorithms.
In view of the fact that very many studies reviewed here considered this theory, we include a brief reminder of its formulation in the one-flavour case.
The Schwinger model is based on the Abelian gauge group $U(1)$, i.e.\ it is quantum electrodynamics in 1+1 dimensions (QED$_2$).
Its wide usage results from its simplicity, but also from the fact that it shares certain properties with QCD in 3+1 dimensions, most notably the non-perturbative generation of a mass gap, fermion confinement and chiral symmetry breaking.
In the massless limit, it is analytically solvable.
As we discuss below, this provided the possibility of precision tests of TN methods, leading to results unprecedentedly close to the analytical solution.
We start with Hamiltonian tensor network methods (variants of MPS) and later, we move on to Lagrangian methods, such as TRG.

MPS is a Hamiltonian approach, hence the QED Lagrangian needs to be subjected to a Legendre transform. Then, the continuum Hamiltonian needs to be discretized.
Since the naive discretization leads to a doubling of fermion species, the MPS studies employed the staggered discretization of Kogut and Susskind~\cite{Kogut1975}.
The fermionic (matter) fields are put on lattice sites and the bosonic (gauge) fields on the links connecting the sites.
For numerical implementation, it is convenient in one spatial dimension to perform a Jordan-Wigner transformation \cite{Jordan:1928wi} to switch from fermionic fields to spin variables represented by the standard Pauli matrices.
Moreover, the gauge invariance, expressed by the Gauss law, makes it possible to integrate out all of the gauge fields, leaving only the incoming flux representing the background electric field.
After all these steps, the (dimensionless) Hamiltonian for an $N$-site system with open boundaries, in the temporal gauge, is cast into the following form:
\begin{align}
H=& x\sum_{n=0}^{N-2} \left [ \sigma_n^+\sigma_{n+1}^- +  \sigma_n^-\sigma_{n+1}^+ \right ]+\frac{\mu}{2}\sum_{n=1}^N \left [ 1 + (-1)^n \sigma_n^z \right ]
\nonumber \\
&+\sum_{n=0}^{N-2} \left [ \ell +\frac{1}{2}\sum_{k=0}^n ((-1)^k+\sigma_k^{z})\right ]^2,
\label{eq:Hsigma}
\end{align}
where: $x=1/(ag)^2$, $a$ -- lattice spacing, $g$ -- coupling, $\mu = 2\sqrt{x}m/g$, $m$ -- fermion mass, $\ell$ -- left boundary electric field (background field).
The Schwinger model Hamiltonian, thus, is formally a spin model in an anisotropic (site-dependent) magnetic field and containing long-range interactions of the form $\sigma^z_n \sigma^z_{n'}$ (with arbitrary site indices $n,n'$), generated by the gauge symmetry.

We begin our review of tensor network results for this model with a study from 2002 within the DMRG method, by Byrnes et al.\ \cite{Byrnes2002}.
This was performed yet before the advent of the TN formalism.
The authors demonstrated for the first time the usage of DMRG for an LGT, improving significantly upon the previous most precise results from the strong coupling expansion, by two-three orders of magnitude for quantities like the vector mass gap.
Another purpose of the paper was to explore the $\theta$-vacuum physics of the model in a background electric field.
For a particular value of the background field, $\ell/g=1/2$ ($\theta=\pi$) in our notation, Coleman predicted in a seminal paper \cite{Coleman:1976uz} the occurence of a special phenomenon of quarks appearing as ``half-asymptotic'' particles and the presence of a phase transition in the quark mass $m/g$.
Using DMRG, the authors confirmed this picture and determined the phase transition point with a precision better than one part per mille.
Note that such study within an MC framework encounters a complex action problem, totally absent in the DMRG Hamiltonian language.

The first investigation fully in the MPS formalism was performed by Ba\~nuls et al.\ in 2013 \cite{Banuls2013}.
The aim of this paper was to extract the ground state and the vector and scalar mass gaps of the massless and massive one-flavour theory.
The excitations of the model were targeted by constructing projectors onto subspaces orthogonal to the ones of the ground state and the lower-lying excitations.
Broad regimes of parameters were explored to robustly establish the connection between the lattice model and the continuum theory.
In order to obtain predictions for the latter, several extrapolations are necessary.
First, one needs to remove the effect of simulating with a finite bond dimension by checking results from a few sufficiently large values of $D$.
When convergence is observed, one can use the results from two-three largest bond dimensions to extrapolate to the infinite-$D$ limit.
The error from this extrapolation can serve as an estimate of the uncertainty introduced by the MPS technique.
Note that within this uncertainty, the MPS result should agree with the result from exact diagonalization (ED), which was explicitly checked for small system sizes where ED is still practical.
Second, the MPS results in the formulation of Ref.\ \cite{Banuls2013} are subject to finite volume effects, which need to be extrapolated away by a fit including a few large enough system sizes.
Finally, the infinite-volume results corresponding to a finite lattice spacing need to have the discretization effects removed via a continuum extrapolation.
It is interesting to observe that the precision reachable with MPS is such that the latter extrapolation is sensitive not only to the leading terms in the lattice spacing expansion, but also to quadratic and cubic terms, something typically not happening in MC simulations with commonly reached statistical errors.
The final precision of the mass gaps extracted in Ref.\ \cite{Banuls2013} amounted to $\mathcal{O}(10^{-4})$/$\mathcal{O}(10^{-3})$ for the vector/scalar mass gap, in accordance with the analytical results for the massless case and the less precise ealier numerical results\footnote{The most precise results for these mass gaps, with relative errors of $\mathcal{O}(10^{-9})$/$\mathcal{O}(10^{-11})$, were obtained in Refs.\ \cite{Cichy2013,Szyniszewski:2014uta} using the strong coupling expansion combined with ED. However, this method was not found to be prospective for non-spectral observables.}.

In a parallel investigation published in 2014, Buyens et al.\ \cite{Buyens2013} performed a complementary study of the spectral properties of the one-flavour Schwinger model.
While Ref.\ \cite{Banuls2013} relied on finite MPS, Buyens et al.\ considered the formulation of MPS in the thermodynamic limit (uMPS).
Moreover, they restricted the family of MPS to explicitly gauge-invariant ones and used time-dependent variational principle (TDVP) \cite{Haegeman2011} to find ground states.
Obviously, the details of the formulation and the algorithm for finding ground states can not matter for the final result, as explicitly demonstrated by the full consistency of the results obtained in the two setups.
Apart from the first vector and scalar mass gaps, Buyens et al.\ also accessed the heavy vector boson, obtaining first estimates of its mass.
Moreover, they investigated non-equilibrium dynamics, using real-time TDVP, in the situation of applying a uniform electric field on the ground state, which leads to the Schwinger pair creation.
Consistently with expectations, they observed plasma oscillations damped in time as manifestation of thermalization.
As mentioned above, real-time evolution necessarily involves growth of entanglement and hence, the fixed-$D$ MPS ansatz, at some point, ceases providing good description of the state.
The authors checked this by comparing results from two values of the bond dimension and found the real time when their largest $D$ is no longer good enough.
Again as expected, this happens before new equilibrium is reached.
Nevertheless, even the restricted range of real times that can be reached with MPS provides important information about the system and this information cannot be obtained with standard MC formulations.
Further results for the spectrum, in the presence of a background electric field, where shown in Ref.\ \cite{Buyens2015a}.

\begin{figure}[h!]
\begin{center}
\hspace*{-1mm}
\includegraphics[width=0.49\textwidth]{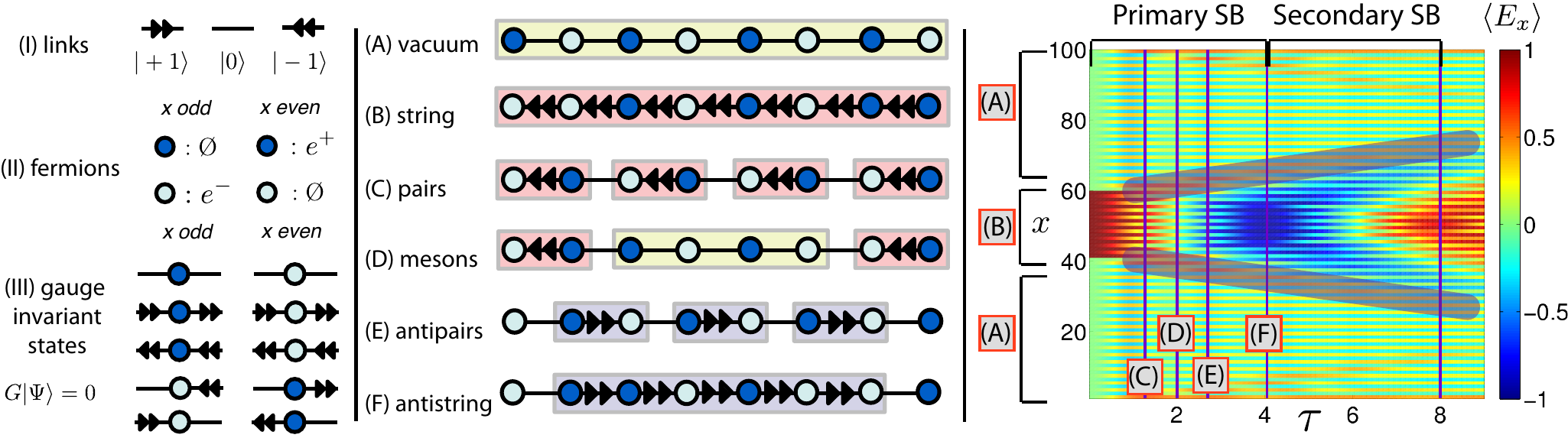}
\vspace*{-5mm}
\end{center}  
\caption{(left and middle) Cartoon illustration of building blocks and the states of the Schwinger model string breaking simulation of Ref.~\cite{Pichler2015}. (right) Typical result of the real-time simulation of dynamical string breaking. The string is initially embedded in a vacuum state and in the course of evolution it undergoes primary string breaking, leading to the formation of an antistring. The latter breaks in the secondary string breaking phase. 
Source: Ref.~\cite{Pichler2015}, reprinted with permission by the authors (article published under the
terms of the Creative Commons Attribution 3.0 license).}
\label{fig:Pichler}
\end{figure}

Another real-time study of the Schwinger model was performed in 2015 by Pichler et al.\ \cite{Pichler2015}.
The focus was on the dynamics of string breaking for the Schwinger model (with staggered fermions) in the quantum link formulation, where the gauge fields are represented by spin-1 operators and hence, their Hilbert space is finite-dimensional.
The authors used the MPS ansatz with gauge symmetry encoded in the tensors, i.e.\ the TN state consists of an MPS part contracted with an MPO that imposes the gauge invariance.
The real-time evolution was computed by means of a Suzuki-Trotter decomposition of the evolution operator.
An example of the results for the dynamical string breaking is shown in the right part of Fig.~\ref{fig:Pichler}.
The initial setup included a string embedded in a vacuum.
The string then undergoes primary string breaking and an antistring (with opposite sign of the electric flux) forms and is broken in the phase of secondary string breaking.
The authors also computed the entanglement entropy and showed that string breaking is intimately related to entanglement propagation, discussing also potential quantum simulation realizations.
In addition, they considered scattering processes between mesonic bound states, showing again how this affects the entanglement properties.

\vspace*{2mm}

\begin{figure}[h!]
%\begin{center}
\hspace*{-3mm}
\includegraphics[width=0.27\textwidth]{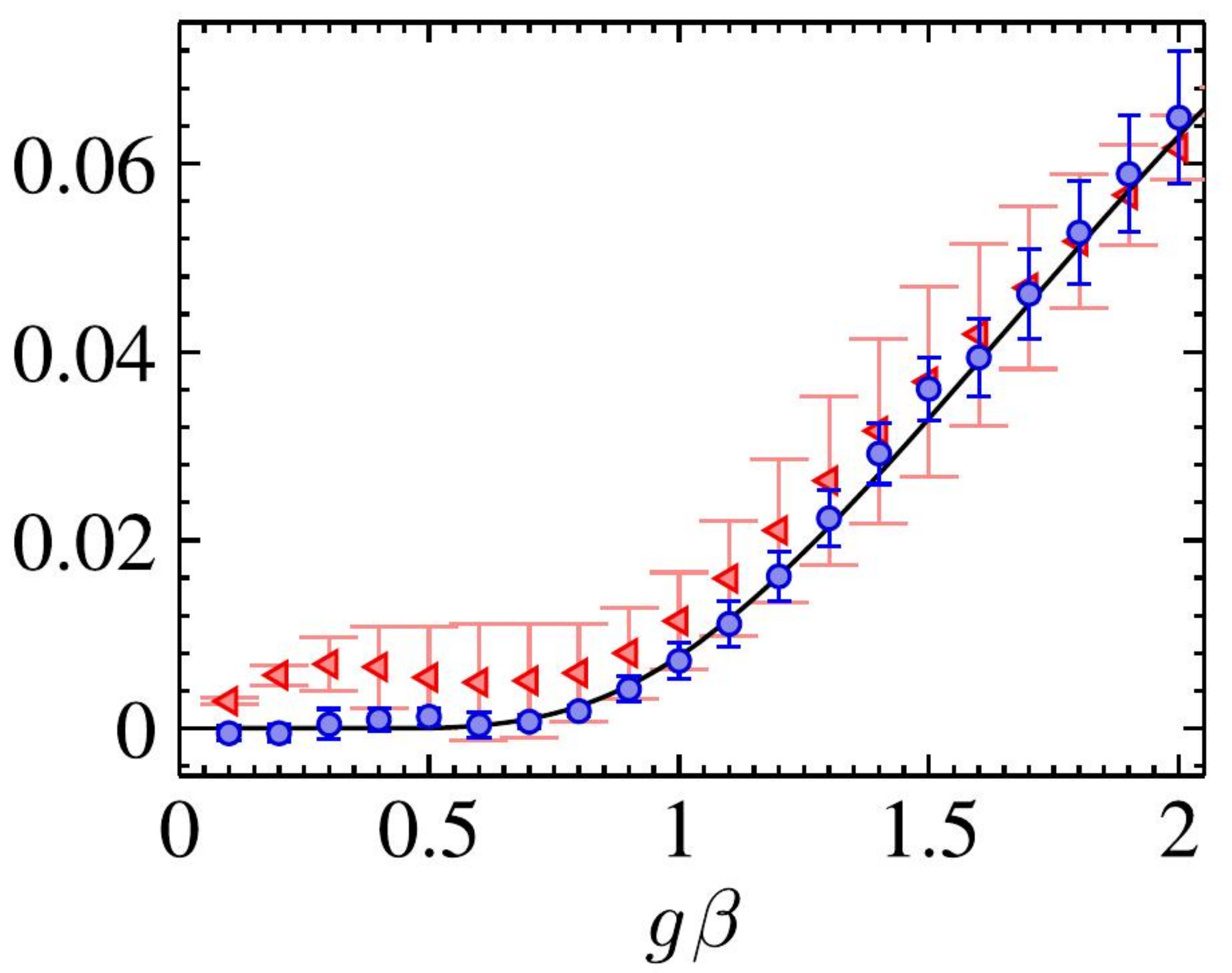}
\hspace*{-1mm}
\includegraphics[width=0.2\textwidth]{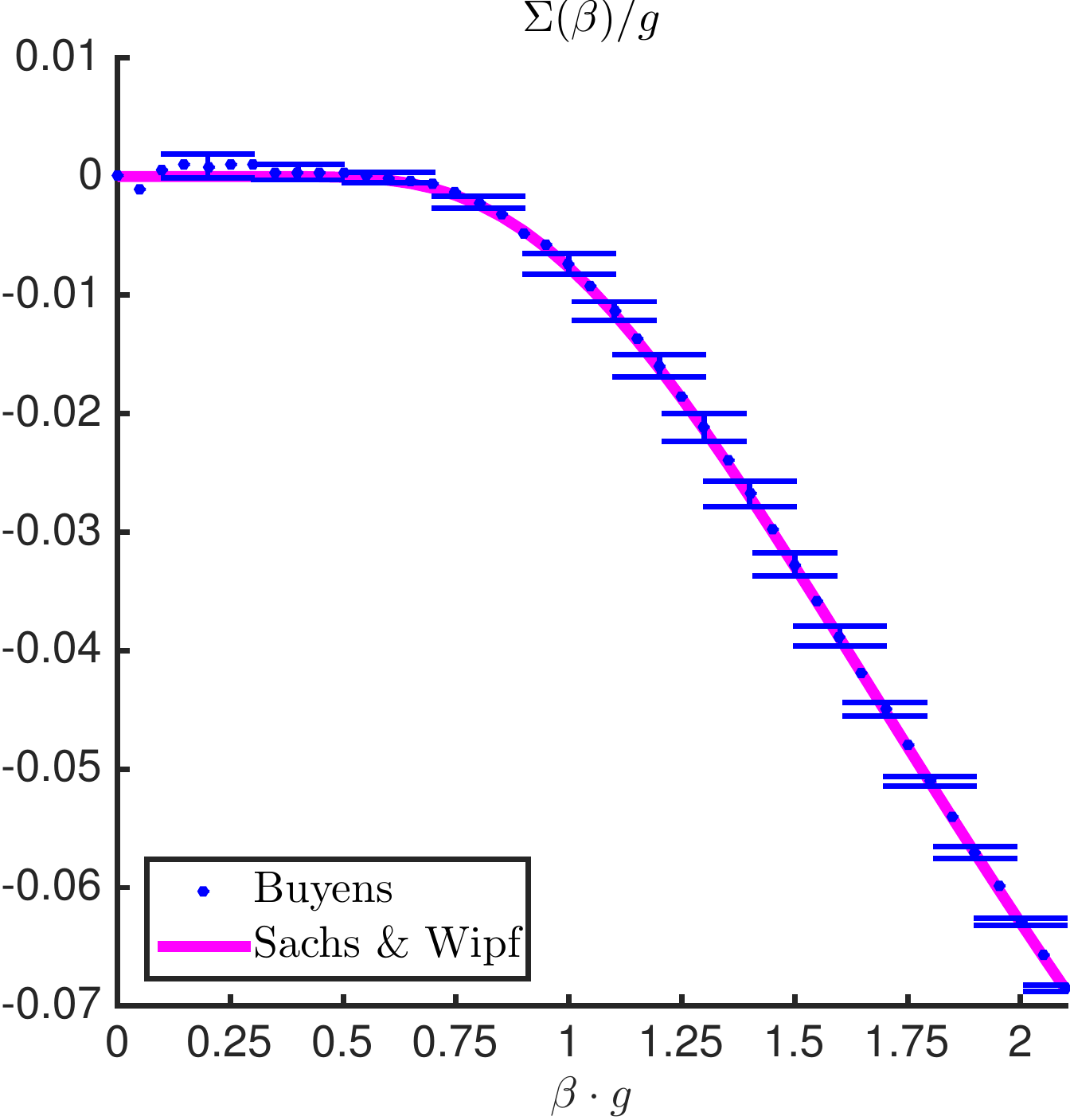}
\vspace*{-2mm}
%\end{center}  
\caption{Inverse temperature dependence of the chiral condensate in the Schwinger model. 
(left) Data from Ba\~nuls et al.\ \cite{Banuls2015}. The red (blue) data points correspond to results from the Taylor expansion (flux truncation), see text for more details, the black line is the analytical result from Sachs and Wipf \cite{Sachs:1991en}.
(right) Data from Buyens et al.\ \cite{Buyens2016}. The blue data points are the results from MPS and the magenta line is the analytical result \cite{Sachs:1991en}.
Note the right plot uses the opposite convention for the sign of the condensate.
Source: Refs.~\cite{Banuls2015,Buyens2016}, reprinted with permission by the authors and the American Physical Society.}
\label{fig:condT}
\end{figure}

A follow-up investigation by the group of Ba\~nuls et al.\ concentrated further on the feasibility study of thermal evolution with TN methods \cite{Saito2014,Banuls2015,Saito2015,Banuls2016}.
Chiral condensate was computed for a wide range of temperatures\footnote{Precise zero-temperature results for the chiral condensate at zero and non-zero fermion masses were also reported -- by Ba\~nuls et al.\ \cite{Banuls2013a,Banuls2016} and by Buyens et al.\ \cite{Buyens2014}.}.
Technically, this involves calculating the MPO approximation to the thermal density operator, $\exp(-\beta H)$, using imaginary time evolution acting on the identity matrix.
Since different terms of the Hamiltonian do not commute with the full operator, one uses a Trotter expansion, dividing the imaginary time into small steps $\delta$.
As it turns out, the hopping and mass terms can be written exactly as an MPO of dimension four.
However, the long-range gauge terms require MPO bond dimension of $N+1$, thus making the computation impractical.
The initial way of treating the gauge part consisted in using a Taylor expansion \cite{Saito2014,Banuls2015,Saito2015}.
The Trotter and Taylor expansions lead to the need of an additional extrapolation of the result, to the limit $\delta=0$.
The full procedure, i.e.\ extrapolations to $\delta=0$, infinite $D$, infinite volume limit and continuum limit led to good agreement with the analytical calculation for the massless model \cite{Sachs:1991en} for temperatures $T/g\lesssim2$ (all they way down to $T/g\approx1/6$ where the condensate value becomes consistent with the zero-temperature result), but a sizable tension for higher temperatures.
The reason for the latter was identified to be enhanced cut-off effects.
To deal with them, smaller lattice spacings needed to be approached.
This was done using another way of treating the gauge term, involving a truncation of the electric flux at some maximum value, $L_{\rm cut}$ \cite{Banuls2015}.
This is, potentially, another source of systematic uncertainty, but numerical evidence coming from investigating the $L_{\rm cut}$ dependence showed that values of $L_{\rm cut}=\mathcal{O}(8)$ are enough to obtain saturation of this dependence.
Moreover, the leading $\delta$-dependence becomes $\mathcal{O}(\delta^2)$ (from the second order Trotter expansion), thus enabling simulations at larger values of $\delta$.
Finally, full agreement with the analytical curve was reached, moreover with up to an order of magnitude smaller uncertainties than when using the Taylor expansion.
This is demonstrated in the left plot of Fig.\ \ref{fig:condT}, where the red (blue) data points correspond to results from the Taylor expansion (flux truncation).
Physically, the results signify smooth restoration of chiral symmetry in the limit of infinite temperature.
The massive Schwinger model was investigated in Ref.\ \cite{Banuls2016}, following again the truncated flux approach.
For this case, there is no exact solution and the only analytical predictions come from bosonization combined with a generalized Hartree-Fock approximation, valid for small fermion masses \cite{Hosotani1998}.
Thus, the agreement between the \emph{ab initio} treatment via tensor networks and the approximated solution was found to be only qualitative.

\vspace*{2mm}

The temperature dependence of the chiral condensate was also computed by Buyens et al.\ \cite{Buyens2016}.
Similarly to the spectrum calculation described above, this was also done directly in the thermodynamic limit, using the infinite version of time-evolving block decimation (TEBD) \cite{Vidal2007infinite} and gauge invariance was imposed at the level of the variational ansatz, i.e.\ when constructing the MPS purification of the MPO ansatz for the thermal density operator.
Trotter expansion was also used, combined with a truncation in the Schmidt spectrum, corresponding effectively to electric flux truncation.
This different setup of the calculation led to results compatible with Ref.\ \cite{Banuls2015}, shown in the right panel of Fig.\ \ref{fig:condT} for the massless case.
Similar consistency of the results was found also for the massive case.
An additional finite-temperature issue, related to asymptotic confinement, was also investigated.
The authors considered an infinitely-separated heavy fermion-antifermion pair with fractional charge. 
They found confinement for all fermion masses, but an exponential decay of the string tension at high temperatures.
For the special case of the background field approaching $\ell/g=1/2$, they also analyzed the spontaneous breaking of the CT symmetry.
At zero temperature, this symmetry is broken at the above mentioned phase transition point found by Byrnes et al.\ \cite{Byrnes2002}.
For any non-zero temperature, Buyens et al.\ found indications that the CT symmetry is always unbroken.

\begin{figure}[h!]
\begin{center}
\hspace*{-3mm}
\includegraphics[width=0.24\textwidth]{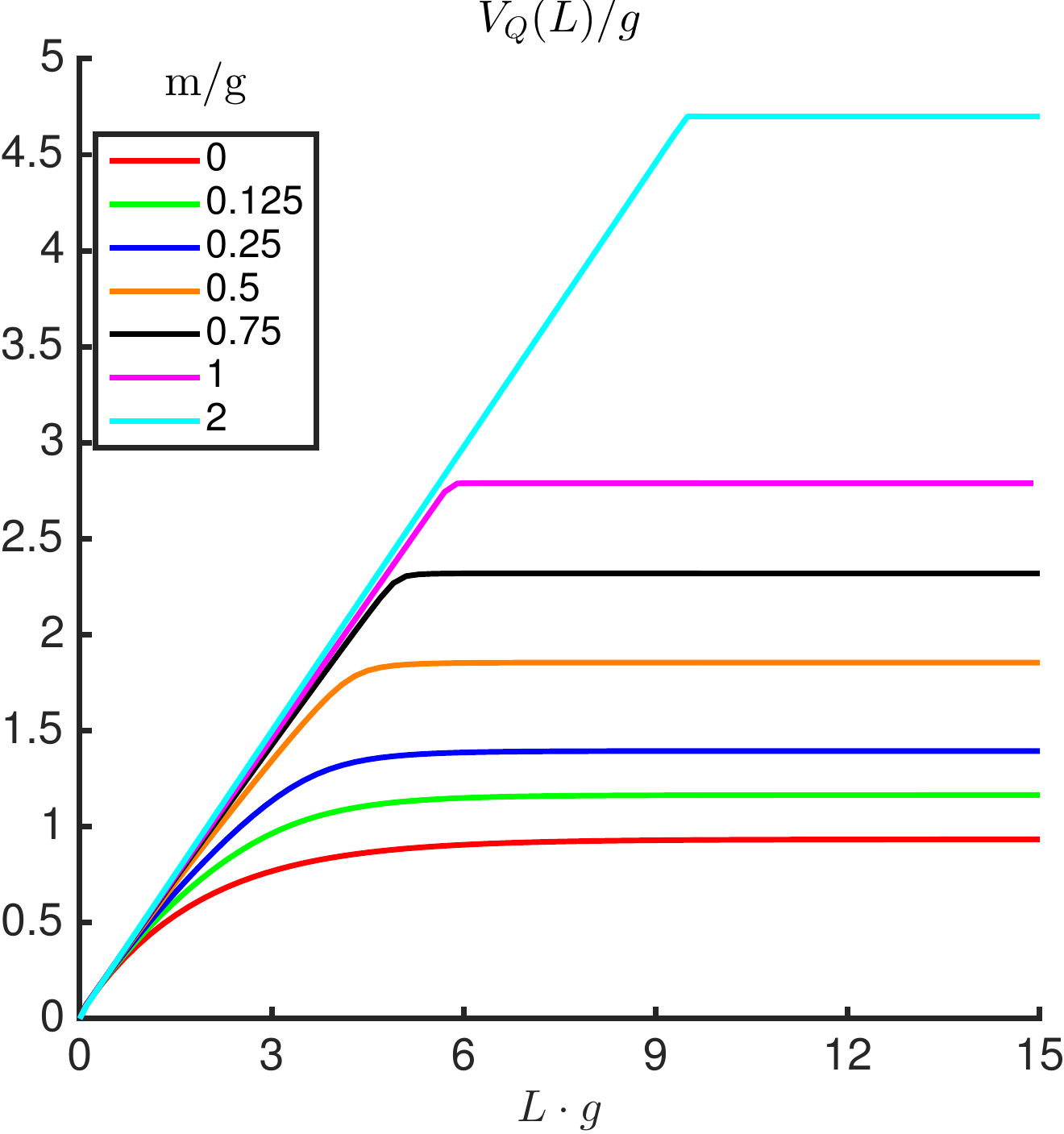}
\hspace*{-1mm}
\includegraphics[width=0.24\textwidth]{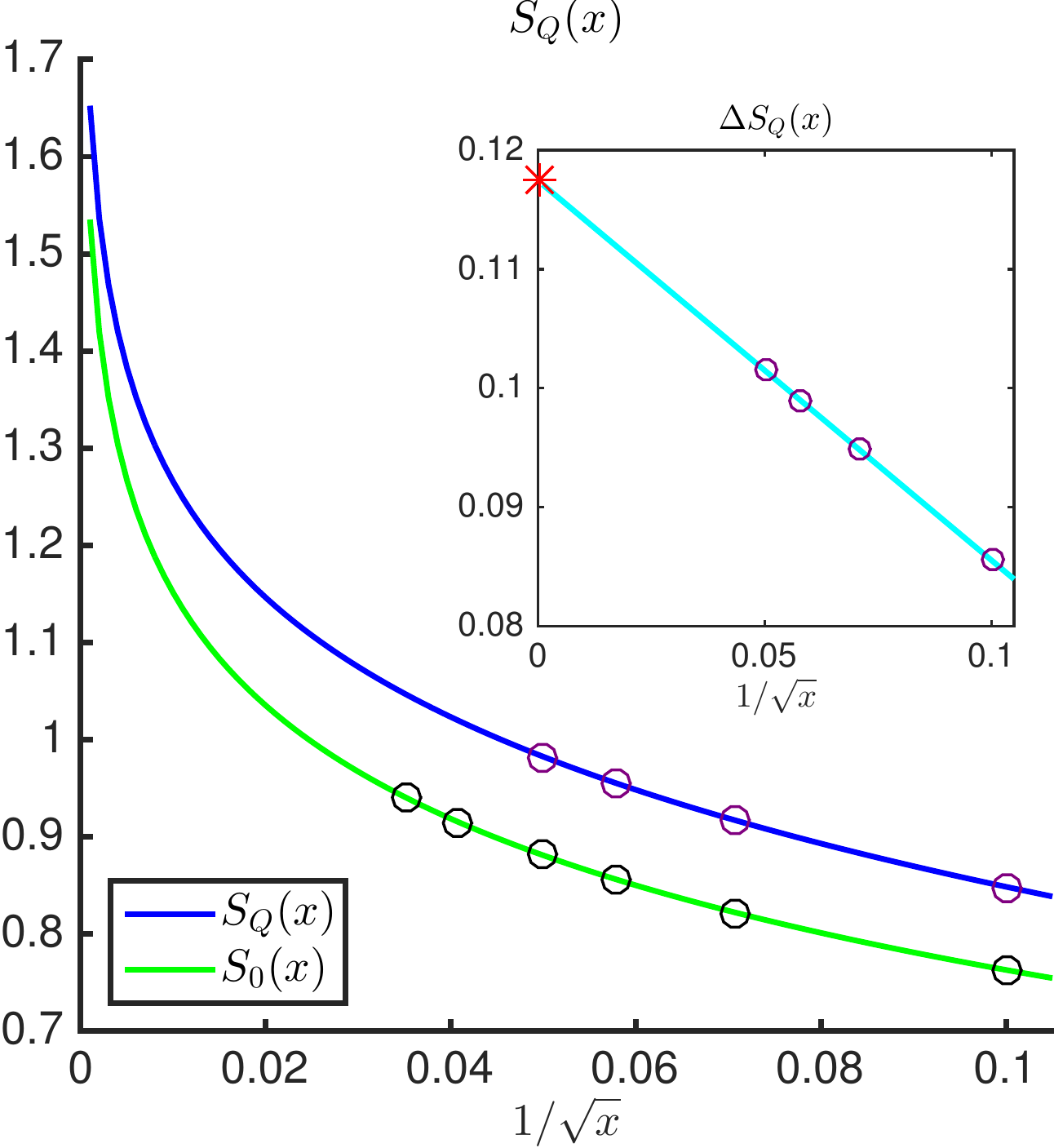}
\vspace*{-5mm}
\end{center}  
\caption{(left) Dependence of the quark-antiquark potential on their separation. The different curves correspond to different quark masses.
(right) Scaling of bipartite entanglement entropy towards the continuum limit. The two curves are the entropy for different probe charges: $Q=0$ (green) and $Q=0.45$ (blue).
Source: Ref.~\cite{Buyens2015}, reprinted with permission by the authors (article published under the
terms of the Creative Commons Attribution 3.0 license).}
\label{fig:string}
\end{figure}

Shortly after Ref.\ \cite{Buyens2016}, Buyens et al.\ published a more detailed investigation of confinement and string breaking in the Schwinger model, in the presence of an external static ``quark'' and ``antiquark'', both with integer and fractional charges \cite{Buyens2015}.
The methodology followed the setup used for previous computations, with gauge invariance built into the MPS ansatz and the latter in the thermodynamic limit.
At short distances between these static ``colour'' sources, the system is in the confining phase (with linearly rising potential, see the left plot of Fig.\ \ref{fig:string}) and as the separation is increased, the phenomenon of string breaking appears (constant potential) and the system undergoes ``hadronization''.
Such behaviour mimics the scenario in full 3+1-dimensional QCD, where there is a long history of such investigations with lattice MC methods, however without the possibility of looking into the real-time aspects of the underlying dynamics as a severe limitation related to the formulation in a Euclidean spacetime.
For the case with fractional probe charges, the authors observed for the first time the phenomenon of partial string breaking.
A heavy meson and a heavy antimeson form from the probe charges surrounded by quarks/antiquarks created dynamically from the vacuum, but there is a remaining unscreened electric field that makes the whole meson configuration confined asymptotically, in contrast to the integer probe charge case.
In addition, the authors computed the bipartite entanglement entropy and showed the changes in this quantity as the string breaking sets in.
As an example, we show in the right plot of Fig.\ \ref{fig:string} the continuum limit scaling of the entanglement entropy for two different values of the probe charge, $Q=0$ and $Q=0.45$.
Both evince the same kind of logarithmic divergence, parametrized by the central charge $c$ of the conformal field theory describing the system at the critical point, according to the Calabrese-Cardy theory \cite{Calabrese2004} (here, there are two fermionic degrees of freedom and thus, $c=1$).
Based on this observation, the authors defined a UV-finite difference of entanglement entropies and showed that this quantity is peaked in the $m-Q$ plane around the critical point of the Coleman's phase transition, $Q=1/2$ and $m/g\approx0.33$.

\vspace*{2mm}

\begin{figure}[h!]
\begin{center}
\hspace*{-3mm}
\includegraphics[width=0.23\textwidth]{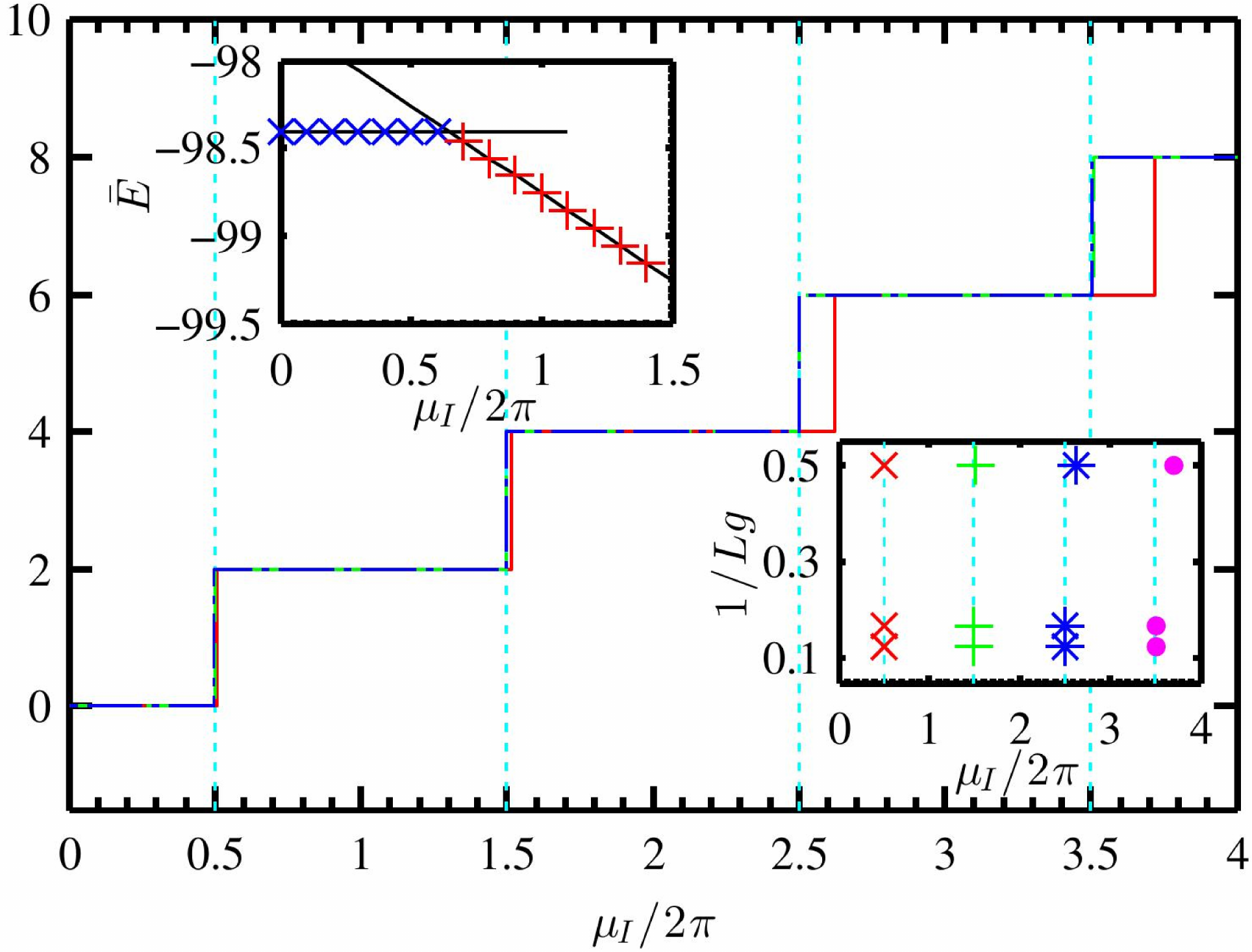}
\hspace*{-1mm}
\includegraphics[width=0.255\textwidth]{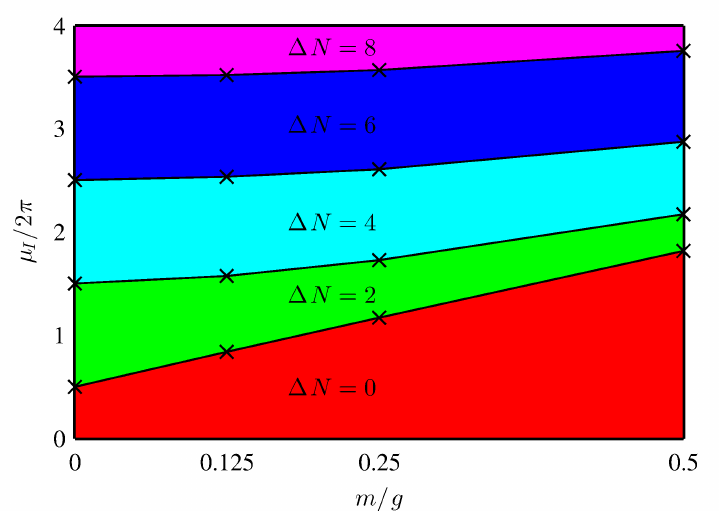}
\vspace*{-5mm}
\end{center}  
\caption{(left) The isospin number, $\Delta N$, as a function of the isospin chemical potential, $\mu_I/2\pi$, in the massless two-flavour Schwinger model.
The different lines correspond to different volumes, $Lg=2$ (red solid), $Lg=6$ (green dashed) and $Lg=8$ (blue dash-dotted line). The vertical lines are analytical predictions from Refs.\ \cite{Narayanan2012,Lohmayer2013}. Upper inset shows the energies of phases corresponding to $\Delta N=0$ (blue crosses) and $\Delta N=2$ (red x's). Lower inset presents the volume dependence of the transition points for the first four transitions.
(right) The phase diagram in the fermion mass --- isospin chemical potential plane.
Source: Ref.~\cite{Banuls:2016gid}, reprinted with permission by the authors and the American Physical Society.}
\label{fig:mu}
\end{figure}

In 2017, Ba\~nuls et al.\ \cite{Banuls:2016gid} investigated for the first time the case of a non-zero chemical potential, where MC studies would encounter a sign problem.
Since chemical potential has no physical effect in the one-flavour model, the simplest non-trivial case was chosen, i.e.\ two fermion flavours.
The chemical potential is a parameter of the Hamiltonian that can be arbitrarily chosen without adding any complications, unlike in MC simulations.
The MPS setup was similar to other studies by these authors and the analyzed cases concerned the massless and massive model.
For the massless one, an analytical calculation is available for the observable of interest \cite{Narayanan2012,Lohmayer2013} -- the isospin particle number, i.e.\ the difference of particle numbers of the two flavours, $\Delta N$.
The system undergoes a series of first-order phase transitions when the difference in chemical potentials of the two flavours (the isospin chemical potential), $\mu_I$, is increased, at $\mu_I/2\pi=1/2,3/2,5/2,\ldots$ in appropriately chosen dimensionless units.
Each transition corresponds to a level crossing -- phases with different isospin numbers have the lowest energy in a certain range of $\mu_I$.
The level crossings were determined from fits to the dependence of the energy on the isospin chemical potential, as exemplified in the upper inset of the left plot of Fig.\ \ref{fig:mu}.
The main plot shows the isospin number of the phase that is stable at a given $\mu_I$.
The analytical result is reproduced when the volume is large enough.
Additionally, the massive case was also considered, which allowed to find the phase diagram in the fermion mass -- isospin chemical potential plane.
As can be seen in the right plot of Fig.\ \ref{fig:mu}, increasing the mass stabilizes the phase with equal particle numbers and shifts all phase transitions to larger values of $\mu_I$.

\vspace*{2mm}

\begin{figure}[h!]
\begin{center}
\hspace*{-3mm}
\includegraphics[width=0.245\textwidth]{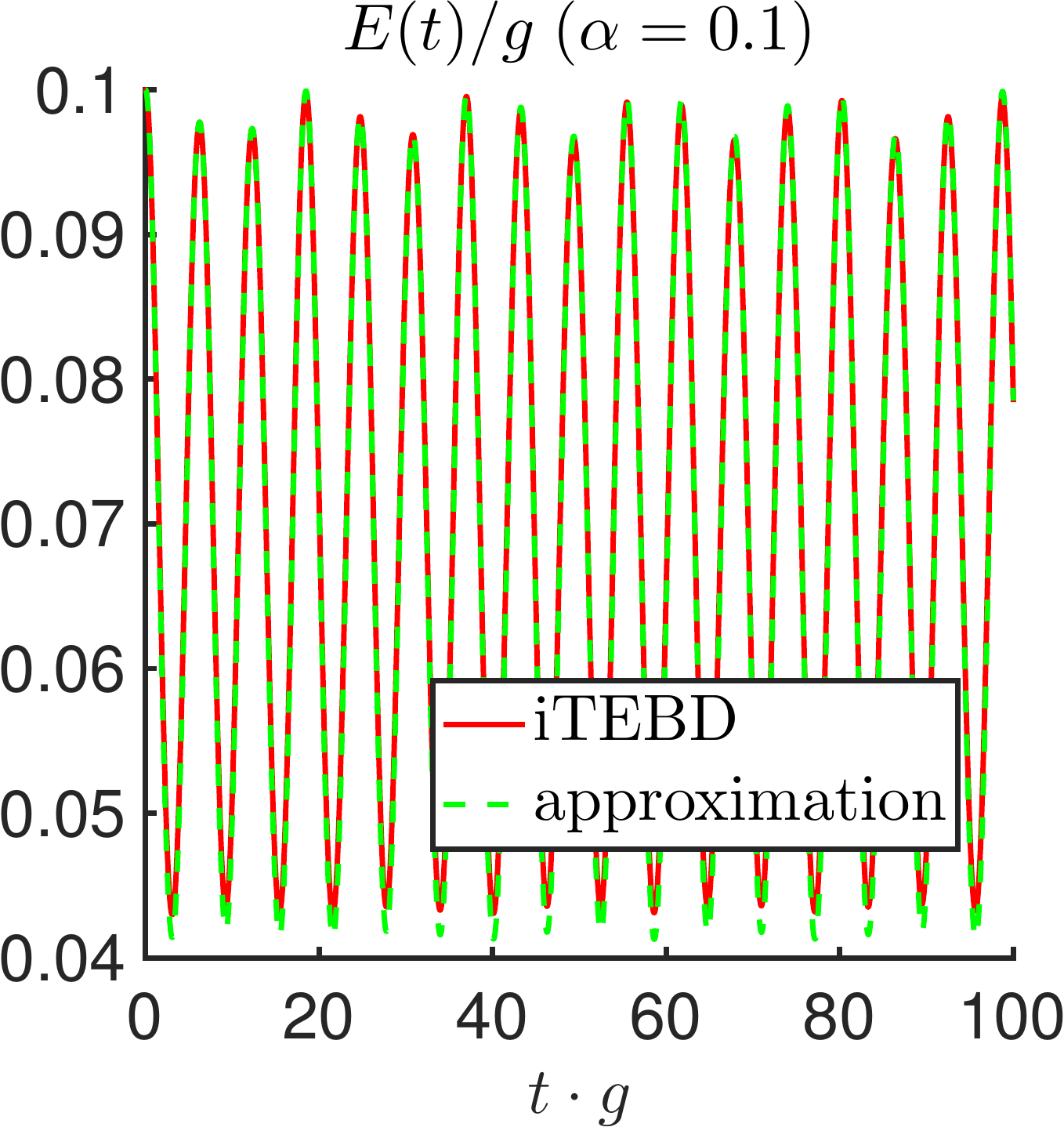}
\hspace*{-3mm}
\includegraphics[width=0.245\textwidth]{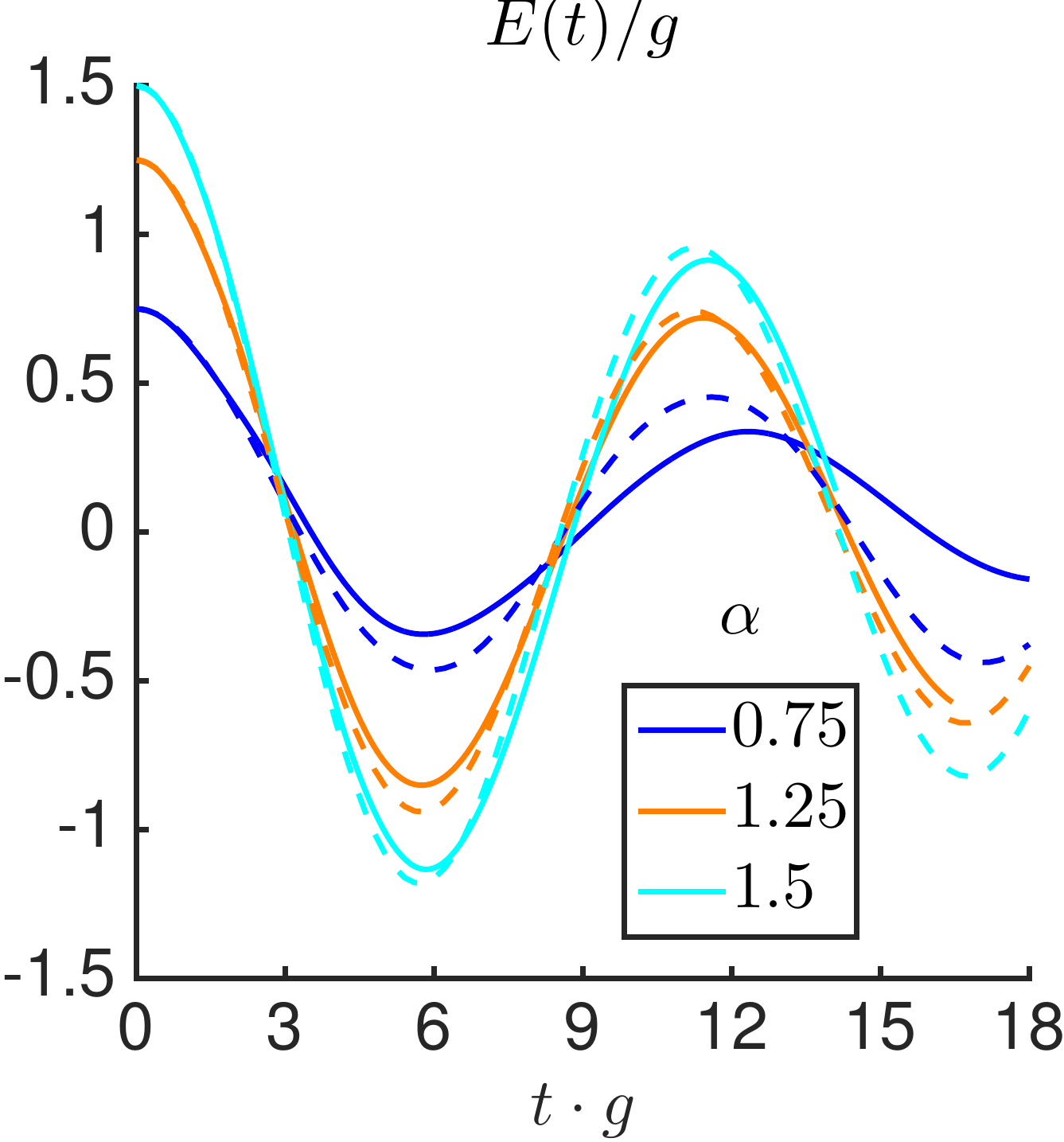}
\vspace*{-5mm}
\end{center}  
\caption{The real-time dynamics of the average electric field for $m/g=0.25$. (left) Weak-field regime ($\alpha=0.1$) -- comparison of the MPS results from the iTEBD algorithm with an approximation retaining two lowest single-particle excitations. (right) Strong-field regime ($\alpha=0.75,1.25,1.5$) -- comparison of MPS (solid lines) with a semi-classical approximation (dashed lines).
Source: Ref.~\cite{Buyens:2016hhu}, reprinted with permission by the authors and the American Physical Society.}
\label{fig:Et}
\end{figure}

A more detailed investigation of the real-time dynamics, following the earlier work and setup of Ref.\ \cite{Buyens2013}, was performed by Buyens et al.\ in 2017 \cite{Buyens:2016hhu}.
The authors considered the effects of an electric quench on the vacuum state for the massive Schwinger model in two regimes -- weak field and strong field.
For the former case, they checked the MPS result, obtained with the infinite TEBD method, against an approximation assuming the dynamics comes from the two lowest single-particle excitations.
The agreement for an electric field $\alpha=0.1$ (where $\alpha$ corresponds to $\ell/g$ in earlier notation) is almost ideal, see the left panel of Fig.\ \ref{fig:Et}, validating the method and the physical explanation.
This agreement persists until slightly larger values of the quench, with the approximation first overestimating the amplitudes and then also failing for the distances between extrema.
In the strong-field regime, the MPS iTEBD results were compared against the semi-classical approximation, see the right panel of Fig.\ \ref{fig:Et}.
As expected, the results become increasingly consistent for larger fields.
The observed damping of plasma oscillations was interpreted as an onset of thermalization.
Checking this hypothesis would require running the simulation until larger real times.
However, this becomes very difficult because of the approximately linear growth of entanglement entropy with time.
This is, again, consistent with physical expectations and corresponds to the separation of electron-positron pairs created via the Schwinger mechanism at earlier times.
The large separation between the created particles and antiparticles entangles distant regions of the system.
Thus, the bond dimension required to model such a situation increases exponentially with time, preventing simulations of thermalization in practice.
This illustrates a generic limitation of MPS real-time simulations.
Nevertheless, due to the sign problem in MC calculations, the information provided by MPS real-time simulations is already unique and powerful.

\vspace*{2mm}

In 2017, another study of the chiral condensate in the Schwinger model appeared, by Zapp and Or\'us \cite{Zapp2017}.
The setup was similar to the one employed by Buyens et al., infinite MPS with gauge invariance encoded in the tensors.
However, the difference with respect to the latter was to work explicitly with gauge variables, instead of integrating them out via the Gauss law.
This keeps the Hamiltonian manifestly local and was motivated by a closer analogy of such a procedure with anticipated application for higher-dimensional models.
A proposal for the usage of infinite PEPS to simulate 2+1-dimensional QED was thoroughly discussed.
For the chiral condensate, the authors found results fully compatible with the earlier studies of Ba\~nuls et al.\ and Buyens et al.

In 2019, Funcke et al.\ \cite{Funcke:2019zna} investigated the Schwinger model with a $\theta$-term, using a setup of earlier simulations by Ba\~nuls et al.
They computed several observables such as the ground-state energy, the chiral condensate and the topological susceptibility.
In the continuum limit, they showed the model is CP-invariant in the continuum limit and also confirmed that negative fermion masses are equivalent to positive masses with a shift in the background electric field.

We discuss now the progress in Lagrangian TN methods applied to the Schwinger model.
The most widely used of such methods is the tensor renormalization group (TRG), which is a coordinate-space technique for coarse graining (blocking) of lattice models.
TRG appeared in the context of lattice field theories in 2013, in a paper by Liu et al.\ \cite{Liu:2013nsa}, where the authors introduced exact blocking formulae for several systems, including the pure gauge Schwinger model (without fermions).
The first numerical TRG study came in 2014, by Shimizu and Kuramashi \cite{Shimizu2014}.
The authors applied Grassman TRG to the one-flavour Schwinger model regularized with Wilson fermions, studying the phase diagram in the mass-coupling plane.
They investigated the chiral susceptibility and Lee-Yang zeros in the complex plane of the hopping parameter, finding that the phase transition in this model belongs to the universality class of the 2D Ising model.
In a follow-up study and using the same approach, Shimizu and Kuramashi analyzed in 2014 the case of the Schwinger model with a topological $\theta$-term \cite{Shimizu:2014fsa}, with $\theta=\pi$.
Peforming Lee-Yang and Fisher zero analyses, they again found the Ising model universality class for the phase transition predicted by Coleman and provided a proof of concept that Grassman TRG can successfully handle cases subject to a sign problem in MC simulations.
In 2018, Shimizu and Kuramashi presented one more investigation of the Schwinger model with one flavour of Wilson fermions, concentrating on the phase structure arising with this lattice discretization at an odd number of flavours \cite{Shimizu:2017onf}.
This phase structure is not fully understood in QCD and the Schwinger model can, thus, shed some light on this aspect.
The setup of this paper combined Grassman TRG and ideas of decorated TRG \cite{Dittrich_2016}, which allow to explicitly preserve the gauge symmetry of the system under coarse graining.
This was used to study the phase structure along the line of large negative fermion mass ($m=-2$) and revealed the existence of a Berezinskii-Kosterlitz-Thouless (BKT) type transition and an emergence of a conformal field theory with the $SU(2)$ symmetry at strong coupling.

\subsubsection{Non-Abelian $SU(2)$ and $SU(3)$ gauge theories}
\label{sec:SU2}
\begin{figure}[h!]
\begin{center}
\hspace*{-3mm}
\includegraphics[width=0.5\textwidth]{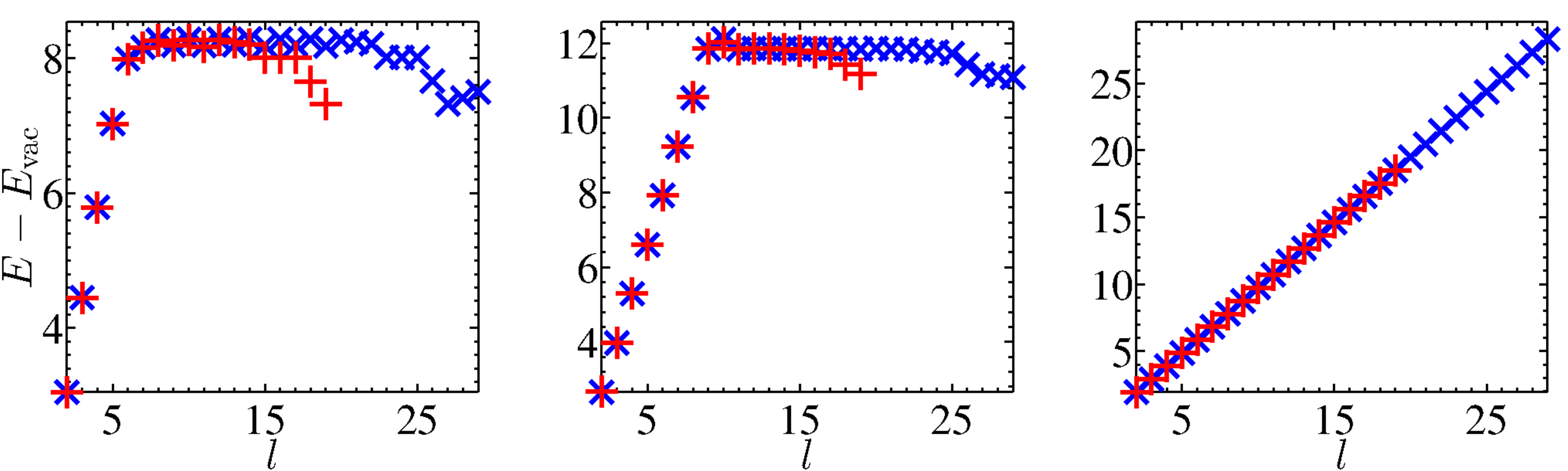}
\vspace*{-5mm}
\end{center}  
\caption{The dependence of the ground state energy of the $SU(2)$ gauge theory with two external static quarks on the initial separation of the quarks. The left panel corresponds to a fermion mass $m=3$, the middle panel to $m=5$ and the right panel to $m=10$.
Source: Ref.~\cite{Kuhn:2015zqa}, reprinted with permission by the authors (article published under the terms of the Creative Commons Attribution 4.0 International license).}
\label{fig:static}
\end{figure}

After successful explorations of the Abelian Schwinger model with TN techniques, one step that was necessary in the quest to apply them to QCD was to test the performance of such methods in a non-Abelian gauge theory.
The first investigation of this kind was performed by K\"uhn et al.\ in 2015 \cite{Kuhn:2015zqa}, for the (1+1)-dimensional $SU(2)$ gauge theory.
This work concentrated on the string breaking phenomenon, explored by studying the static potential between two heavy quarks and the real-time dynamics.
The $SU(2)$ gauge symmetry was realized exactly using finite-dimensional link variables.
This is a necessary truncation of the theory, boiling down to a cutoff in the maximum possible flux on the link.
In this paper, the authors used the simplest non-trivial truncation, obtaining a five-dimensional Hilbert space of each link.
The finite MPS ansatz was used with imaginary and real-time evolution, to access the statical and dynamical properties, respectively.
The evolution operator was approximated using a Taylor expansion.
In Fig.~\ref{fig:static}, we show an example result for an observable that is traditionally used in MC simulations, i.e.\ the potential between two heavy quarks.
For small separations of the external charges, this potential grows linearly (indicating the presence of a string) and at some point, it becomes energetically more favorable to create a quark-antiquark pair and break the string.
The string breaking point is independent of the system size, but it depends on the external quark mass -- in the right panel, no string breaking appears in the simulated range of initial separations.
This kind of statical investigation is accessible with MC methods, contrary to the dynamical aspects that require the computation of the real-time evolution.
The latter was done for both static external charges, as well as for dynamical ones, leading to more insight on the phenomenon of string breaking.
In particular, the authors observed the creation of dynamical fermions around the heavy static charges, i.e.\ the screening of these charges.

\vspace*{2mm}

\begin{figure}[h!]
\begin{center}
\hspace*{-3mm}
\includegraphics[width=0.5\textwidth]{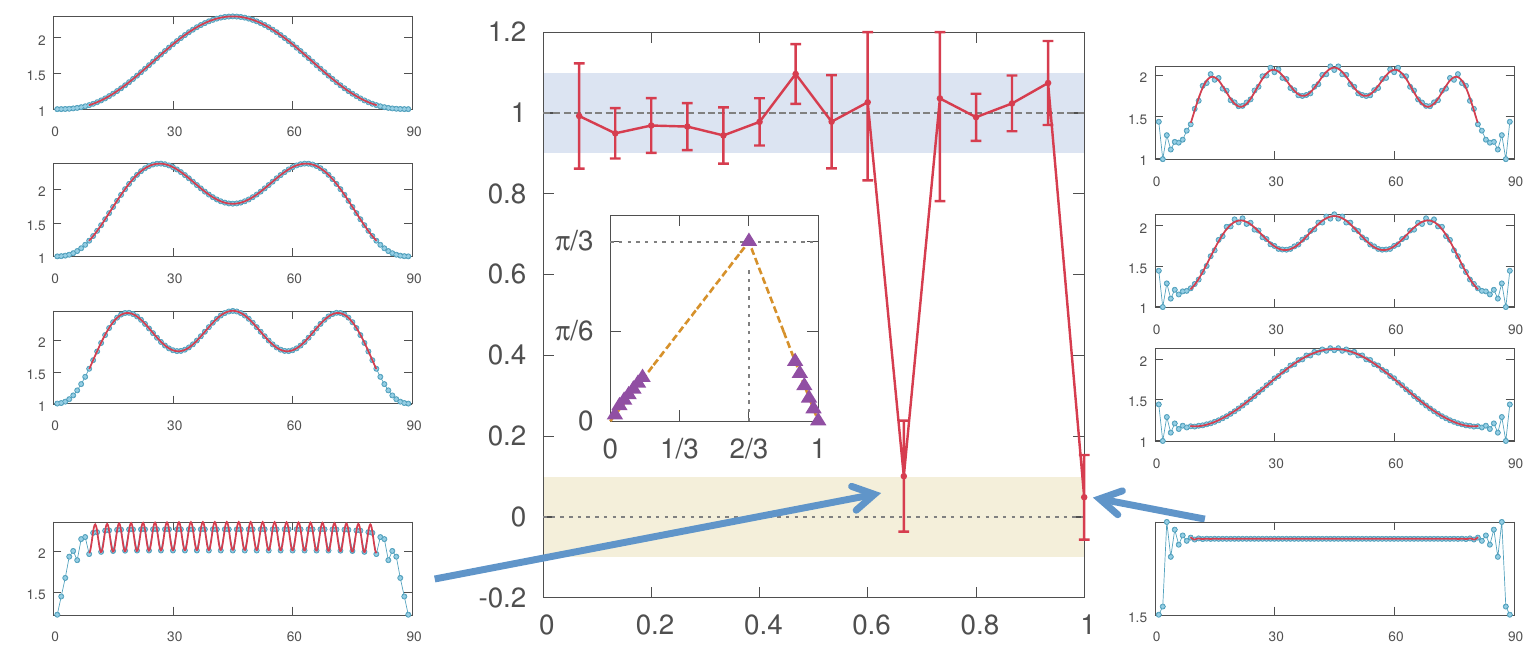}
\vspace*{-5mm}
\end{center}  
\caption{(middle) Dependence of the extracted central charges $c$ in the strong coupling regime on the filling $f_M$ for a $90$-site system. (left and right) Examples of entanglement  profiles (blue dots) vs.\ the partition point $\ell$ along the chain, at different fillings. 
Source: Ref.~\cite{Silvi2016}, reprinted with permission by the authors (article published under the terms of the Creative Commons Attribution 4.0 International license).}
\label{fig:Silvi}
\end{figure}

Another study of the $SU(2)$ gauge theory with tensor networks was performed by Silvi et al.\ \cite{Silvi2016} in 2016, with the aim of investigating the finite density phase diagram in the plane filling vs.\ coupling.
The authors used the quantum link formulation to achieve finite-dimensional link Hilbert space while preserving the exact gauge symmetry.
They applied finite MPS with gauge and global symmetries encoded in the tensors and found the ground state via TEBD.
Results from different system sizes were extrapolated to the infinite volume limit.
By analyzing the entanglement entropy for different fillings, they found two insulating phases (different charge density wave (CDW) orderings), analogues of Mott insulators in Hubbard-type models, at large couplings and fillings 2/3 and 1.
For gapless systems, the entropy diverges logarithmically, proportionally to the central charge of the underlying conformal field theory.
Instead, at these specific couplings, the entropy saturates to a constant (the central charge is effectively 0), indicating a gap in the spectrum.
Away from fillings 2/3 and 1, the system is in a liquid phase ($c=1$).
At weak coupling, it is a BCS state, where quarks form analogues of Cooper pairs.
When the coupling is increased, these pairs break and a simple liquid (metallic) phase is energetically favored.
The entropy analysis is illustrated in Fig.~\ref{fig:Silvi}.
The main panel contains the fitted central charges for different fillings and the side panels display examples of entanglement profiles for selected fillings.
Apart from entanglement entropy, the authors also looked at other observables characterizing the phases, in particular the CDW order parameter and the correlation length of the meson superfluid order.
In this way, they found the location of the phase transitions between the different regions of the phase diagram.

\vspace*{2mm}

\vspace*{2mm}
\begin{figure}[h!]
\begin{center}
\hspace*{-3mm}
\includegraphics[width=0.25\textwidth]{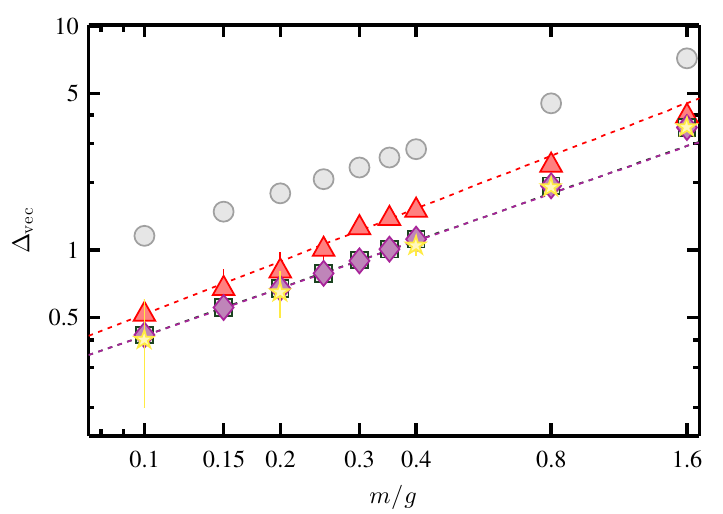}
\hspace*{-4mm}
\includegraphics[width=0.25\textwidth]{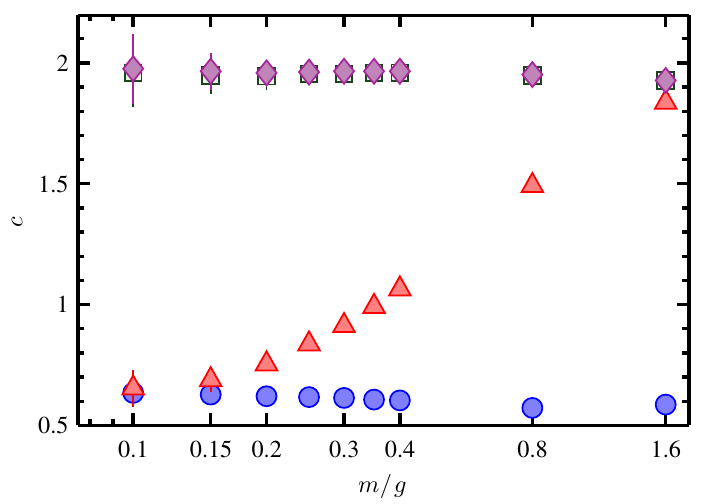}
\vspace*{-5mm}
\end{center}  
\caption{(left) Scaling of the vector mass gap vs.~the fermion mass for three truncations of the $SU(2)$ theory, corresponding to $j_{\rm max}=1/2$ (gray circles), $j_{\rm max}=1$ (red triangles), $j_{\rm max}=3/2$ (green squares) and $j_{\rm max}=2$ (magenta diamonds), compared to values from the strong coupling expansion \cite{Hamer:1981yq}. The fits (for $j_{\rm max}=1,3/2,2$) correspond to a power-law fitting ansatz and yield values of the power-law exponent compatible with 2/3, a prediction in the large-$N_c$ limit \cite{Steinhardt:1980ry}.
(right) Fermion mass dependence of the central charges extracted from the Calabrese-Cardy scaling of the
entanglement entropy, $j_{\rm max}=1/2$ (blue circles), $j_{\rm max}=1$ (red triangles), $j_{\rm max}=3/2$ (green squares) and $j_{\rm max}=2$ (magenta diamonds).
Source: Ref.~\cite{Banuls:2017ena}, reprinted with permission by the authors (article published under the
terms of the Creative Commons Attribution 4.0 International license).}
\label{fig:SU2PRX}
\end{figure}

A follow-up investigation of the $SU(2)$ gauge theory from Ref.~\cite{Kuhn:2015zqa} was performed by Ba\~nuls et al.\ \cite{Banuls:2017ena} in 2017.
In this paper, an efficient basis was found for the physical subspace of the theory, considerably reducing the link Hilbert space dimension and making it more amenable for an MPS investigation, as well as for quantum simulations.
Using this basis, one can truncate the colour-electric flux at an arbitrary value.
The operator for the colour-electric flux in such a formulation of the $SU(2)$ theory is the total angular momentum operator $J$, with eigenvalues $j=1/2$ (simplest non-trivial truncation of Ref.\ \cite{Kuhn:2015zqa}), $j=1,3/2,\ldots$.
The truncation is described by the maximum angular momentum, $j_{\rm max}$, with link Hilbert space dimension of $d_{\rm link}=2j_{\rm max}+1$.
It was checked that all results from $j_{\rm max}=3/2$ and $j_{\rm max}=2$ are compatible with each other and thus, they effectively correspond to the full $SU(2)$ theory, accessed at a much reduced cost with respect to the basis used in Ref.\ \cite{Kuhn:2015zqa} (where already $d_{\rm link}=5$ for the simplest non-trivial truncation and with $j_{\rm max}=2$ leading to $d_{\rm link}=55$).
The efficiency of this basis was demonstrated with spectral computations of the vector mass gap and the determination of the critical exponent describing the scaling of this mass gap with respect to the quark mass (see the left panel of Fig.~\ref{fig:SU2PRX}).
The authors also analyzed the entanglement properties in the ground state, via Calabrese-Cardy scaling, confirming the expectation that the continuum limit is described by a $c=2$ conformal field theory (right panel of Fig.~\ref{fig:SU2PRX}), if the full $SU(2)$ theory is considered.
It is clear from this investigation that the theories truncated at $j_{\rm max}\leq1$ do not correspond to the full theory, as their continuum limit is described by conformal field theories with a different central charge\footnote{Actually, a further investigation would be needed to assess whether such truncated theories posess a continuum limit at all.}.

\begin{figure}[h!]
\begin{center}
\includegraphics[width=0.5\textwidth]{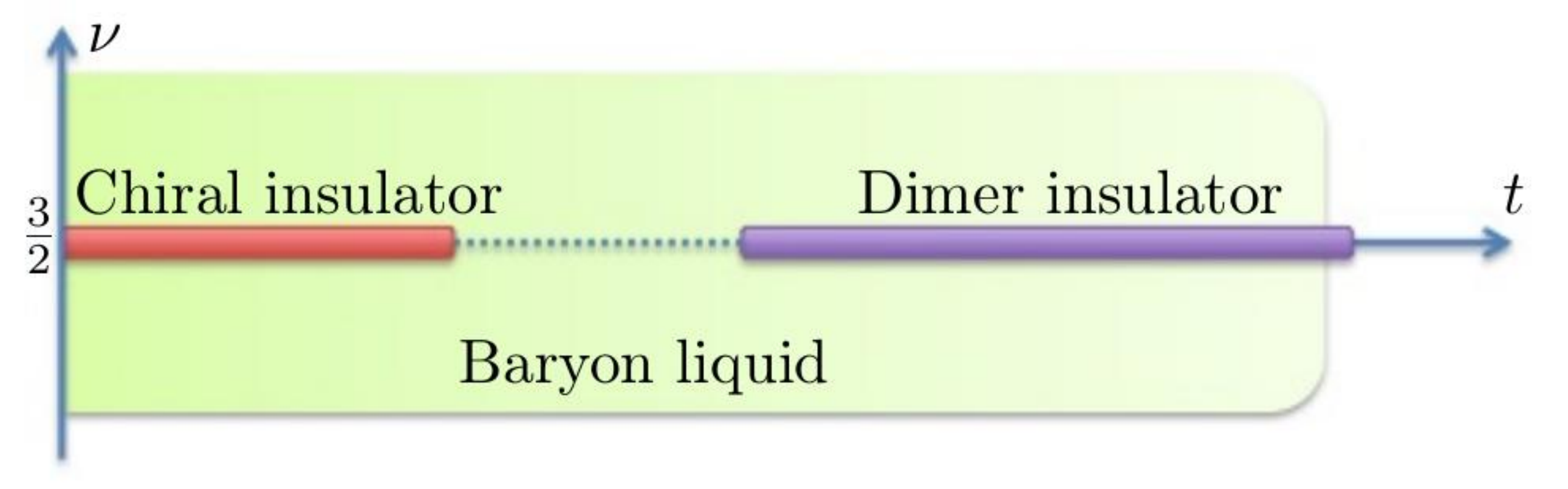}
\vspace*{-5mm}
\end{center}  
\caption{Schematic phase diagram of the $SU(3)$ theory in the quantum link formulation. Bare quark mass is $m=0$.
Source: Ref.~\cite{Silvi:2019wnf}, reprinted with permission by the authors.}
\label{fig:SU3Silvi}
\end{figure}

In 2019, the first TN investigation of a 1+1-dimensional QCD-like theory appeared, by Silvi et al. \cite{Silvi:2019wnf}.
The authors concentrated on the ground state at finite density, using gauge-invariant MPS with the imaginary time TEBD algorithm.
They used the quantum link formulation for the $SU(3)$ gauge fields and staggered quarks.
The analysis led to qualitative phase diagrams of gauge coupling vs.\ quark filling, for different quark masses.
This is illustrated in Fig.~\ref{fig:SU3Silvi} for zero bare quark mass.
The most robust phase is a baryonic Luttinger liquid, being the only liquid phase in this model, in contrast to a mesonic or quark (coloured) one found in the $SU(2)$ theory \cite{Silvi2016}.
This shows that the theory is strongly confined.
At the special quark filling $\nu=3/2$, there are also two insulating (gapped) phases -- a chiral insulator and a dimer insulator.
The former is found at low matter-gauge coupling and corresponds to a type of CDW ordering, while the latter, found at strong couplings, is a state with entangled dimers.
In between them, a competition between these two phases leads to the Luttinger liquid becoming energetically stable.
The authors also studied binding energies and found that baryons are strongly bound, while two baryons repel each other, disfavoring the creation of atomic nuclei in this theory.

\vspace*{2mm}

In this way, the proof of concept for all three most important gauge theories, i.e.\ the Schwinger model and the $SU(2)$ and $SU(3)$, in 1+1 dimensions, has been carried out.
It remains a difficult task to go to higher dimensions, which we discuss in Sec.\ \ref{subsec:higherD}.
However, by now it is clear that the TN formulation is well-suited for investigations of lattice gauge theories, as convincingly shown in the described papers encompassing many different aspects of these theories.

\subsubsection{Other 1+1 dimensional theories}
\label{sec:scalar}
In this subsection, we discuss applications of TN techniques to other lattice field theories in one spatial dimension. 
We always consider first the Hamiltonian formulation (MPS ansatz) and then the Lagrangian formalism (variants of TRG).
\vspace*{2mm}\\
\textbf{$\lambda\phi^4$ scalar field theory}.
We begin with a DMRG study of the two-dimensional scalar $\lambda\phi^4$ theory, performed in 2004 by Sugihara \cite{Sugihara:2004qr}.
The aim was to study the spontaneous breakdown of the $Z_2$ symmetry.
The critical coupling was determined and its value, upon continuum and infinite volume extrapolation, was found to be close to ealier MC studies and with much smaller uncertainty.
Moreover, the critical exponent $\beta$ was found to be consistent with the analytical result of 1/8 for models in the universality class of the two-dimensional Ising model.
As such, it demonstrated that this class of techniques can successfully tackle problems in lattice field theory and offer competitive precision.

In 2013, Milsted et al.\ applied the uMPS approach to the $\lambda\phi^4$ theory  \cite{Milsted:2013rxa}.
They used the TDVP to find the ground state and the low-lying excitations, as well as to analyze the critical properties of the theory and to extract its critical exponents and the central charge of the conformal field theory describing the system at criticality.
This confirmed the expectation that this theory is in the universality class of the Ising model in a transverse magnetic field.
The paper of Milsted et al.\ was one of the first exploratory studies of TN methods applied to lattice field theories and the very promising results obtained by the authors were an important confirmation of the suitability of these methods.

Topological defects in the $\lambda\phi^4$ theory were studied in 2017 in two papers by Gillman and Rajantie \cite{Gillman2017,Gillman:2017ycq}, using uMPS.
In the first paper, they calculated the scalar and kink masses and studied the contribution of kink-antikink excitations to the ground state. 
In the follow-up study, they examined the Kibble-Zurek mechanism of defects formation by computing the equal-time momentum-space two-point correlator and analyzing it when the system was driven through a quantum phase transition.
This provided clear evidence of the defects formation and showed the feasibility of TN methods for this type of real-time phenomena.

In the Lagrangian formalism, the first application of TRG to a scalar field theory, based on the truncated singular value decomposition of a compact operator, was performed in 2012 by Shimizu \cite{Shimizu:2012zza}. Critical points were evaluated and found to agree with MC results.

A more complete TRG analysis of the real $\lambda\phi^4$ theory was performed recently by Kadoh et al. \cite{Kadoh:2018tis}, using their TRG formulation described in an earlier paper \cite{Kadoh:2018hqq} (see below).
They determined the critical coupling in the continuum and infinite volume limits for the spontaneous breakdown of the $Z_2$ symmetry, achieving a sub-percent precision and finding consistency with other calculations in the literature.
\vspace*{2mm}\\
\textbf{$Z_2$ lattice gauge theory}.
In 2005, Sugihara \cite{Sugihara:2005cf} applied for the first time the MPS formalism to a lattice gauge theory, the $Z_2$ model on a spatial ladder.
He demonstrated that accurate predictions for the lowest-lying states can be obtained reliably with very small bond dimensions, $D=2-4$, on lattices with 500 sites.
Thus computed spectrum contains both gauge-invariant and gauge-variant states and he showed that the former can be identified by calculating expectations values of the Gauss law operator.
\vspace*{2mm}\\
\textbf{$O(N)$ sigma models and $CP(N-1)$ models}.
In 2016, Milsted studied the 2D $O(2)$ and $O(4)$ sigma models with uMPS, determining the ground states, mass gaps and beta functions of both models \cite{Milsted2016}. 
The infinite local Hilbert space was truncated to a finite one by restricting the number of considered Fourier modes. 
For the $O(2)$ model, the BKT transition was identified and the entanglement entropy scaling gave a central charge compatible with $c=1$, as expected for a theory equivalent to the classical 2D XY model.
In the case of the $O(4)$ model, the asymptotic weak-coupling regime was approached, but it was found that the higher Fourier modes play an increasingly important role towards weak coupling, where the entanglement also grows rapidly, making the MPS description problematic.
Nevertheless, it is important to emphasize that these effects can be controlled in the simulations.

An investigation of the $O(3)$ model appeared in 2019, by Bruckmann et al.\ \cite{Bruckmann:2018usp}.
They calculated the ground state energy and the mass gap using finite MPS and analyzed also the entanglement entropy scaling towards the continuum limit, finding a central charge of 2, according to expectations.
However, this value is obtained only if keeping sufficiently many basis states of the angular momentum operator ($l_{\rm max}\geq3$)\footnote{Note that such behaviour of the entanglement entropy was observed also in other theories, e.g.\ the $SU(2)$ gauge theory \cite{Banuls:2017ena}}.
Finally, the authors investigated also the phase structure at non-zero chemical potential, locating the transition points between different charge sectors of the Hamiltonian.

In 2014, Unmuth-Yockey et al.\ \cite{Unmuth-Yockey:2014afa} investigated the $O(3)$ model with TRG. They showed how to construct correlation functions in this approach and computed the average energy, entropy and the two-point correlator, comparing to other results in the literature.

One year later, Yang et al.\ \cite{Yang:2015rra} considered the $O(2)$ model with finite chemical potential, comparing TRG and the worm algorithm.
They calculated several observables -- the superfluid particle number density, the thermodynamic entropy and entanglement entropy, getting in general good agreement for both approaches.
They also identified open questions that can be accessed in the future.

In 2016, Kawauchi and Takeda  \cite{Kawauchi:2016xng} used HOTRG to analyze the $CP(N-1)$ model.
They confirmed consistency of results for the $N=2$ model with the ones for the $O(3)$ model \cite{Unmuth-Yockey:2014afa}.
Agreement with earlier MC simulations was also concluded. 
They also derived the formulation that can be used for including the $\theta$-term in this model.

Another TRG investigation (with a cross-check using also MPS) of the $O(2)$ model with isotropic and anisotropic couplings was performed in 2017 by Bazavov et al.\ \cite{Bazavov:2017hzi}.
The authors extracted the central charge from Renyi entanglement entropies.
Furthermore, they proposed a mapping of this model to a single-species Bose-Hubbard Hamiltonian, which could enable observing the Calabrese-Cardy scaling in a quantum simulation.
\vspace*{2mm}\\
\textbf{Models with four-fermion interactions}.
Very recently, a finite MPS setup was applied to the Thirring model \cite{Banuls:2019hzc}.
Ba\~nuls et al.\ investigated its phase structure by looking at entanglement entropy, the chiral condensate and two kinds of correlation functions, finding a conformal (critical) phase and a massive (gapped) phase, separated by a BKT-type transition.
As a further line of research, real-time phenomena and scaling of mass gaps in a mass-deformed conformal field theory are planned.

In 2014, Takeda and Yoshimura \cite{Takeda:2014vwa} applied Grassmann TRG to the one-flavour Gross-Neveu model at finite chemical potential, a system with a sign problem in MC simulations. 
The Gross-Neveu model is another toy model of QCD, sharing with it, in particular, the property of asymptotic freedom.
The authors computed the quark number density and susceptibility and found a crossover between the regimes of small and large chemical potentials.
They also observed that the convergence of TRG decreases in the crossover region, even though it is not a phase transition.
The results were validated against known exact results at zero coupling.
They also introduced an analogue of reweighting in MC simulations, consisting in approximately coarse graining tensors for certain parameters from other parameter values.
%
%In 2017, Sakai et al.\ \cite{Sakai:2017jwp} introduced Grassmann HOTRG (GHOTRG) for the study of relativistic fermion systems.
%They tested the method for systems with a chemical potential: 2D and 3D free Wilson fermions and the Thirring model, calculating the fermion number density in the latter.
%The results were compared to analytical formulae or previous results with other methods, and agreement was found.
%We also mention a follow-up work by an extended group of authors \cite{Yoshimura:2017jpk} developing a method for the computation of fermionic Green functions in the framework of GHOTRG, with a test for 3D free Wilson fermions.
\vspace*{2mm}\\
\textbf{Other systems}.
Finally, we review some applications of TN methods to other field-theoretical systems.
The (1+1)-dimensional $\mathcal{N}=1$ Wess-Zumino model was considered in 2018 by Kadoh et al.\ \cite{Kadoh:2018hqq}, as a first application of TN techniques to supersymmetry.
A TN representation for the partition function of this model was found and the formulation was tested in TRG and Grassmann TRG for bosons and fermions, respectively.
The test case was the free theory, which allowed comparisons with analytical results and consistency was found.

Also in 2018, Unmuth-Yockey et al.\ studied the 2D Abelian Higgs model (compact scalar electrodynamics) with the HOTRG approach \cite{Unmuth-Yockey:2018ugm}.
They calculated the Polyakov loop and checked the results against MPS and MC simulations, observing universal finite-size scaling of this observable, both in the (1+1)-dimensional lattice model and in the limit of continuous time.
In the following year, the study was extended to the non-Abelian $SU(2)$ case by Bazavov et al.\ \cite{Bazavov:2019qih}.
The authors derived the HOTRG formulation for this case and calculated different observables and compared them to MC simulations, finding consistency and clear advantages of using the TRG approach, e.g.\ for determining the static quark potential from the Polyakov loop correlation function.

%Other recent works have extended the TRG methods to deal with specific field theories. 
%Sakai et al.~\cite{Sakai:2017jwp} presented a modification of HOTRG to work with Grassmann variables in any number of spatial dimensions and applied it to relativistic fermions.
In another recent work, Campos et al.~\cite{Campos2019boson} implemented a TRG scheme for the partition function of a free bosonic field theory in which the usual discrete SVD was 
modified to handle the non-compact bosonic (infinite-dimensional) degrees of freedom. 

In 2019, an attempt of applying TN techniques for 2D Euclidean quantum gravity appeared, by Asaduzzaman et al.\ \cite{Asaduzzaman:2019mtx}.
The theory was recast as a gauge theory and the authors derived a suitable TN formulation for the case with positive cosmological constant.
Such a lattice model is exactly solvable and this allowed them to study the strong coupling regime of the theory without running into a sign problem and to locate first-order phase transitions in the space of couplings.

\subsection{Higher-dimensional perspectives}
\label{subsec:higherD}

It is indeed possible to apply TNS methods to higher-dimensional problems,
although the complexity of the corresponding algorithms increases with the dimensionality,
and the implementations are technically more challenging.
The number of projects that apply TNS to LGT in two spatial dimensions is still smaller than in 
the one-dimensional case, but the subject is an active research topic.
In these paragraphs, we review the progress and the different alternatives being developed around this topic.

The first work applying TNS to a LGT in two spatial dimensions was presented in 2010
 by Tagliacozzo and Vidal \cite{Tagliacozzo:2010vk}, who used MERA to study a $Z_2$ LGT.
They proposed a symmetry-preserving numerical coarse graining scheme yielding a low-energy effective description of the model and resulting in a variational ansatz for the ground state and
the low excitations.
The study successfully reproduced the known ground state phase diagram of the model and determined precisely energy gaps and other observables,
and constituted the first instance of an explicitly gauge-invariant TN ansatz.

Given the complexity of higher-dimensional algorithms, reducing the number of variational parameters by restricting 
the ansatz to the physical (gauge-invariant) subspace may be necessary
to achieve competitive numerical results, besides providing a powerful analytical tool to study LGT.
Several groups have formulated gauge invariance in the TNS language.

Tagliacozzo et al.~\cite{Tagliacozzo2014} introduced a TNS framework for pure gauge LGT 
 for any discrete or (compact) continuous group. 
 Besides providing an invariant truncation of the Hilbert space,
 and a systematic construction of gauge-invariant operators, the work proposed
 a variational explicitly invariant TNS ansatz, in which tensors split in one part fixed by the symmetry which ensures the constraints, and another part
 containing the variational parameters.
 The ansatz was probed numerically by exploring a family of states with a single variational parameter that interpolates
 between two different phases.

Haegeman et al.~\cite{Haegeman2015} adopted another strategy to construct gauge-invariant PEPS for any
finite or compact group. 
Starting from a PEPS for the (bosonic) matter degrees of freedom (living on the vertices of the lattice) which is globally invariant, 
the state is \emph{gauged} by introducing gauge degrees of freedom on the links and 
applying a projector onto the physical subspace.
The procedure was illustrated numerically by applying the construction to a 
$\mathbb{Z}_2$ symmetry for Higgs matter,
constructing a PEPS with only two variational parameters, and exploring the phase diagram.

Zohar and Burrello~\cite{Zohar2015} presented a formulation of LGT oriented to its potential realization in atomic quantum simulators.
The proposal started from fermionic matter with the global symmetry, which is then promoted to a local character with additional gauge degrees of freedom.
Using the representation basis for the gauge yielded a gauge-invariant truncation scheme for continuous groups, simular to the one in Ref.~\cite{Tagliacozzo2014}.
Building on this formalism, the authors~\cite{Zohar2015b} constructed gauge-invariant PEPS for arbitrary symmetry groups
including fermionic or bosonic matter.

Although not fully variational calculation has yet been reported with PEPS for a LGT,
there is no fundamental limitation, and all the required technology is already available, as summarized in Ref.~\cite{Zapp2017}. 
Meanwhile, restricted PEPS constructions have been defined that allow for more efficient calculations, 
and their performance has been explored in different models.
In particular, Zohar et al.~\cite{Zohar2015c},
starting from a Gaussian fermionic PEPS for matter, constructed the simplest $U(1)$ gauge-invariant states by the gauging procedure, consistent with 
all symmetries of the problem. The resulting PEPS, named gauged Gaussian PEPS (GGPEPS), were specified in the simplest case by three parameters,
whose phase diagram was explored numerically.
The same authors~\cite{Zohar2016su2} constructed the $SU(2)$ case using the same framework, and, by 
exactly contracting the GGPEPS on a narrow cylinder,
they studied the phase diagram of the pure gauge case, 
where they found a Higgs and a Coulomb phase, and the case with matter, in which no phase transition was visible, but different screening and non-screening behaviours were observed.
Extending such calculations to larger systems or more general PEPS is computationally very demanding, but
Zohar and Cirac~\cite{Zohar2018mc} demonstrated that GGPEPS can be rewritten as a sum
which allows the efficient computation of expectation values for gauge-invariant operators using MC methods.
Since the probability density is given by the norm of a state, it is always positive, and the method is free from the sign problem.

The MPS ansatz can also be used to study two dimensional problems, although it is not a scalable alternative.
Tschirsich et al.~\cite{Tschirsich2019spinice} used this strategy for a $U(1)$ quantum link model, namely spin ice.
A narrow cylinder was mapped to a chain, by grouping together sites on the same transverse circle, and 
 gauge-invariant MPS were used to perform numerical calculations.
They studied the phase diagram (in which a N\'eel and a resonating valence bond state phase are present), and the entanglement properties of ground states and low excitations.

Tensor renormalization strategies can also be applied in more than one spatial dimension, and there have already been studies for (2+1)-dimensional LGT.
One of the first attempts was presented in 2018 by Kuramashi and Yoshimura~\cite{Kuramashi:2018mmi}.
They applied HOTRG at finite temperature and performed a finite size scaling analysis, determining the transition temperature
and the critical exponent $\nu$.
These results improved upon older MC estimates and, perhaps even more importantly, showed that TN methods can also be used beyond one spatial dimension.

Also in the context of TRG techniques, modifications of the algorithms are being proposed to deal with the
specific features of LGT.
In 2017, Sakai et al.~\cite{Sakai:2017jwp} introduced Grassmann HOTRG (GHOTRG) for the study of relativistic fermion systems.
They tested the method for systems with a chemical potential: 2D and 3D free Wilson fermions and the Thirring model, calculating the fermion number density in the latter.
The results were compared to analytical formulae or previous results with other methods, and agreement was found.
In a related follow-up work, an extended group of authors \cite{Yoshimura:2017jpk} developed a method for the computation of fermionic Green functions in the framework of GHOTRG, with a test for 3D free Wilson fermions.

The works we just discussed attest the interest of the topic and the need for further explorations in order to decide the most efficient strategies 
to apply TNS to LGT problems. In this direction, it may be that new formulations, such as the recent elimination of fermionic fields proposed by Zohar and Cirac~\cite{Zohar2018fermions,Zohar2019fermions} also lead to more convenient numerical strategies.

\section{Quantum resources for LGT}
\label{sec:quantum}

\subsection{Quantum Simulation}
\label{subsec:QSim}
The techniques and results reported in the previous section take advantage of quantum information
concepts, but they are intrinsically classical.
For some problems, for instance for the most general out-of-equilibrium dynamics,
it is most likely that the scaling of the computational cost
will exceed the capabilities of these solutions, and ultimately a
truly quantum alternative will be needed.

The concept of quantum simulator was first introduced by
Feynman~\cite{Feynman1982}
by noting that the difficulty of classically simulating quantum problems
stems from the very laws of quantum mechanics, and
the best strategy should instead be to use quantum resources, governed
by the quantum principles.
A quantum simulator is a controllable quantum system that is engineered to
mimic the dynamics of the problems of interest. In order to study some 
physical quantity, then, one needs to perform an experiment on the surrogate: 
preparing, evolving and measuring it.

From the perspective of information theory, quantum simulators are dedicated quantum computers.
As such, they offer only a 
fraction of the computational capabilities of the latter, but also at a lower cost
in terms of requirements.
In particular, a quantum simulator is expected to be resilient to small errors, and thus it
does not need to implement error correction~\cite{Cirac2012} (or do it to a less demanding extent),
which makes it more scalable than a fully-fledged quantum computer.

A quantum simulation can be performed in a digital~\cite{Lloyd96qsim} 
or analog~\cite{Jaksch98cold} way.
In the first case, the evolution is discretized and expressed as a sequence of quantum gates, 
typically involving only a few sites.
In the second, the system is evolved continuously under a properly designed Hamiltonian that contains the
interactions of the model to be simulated.

Since Feynman's suggestion, the first
quantum simulations have become a reality.
A most notable milestone was the simulation of the
phase transition between a Mott insulator and a superfluid in
the Bose-Hubbard model, performed
with ultracold atoms on an optical lattice~\cite{Greiner2002}.
Nowadays, different experimental platforms are investigated and used as 
potential quantum simulators~\cite{qsim2013AnnPhys,Georgescu2014qsim}. The most advanced ones include trapped ions, ultracold atoms in optical lattices and superconducting qubits,
and other systems are also investigated (quantum dots, photons, NV-centers, etc.).

The goal is for a quantum simulator to realize a calculation beyond the 
capabilities of classical computers.
Although this has not yet been achieved,
except arguably in the case of real time evolution \cite{Trotzky2012},
the experiments have reached highly significant milestones,
and simulated a large variety of many-body phenomena, in and out of equilibrium.

The field of LGT offers plenty of problems amenable to quantum simulation~\cite{Wiese2013,Zohar2013,Dalmonte2016}.
Simulating LGT with experimental quantum systems involves challenges beyond those of
more usual condensed matter problems.
Among them, there is the necessity of representing both fermionic and bosonic degrees of 
freedom (at least in the high energy models)
and the crucial requirement of maintaining the gauge symmetry,
which is not a fundamental one in the experimental platforms.

A related, but different, set of quantum simulations has targeted and achieved the realization
of classical (or static) gauge fields, often termed synthetic gauge fields~\cite{Dalibard2011artificial,Goldman2014ropp,Aidelsburger2017},
for instance realizing the Hofstadter model~\cite{Aidelsburger2013hofstadter}.
Some proposals have managed to make these gauge fields sensitive to 
the atomic matter~\cite{Edmonds2013,Bermudez2015,Raventos2016topo},
but they are not fully dynamical.
Another group of proposals have focused on the simulation
of quantum field theories, as formulated in the continuum, using different platforms
(e.g.~\cite{Cirac2010qft,Casanova2011qft,Longhi2012qft,GarciaAlvarez2015fermion-fermion}).
Although these projects have indeed points in common with the LGT simulation, for the sake of completeness, in the rest of the section 
we will focus on proposals that specifically deal with the issues of gauge theories, with
or without matter, where the gauge degrees of freedom are fully dynamical.

Formulating a proposal for a full quantum simulation of a LGT requires, in the first place, mapping the 
gauge and matter
degrees of freedom of the theoretical model
 onto the physically available ones, which depend on the platform.
% may be fermionic or bosonic. 
 Then it is necessary to engineer the interactions to obtain the effective Hamiltonian.

The rich possibilities offered by the various available experimental platforms
are reflected in the large number of theoretical proposals to experimentally realize
the simulation of different aspects of LGT
 that have been put forward in the last few years.
Since an exhaustive review of every existing proposal would exceed the 
scope of this article, we will aim to present an overview of 
the most advanced alternatives, 
and so give the reader a window onto this fast developing world.

There are tight connections between the quantum simulation 
proposals for LGT and TNS investigations of the same models. The reason is that 
both approaches can have similar requirements on the 
formulation of the problem, namely a Hamiltonian form and finite 
dimensional local degrees of freedom.\footnote{In the case of the approaches based on the partition function, the Hamiltonian requirement can be relaxed.}
This makes TNS algorithms a useful tool to help the design and validation of quantum simulation protocols.

\subsubsection{Ultracold atoms in optical lattices}
\label{subsec:ultracold}
% ultracold atoms
% QSIm with Ultracold atoms in OL

\begin{figure}
\centering
\includegraphics[width=.5\columnwidth]{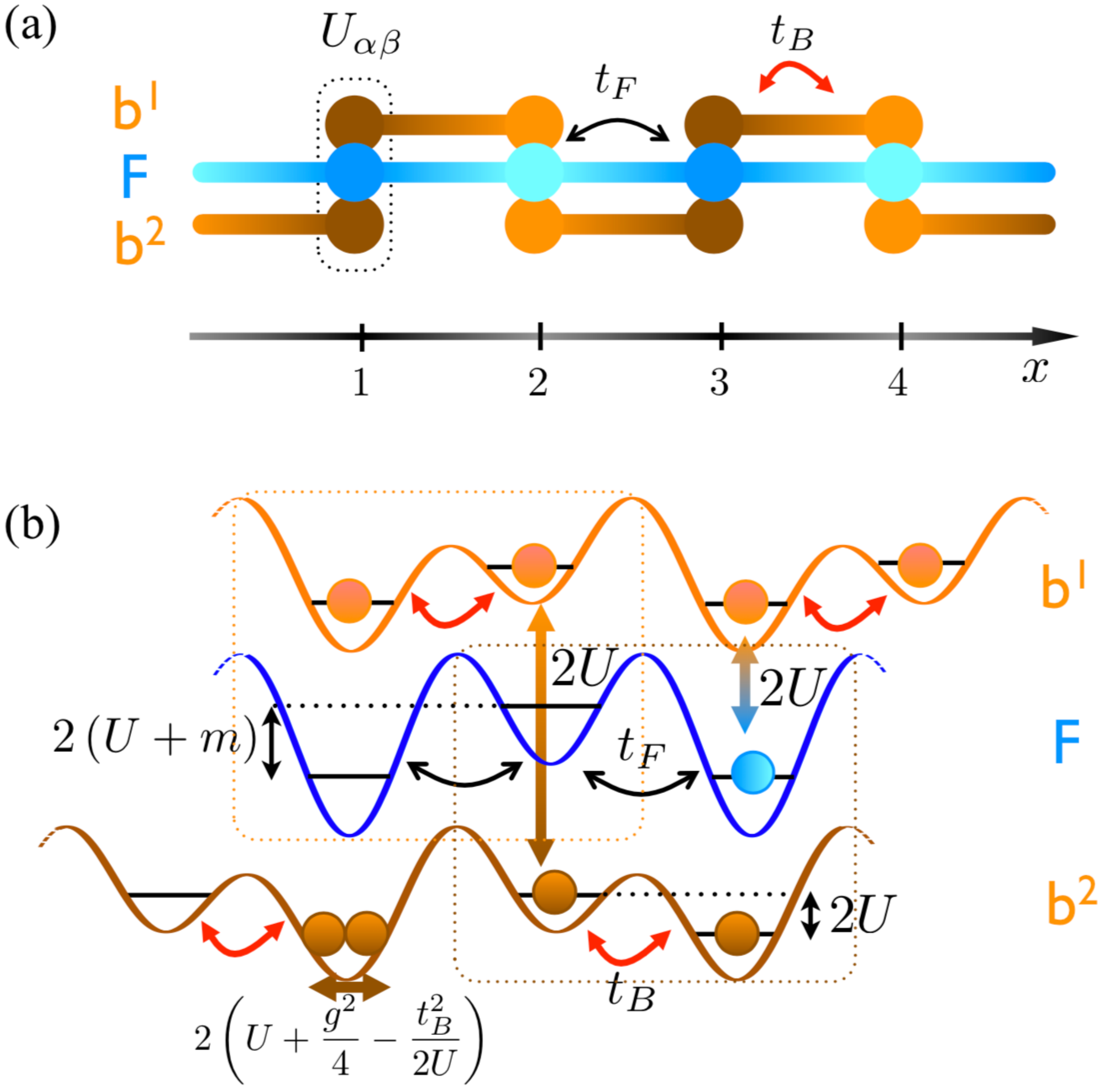}
\caption{Scheme of the triple layer optical lattice for the simulation of $U(1)$ quantum link model proposed in Ref.~\cite{Banerjee2012}.
Two different bosonic species (orange) represent the link variables for links originating on even and odd sites.
The correlated hopping is enforced by energy conservation.
Source: Ref.~\cite{Banerjee2012}, reprinted with permission by the authors and the American Physical Society. }
\label{fig:ultracold1}
\end{figure}

\begin{figure}
\centering
\includegraphics[width=.65\columnwidth]{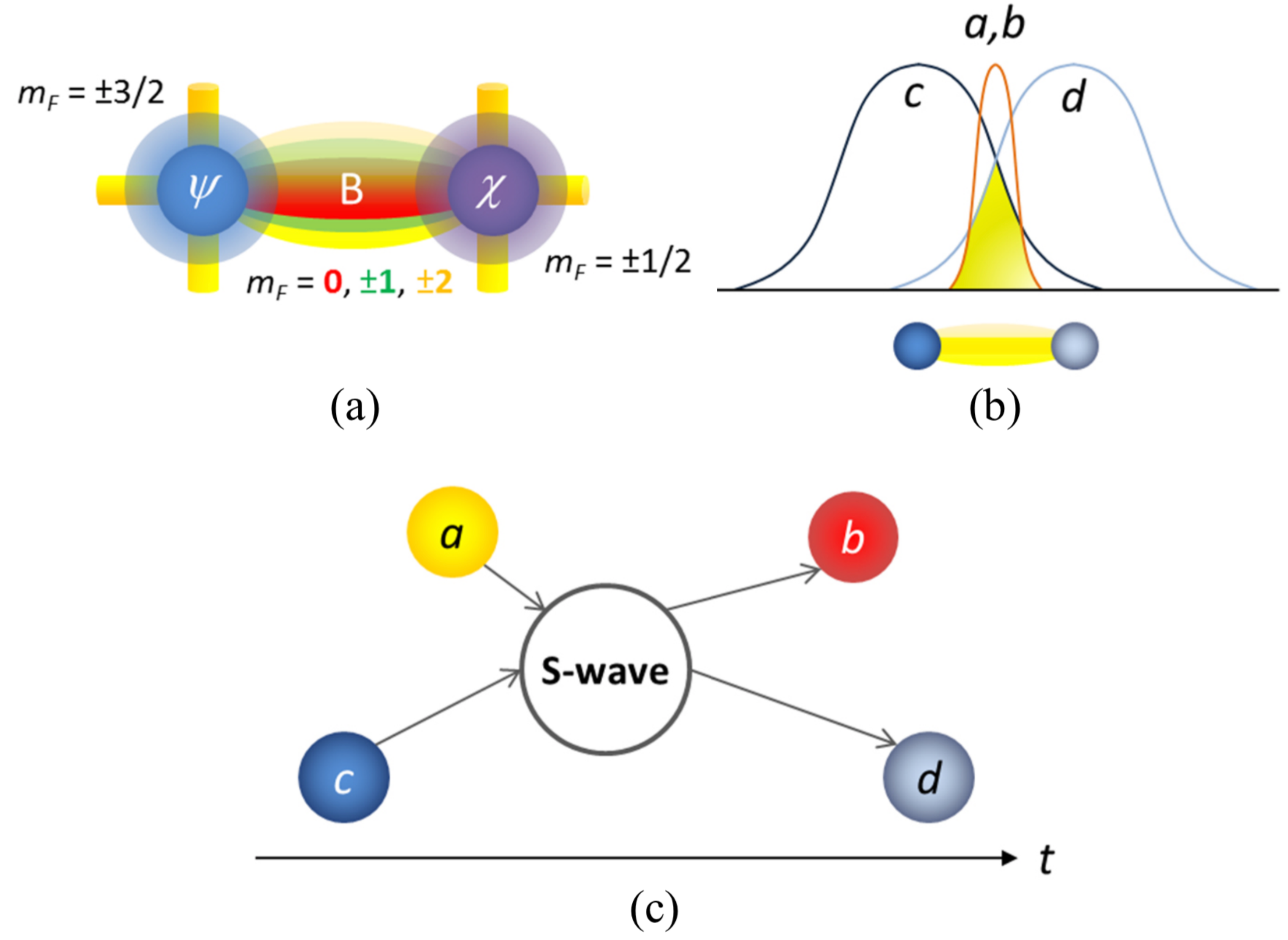}
\caption{Scheme of the proposal introduced in Ref.~\cite{Zohar2013} to simulate the Schwinger model,
which ensures gauge invariance by
mapping it to hyperfine angular momentum conservation. 
Two bosonic species ($a$, $b$) realize the link variables in the Schwinger representation.
Their third components $m_F$, and those of the fermionic species representing the fermions on the vertices,
are properly chosen to ensure only the desired processes take place.
Source: Ref.~\cite{Zohar2015a}, reprinted with permission by the authors.
}
\label{fig:ultracold2}
\end{figure}

%\noteMC{One or two more figures here: digital simulator? Scheme of wells from Wiese? Scheme of an OL?}

Cold atoms trapped in optical lattices constitute one of the most versatile experimental platforms for 
quantum simulation~\cite{Bloch2018qsim}.
In an optical lattice, ultracold neutral atoms are trapped in an effective periodic potential
that results from their dipole interaction with the standing wave produced
by counter-propagating laser beams. 
Two atoms that meet in the same site of the lattice may interact, and 
the atoms may as well hop to nearby sites.
In second quantization, the dynamics of the atoms is described by a
Hubbard-like 
Hamiltonian, containing hopping and interaction terms that can be finely tuned 
in the experiment.
In this way, by adjusting the properties of the lasers, it is possible to
engineer the dynamics of the atoms to reproduce the model of interest.
The nature of the trapped atoms can be fermionic (the only platform where this is possible) or bosonic, and using different number and configurations of laser beams, 
lattices of different geometries
can be created, in one, two and three spatial dimensions.
Time-of-flight measurements can be used to reconstruct the momentum distribution of atoms 
in a lattice, but new techniques allow also single-site resolution measurements and addressability
~\cite{Kuhr2016microscope,Gross2017qsim}.

In the following paragraphs, we compile an overview of the broad variety
of proposed schemes to implement LGT with ultracold atoms in optical lattices.
More details about the platform, and the basic ingredients for LGT simulation,
as well as detailed descriptions of the earliest proposals, can be found in the 
excellent review articles by E.~Zohar et al.~\cite{Zohar2015a}
and U.-J.~Wiese~\cite{Wiese2013}.

%In the last few years, the variety of theoretical proposals to map the LGT 
% HEre newest proposals?

% Proposals

\begin{center}
\textbf{Analog proposals: Different strategies for gauge invariance}
\end{center}
One of the biggest challenges for the simulation of LGT with ultracold atoms is the need to 
impose the gauge invariance on the simulated system, for which it is not a fundamental symmetry.
Different strategies  to achieve this goal have been explored in the context of proposals for 
the realization of LGT with ultracold atoms.

\vspace*{2mm}\noindent\textbf{Energy penalty.}
A natural option is to add a penalty term to the Hamiltonian that suppresses 
symmetry-violating transitions,
such that the gauge invariant dynamics is obtained as an effective model at low energies.
This strategy was proposed first by Zohar and Reznik~\cite{Zohar2011}
to simulate pure gauge compact QED (cQED) in 2+1 dimensions.
The first proposal required Bose-Einstein condensates of four atomic species, 
trapped in different sites of the lattice, and 
achieved plaquette interactions as second order perturbative effects.
The scheme was simplified in~\cite{Zohar2012}, 
to require only one bosonic atom per link, in one of $2\ell +1$ possible states,
and thus implement a truncated cQED 
 with maximum electric flux per link $\ell$.
In a further generalization, dynamical charges were included in the model as naive fermions~\cite{Zohar2013c}.
Kasamatsu et al.~\cite{Kasamatsu2013higgs,Kuno2015} demonstrated the strategy to be relatively robust:
for a wide range of parameters in the atomic system, the gauge violating 
terms of the Hamiltonian could be reinterpreted as actually implementing a gauge-Higgs model,
which can also be simulated in 3+1 dimensions~\cite{Kuno2016}.
%~\cite{Park2019glassy} Same team: look for phases, argue for glassy; rather about the model

A penalty term to ensure the gauge symmetry was also included in the proposal by Banerjee et al. ~\cite{Banerjee2012}
to simulate a $U(1)$ quantum link model, using both fermionic and bosonic atoms. In this case, 
$2S+1$ bosonic
atoms needed to be trapped in each link to represent the gauge degrees of freedom for a spin-$S$ representation.
A triple-layer optical lattice was proposed for the (1+1)-dimensional case, each layer holding, 
respectively, the fermions and the bosons in the even or odd  links (see Fig.~\ref{fig:ultracold1}). 

A related case is that of the discrete $Z_N$ gauge symmetry (which approaches Wilson's lattice QED in the large-$N$ limit).
A possible implementation of the (1+1)-dimensional case was discussed by Notarnicola et al.~\cite{Notarnicola2015},
 using one fermion per vertex and one fermion or boson per link, again for gauge invariance being effective at low energies.
 %\note{Somewhere before the distinction between LGT a la Wilson and purely discrete or Qlinks has to be mentioned.}

Another proposal in which the gauge theory is obtained as the low-energy limit of 
the atomic interactions was introduced by Dutta et al.~\cite{Dutta2016toolbox},
who proposed a setup to realize different gauge theories 
using two atomic species trapped in a two-dimensional optical lattice.
One of the atoms acts as an ancilla, while the other is a boson trapped at large filling.
By lattice shaking, the interactions can be suitably modulated, and a Hamiltonian 
is engineered that at low energies corresponds to different Abelian gauge theories.

Nevertheless, imposing the gauge symmetry through an energy penalty carries some limitations.
Since the desired model is obtained in the low-energy limit of the dynamics, only a limited
range of parameters can be explored while keeping the validity of the effective model.
Additionally, for non-Abelian cases, the implementation of the corresponding constraint terms
becomes more complicated.

\vspace*{2mm}\noindent\textbf{Dissipation. }
An alternative to the energy penalty was proposed by Stannigel et al.~\cite{Stannigel2013} 
using dissipation, by a so-called continuous Zeno effect. 
In this proposal, the system is driven by classical noise terms, linearly coupled to the Gauss' law generators, 
which result in an effective evolution in the gauge invariant subspace.
The strength of the noise and of the gauge violating terms in the atomic Hamiltonian determine the
time range of the gauge protection.

\vspace*{2mm}\noindent\textbf{Intrinsic symmetry. }
A more robust way to accomplish gauge invariance is to map the symmetry onto a fundamental
one of the experimental setup. Zohar et al.~\cite{Zohar2013,Zohar2013a} introduced a scheme 
where the conservation of hyperfine angular momentum in the collisions among fermionic and bosonic atoms 
ensures the gauge invariance, when species with suitably chosen third component $m_F$ are used to represent the different degrees of freedom in the theory (see Fig.~\ref{fig:ultracold2}). 
This enables the 
realization of different models, including cQED, $Z_N$ and $SU(N)$. In particular, to realize (truncated) cQED with staggered fermions,
two bosonic species per link, and one fermion per vertex are required. 
The effect of the truncation in this proposal
and the robustness of the scheme against potential gauge-violating errors
 was studied numerically by K\"uhn et al.~\cite{Kuehn2014}.
 A related proposal by Kasper et al.~\cite{Kasper2015,Kasper2016} argued how to simulate real time dynamics
 and observe the pair production mechanism near the untruncated limit.
 The same conservation law can be used as fundamental block to implement $SU(2)$~\cite{Zohar2013a},
 although in this case,
 the link interactions have to be engineered out of two elementary links, with four bosonic species on each of them, and with help of one additional
 fermionic degree of freedom and an energy penalty, such that the $SU(2)$ link is obtained at second order in perturbation theory.
In these schemes, higher dimensions can also be realized by a loop method. To this end, additional species have to be included in the vertices, 
with an energy penalty that selects their configuration,
and the plaquette terms in the Hamiltonian are generated as higher order perturbative terms, in virtual processes that drive such particles around the plaquette.
The same authors presented in~\cite{Zohar2015a} another simulation scheme for $SU(2)$, which avoided the need for decomposing the links by using the 
 invariant truncation scheme introduced in~\cite{Zohar2015}.
More recently Gonz\'alez-Cuadra et al.~\cite{Gonzalez-Cuadra2017} showed how six bosonic species
would be enough to simulate the Abelian-Higgs model (i.e.\ a scalar field with $U(1)$ gauge symmetry) based on the same angular momentum conserving scheme.
%\note{As figure for this, either Fig. 5 of ~\cite{Gonzalez-Cuadra2017} NJP2017, or 8 (virtual hopping), or some from Erez's review? (e.g. 9c\&d) }

Another proposal that exploited intrinsic symmetries of the atomic system 
was put forward by Banerjee et al.~\cite{Banerjee2013} 
to realize $U(N)$ and $SU(N)$ quantum link models
using fermionic alkaline-earth atoms.
In this case, the dynamic fermions and rishon constituents of the quantum link are mapped 
onto different atomic Zeeman levels of a given nuclear spin $I$, which
satisfy a $SU(2I+1)$ symmetry.

The same platform and symmetry was exploited by Laflamme et al.~\cite{Laflamme2016} for a different approach
to realize the continuum limit of 1+1 dimensional $CP(N-1)$ models. 
% toy models for some features of QCD: asymptotic freedom, confinement and theta vacua; dimensional reduction: emerge from SU(N) 
Different from Wilson's LGT, in which the continuum limit is obtained by systematically decreasing the lattice spacing, while
modifying the Hamiltonian parameters in a suitable way,
 in the case of quantum links, the continuum limit can emerge from dimensional reduction~\cite{Chandrasekharan1997}. %\note{** Here a ref to sth in the TN section about cont. limit and maybe shorten}
The proposal in~\cite{Laflamme2016} used the nuclear spin symmetry of the atoms to implement a (global) $SU(N)$ spin model
in two spatial dimensions, on a ladder geometry, which gives rise to $CP(N-1)$ continuum fields as the shortest dimension is increased.

Dehkharghani et al.~\cite{Dehkharghani2017hop} proposed an alternative strategy, 
in which neutral atoms trapped in a superlattice represent matter fields, and 
the hopping is controlled by the internal state of (neutral or charged) impurities, trapped between the sites in a second lattice.
The gauge symmetry is obtained by a resonant condition of the coupling between the impurity and the
atoms.

\vspace*{2mm}\noindent\textbf{Encoding the physical subspace. }
Another strategy that completely avoids the need of imposing the gauge symmetry, 
is working directly in the physical subspace.
Bazavov et al.~\cite{Bazavov2015gaugeinv} suggested a way
to implement the (1+1)-dimensional Abelian Higgs model after integrating out the gauge variables,
in a limit of large Higgs mode mass, reducing it to a classical $O(2)$ model
which can be simulated~\cite{Zou2014o2} using two bosonic species in a one-dimensional 
or ladder-shaped optical lattice.

Also a gauge invariant formulation was used by Rico et al.~\cite{Rico2018so3} to propose a simulation scheme for the $SO(3)$ 
quantum link model, arguably the simplest non-Abelian gauge theory which may be a toy model for QCD.
In the (1+1)-dimensional case, the gauge invariant subspace is mapped onto a spin-$3/2$ Heisenberg chain, 
which can be implemented using two bosonic species trapped in a one-dimensional optical lattice.

For gauge theories with fermionic matter, the lattice formulation needs to make a choice about the discretization of 
fermion fields, as it is impossible to simultaneously satisfy all the continuum symmetries~\cite{Nielsen1981}.
All the proposals mentioned above use a staggered formulation, but Zache et al.~\cite{Zache2018wilson}
recently argued that a Wilson discretization may enable a more efficient simulation of QED in 1+1 dimensions.
The proposal used a tilted optical lattice, with two fermionic species and a bosonic condensate to 
implement the quantum link model.

\begin{center}
\textbf{Digital proposals}
\end{center}
The proposals described above focus on the analog simulation of LGT. Although they make use of existing experimental techniques,
combining all the necessary ingredients in a single experiment is still a challenge. A significant difficulty is the generation of
the many-body plaquette terms for 2+1 or higher-dimensional models. These can be obtained as higher-order perturbative contributions, 
but this limits the range of parameters that the simulator can explore.
Instead, Zohar et al.~\cite{Zohar2016c} introduced a digital simulation scheme
in which the evolution is performed in (Trotter) discrete time steps.
For each of them, the different terms of the Hamiltonian are generated
independently as a sequence of two-body gates.
The implementation, in this way, requires including additional ancillary degrees of freedom, and using a 
multilayer optical lattice, which allows relative displacements of the layers.
Concrete prescriptions were presented for $Z_2$ and $Z_3$ in 2+1 dimensions~\cite{Zohar2016c,Zohar2016b},
and the generalization to 3+1 dimensions and a concrete proposal for the simplest non-Abelian case, the dihedral group $D_3$,
was presented in~\cite{Bender2018}.

\begin{center}
\textbf{Experiments}
\end{center}
A quantum simulation of an LGT with ultracold atoms in optical lattices has not yet been experimentally realized yet.
Nevertheless, 
the implementation of a minimal working instance of matter-gauge $Z_2$ has been 
recently achieved by Schweizer et al.~\cite{Schweizer2019floquetZ2},
following a prior theoretical study~\cite{Barbiero2018z2}.
In the experiment, fermionic and gauge degrees of freedom are mapped onto two atomic species: Zeeman levels
of a hyperfine manifold, trapped in a species-dependent double-well potential,
and the interactions are engineered by periodically driving the system, as an effective Floquet Hamiltonian.
%% Unclass
%% ~\cite{Kosior2014} - propose encoding the "color" indices not in internal states of the atoms but in positions of the lattice and engineer the interactions to realize U(2). However, it seems (said in the abstract, too) U(2) static.

% Trapped Ions 
\subsubsection{Trapped ions}
\label{subsec:ions}

Systems of cold atomic ions contained in electromagnetic traps
provide another one of the most advanced platforms currently investigated for quantum 
simulation and computation.
In a trapped ion system, a qubit is encoded in two internal states of an ion, and can be manipulated
by laser or microwave pulses~\cite{CiracZoller95,Haeffner2008ions,Lanyon2011iondigital,Blatt2012ions}.
In a linear trap, the ions are mostly confined in the transverse plane, forming a string.
At very low temperatures, their movement around the equilibrium positions
is quantized in the collective vibrational modes of the chain.
Using laser beams, and coupling the internal states to the motional 
degrees of freedom, effective interactions, and thus entangling gates, can be generated among the qubits.
Finally, the state of the qubits can be detected via fluorescence measurements.

A first proposal to engineer an analog quantum simulation of an LGT using trapped ions was 
presented by Hauke et al.~\cite{Hauke2013} for the $U(1)$ spin-$1/2$ quantum link model in 1+1 dimensions.
The proposal includes two kinds of qubits (pseudospins) with different
resonance frequencies, which are used to encode, respectively, the gauge and matter (after a Jordan-Wigner transformation) degrees of freedom.
Using an energy penalty term for the gauge violating configurations, 
the LGT Hamiltonian is obtained at second order in perturbation theory, as the effective low energy description.

A different strategy for the analog simulation of the Schwinger model was presented by Yang et al.~\cite{Yang2016}. They proposed 
using a string of ions individually confined in microtraps.
Two electronic states of each ion are used to encode the matter spin degrees of freedom, 
after the Jordan-Wigner transformation.
In this proposal, the gauge link variables are substituted by bosons, that need to be initialized 
at a high occupation number (compared to the length of the chain) in order to approach the 1+1d QED behaviour.
Some of the radial vibrational modes of the ions can be used to encode this bosonic degrees of freedom,
while the others are used to engineer the couplings.
These phonon modes are assumed to be well localized between ions, which can be achieved by controlling the radial frequency,
even with some experimental imperfections.
The same paper~\cite{Yang2016} proposes
a simpler scheme, which could be realized in a linear trap, for the simulation of the $U(1)$ spin-$1/2$ quantum link model.
In this alternative proposal, the link variables are encoded in the 
pseudospin represented by the ions, and the matter degrees of freedom, instead, are represented by axial collective phonons,
ordered by energy, such that the real space lattice is mapped onto an energy one.

Other proposals exist for purely discrete LGT.
For instance, and also using ions individually held in a linear array of microtraps, Nevado and Porras~\cite{Nevado2015rabi} proposed the realization of an
Ising-Rabi model, which exhibits a local $Z_2$ gauge invariance. In this proposal, transverse vibrational modes of the ions
encode the local bosonic degrees of freedom, and the different frequencies of each trap ensures the bosons do not interact with each other.
The interactions between neighbouring spins (internal states of the ions) are mediated by a longitudinal laser field.
Nath et al.~\cite{Nath2015hexagonal} discussed a different proposal to realize hexagonal plaquette interactions 
in an ion crystal, using three internal states, and pinning particular ions
to modify the phonon spectrum adequately.
They proposed this scheme as a building block to realize a spin Hamiltonian 
with effective $Z_2$ gauge symmetry at low energy,
on a kagome lattice.

\begin{center}
\textbf{First experimental simulation}
\end{center}
The first experimental demonstration of a quantum simulation of an LGT 
was recently performed in Innsbruck by Mart{\'{\i}}nez et al.~\cite{Martinez2016,Muschik2016}.
The experiment realized a digital simulation of the Schwinger model, using a chain of 
four $^{40}$Ca$^+$ ions confined in a Paul trap.
After using Gauss law to integrate out the gauge degrees of freedom,
the model reduces to a long-range spin Hamiltonian in the physical subspace (as detailed in Sec.~\ref{sec:schwinger}). %\note{Make this reference more concrete, or cite Hamer}
The spins correspond to the matter fields and were encoded in two electronic states of the ions,
which can be optically manipulated by laser pulses. 
In the experiment, the system was initialized in the strong coupling vacuum, and real time evolution 
was performed as a sequence of discrete (Trotter) time steps, for several values of the fermion mass.
Performing full tomography of the system state and postselecting the zero-charge component (to correct for errors),
several quantities were computed (particle density, entanglement, vacuum persistence) 
to examine the process of pair creation.

% Superconductive
\subsubsection{Superconducting qubits}
\label{subsec:sc}

\begin{figure}
\centering
\includegraphics[width=.8\columnwidth]{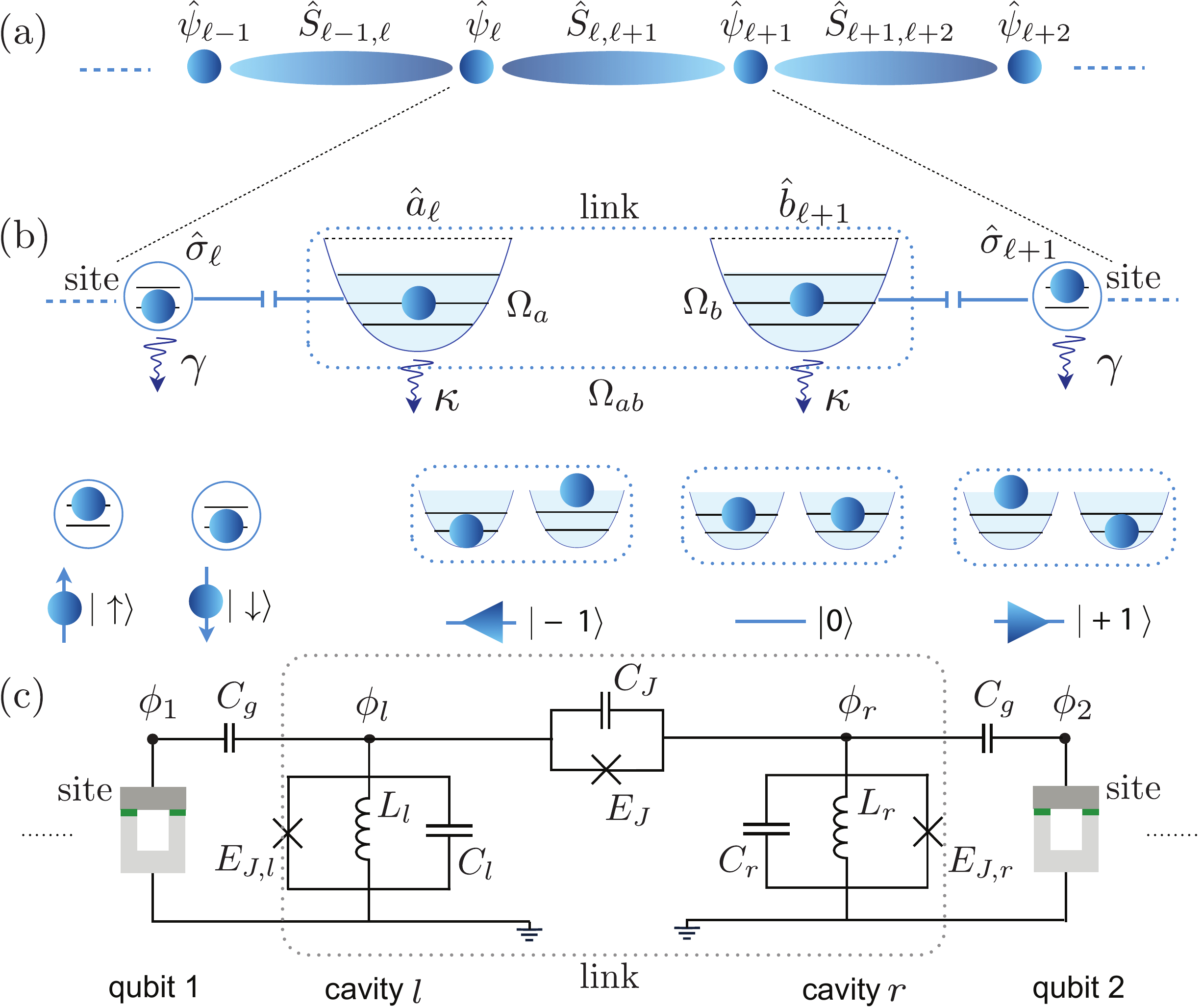}
\caption{Scheme of the unit cell for the simulation of the $U(1)$ quantum link model proposed in~\cite{Marcos2013}.
The model degrees of freedom (a) are represented by basic SC circuit components (c),
with a SC qubit at each (matter) vertex and two LC resonators encoding the Schwinger 
representation of the link spin.
Source: Ref.~\cite{Marcos2013}, reprinted with permission from the authors and APS.}
\label{fig:sc}
\end{figure}

Among the potential platforms for quantum simulation,
superconducting (SC) circuits have seen one of the fastest developments
in the last years~\cite{Devoret2013,Kjaergaard2019}.
A SC qubit is built up from circuit elements and includes one or more 
Josephson junctions that provide a non-linearity. The latter is crucial for its use as a qubit, 
as it enables singling out two levels that are used as the logical states.
The Hamiltonian that describes the SC qubit is that of a non-linear resonator, but
there exist different types of SC qubits, depending on the concrete circuit elements they combine,
which encode the qubit states differently.
A SC qubit can be manipulated using microwave radiation, and
two SC qubits can be coupled through resonators
that can also be designed and implemented to engineer complex
interactions.
One of the big advantages of SC circuits, together with being highly tunable, is that they
can be fabricated using the technology of integrated circuits, 
which favours their scalability. 

Marcos et al.~\cite{Marcos2013} proposed an implementation of a (1+1)-dimensional
$U(1)$ quantum link model, corresponding to the truncated Schwinger model.
After the Jordan-Wigner transformation and using the Schwinger boson representation
for the link spin variables, the unit cell consists of
two spin-$1/2$ matter sites and one link.
The former are encoded using a SC qubit per vertex, while the links require two non-linear LC resonators,
coupled to each other via another Josephson junction (see Fig.~\ref{fig:sc}).
This unit cell was shown~\cite{Marcos2013} to be able to simulate 
the string breaking of the LGT under realistic parameters for the SC circuit.

In a subsequent work, Marcos et al.~\cite{Marcos2014}
considered the pure gauge $U(1)$ case in 2+1 dimensions.
Using a qubit per link to represent the spin-$1/2$ gauge degrees of freedom, 
and establishing direct couplings via Josephson junctions between each pair of 
 qubits that share a vertex, it was shown that the gauge-invariant 
 quantum link Hamiltonian, including a plaquette term, can be obtained as a low-energy description.
 A modification of the setup should also give access to other gauge-invariant terms, in particular 
 a four-body Ising interaction, and thus allows the simulation of quantum dimers.
 
 Brennen et al.~\cite{Brennen2016} proposed a different SC architecture for the simulation of the
 same model, based on fluxonium devices, which can operate as qutrits to represent the link variables of 
 the $U(1)$ spin-$1$ quantum link model, or as ancillary qubits, which are included in the vertices to impose the Gauss law.
 This is achieved by an inductive coupling between the ancillas and the neighbouring links which results in 
 an effective projection,  for low energies, onto the subspace conforming with the Gauss law.
 The plaquette terms are obtained at second order in perturbation theory, together with
 an extra gauge-invariant contribution.
 The setup allows the observation of different regimes of the model.
 In particular, the proposal allows for the implementation
 and non-destructive measurement of Wilson loop and 't Hooft string non-local operators (order and disorder parameters of the gauge model), 
 which can detect the presence of a confining phase.
 
 The first proposals worked in the framework of analog simulation, where, by suitable tuning of the
 design parameters of the circuit, the effective Hamiltonian 
 of the system coincides with the target one.
 Instead, the proposal for the non-Abelian case by Mezzacapo et al.~\cite{Mezzacapo2015}
 is based on a digital scheme.
More concretely, an implementation scheme is studied for
 the minimal unit for the pure gauge (2+1)-dimensional $SU(2)$ quantum link model on a triangular lattice.
 The gauge variables, sitting on the links, are represented by two qubits each,
 which can be implemented with a total of six SC qubits (two per link) coupled to a common resonator.
 In this setup, collective gates can be applied, and the Trotter steps corresponding to the
 many-body plaquette terms in the LGT Hamiltonian can be implemented as a sequence of 
 them, interspersed with single qubit rotations.

%%  A related proposal was presented by Sameti et al.~\cite{Sameti2017topo} 
%% Really, this one is for the toric code, not a dyn LGT per se, although they point out the possibility to generalize it in the conclusions.
%% The proposal uses transmon qubits on links, coupled with SQUIDs via auxiliary nodes: the four qubits in the plaquette are coupled  in two pairs via an auxiliary note, and then these two are coupled to a SQUID. Driving this one with oscillating magnetic pulses and eliminating the auxiliary nodes, realizes the many-body interactions
 
 %% Related but not directly: ~\cite{Alaeian2019cqed} Via a periodic modulation of two transmon frequencies, the coupling is modified to have non-trivial phase. This is used to propose a realization of the bosonic- Creutz ladder. The model is interesting for topological features, chiral modes, but not really a LGT.

%  Rydberg
\subsubsection{Rydberg atoms}
\label{subsec:Rydberg}
% Rydberg atoms

\begin{figure}
\centering
\includegraphics[width=.65\columnwidth]{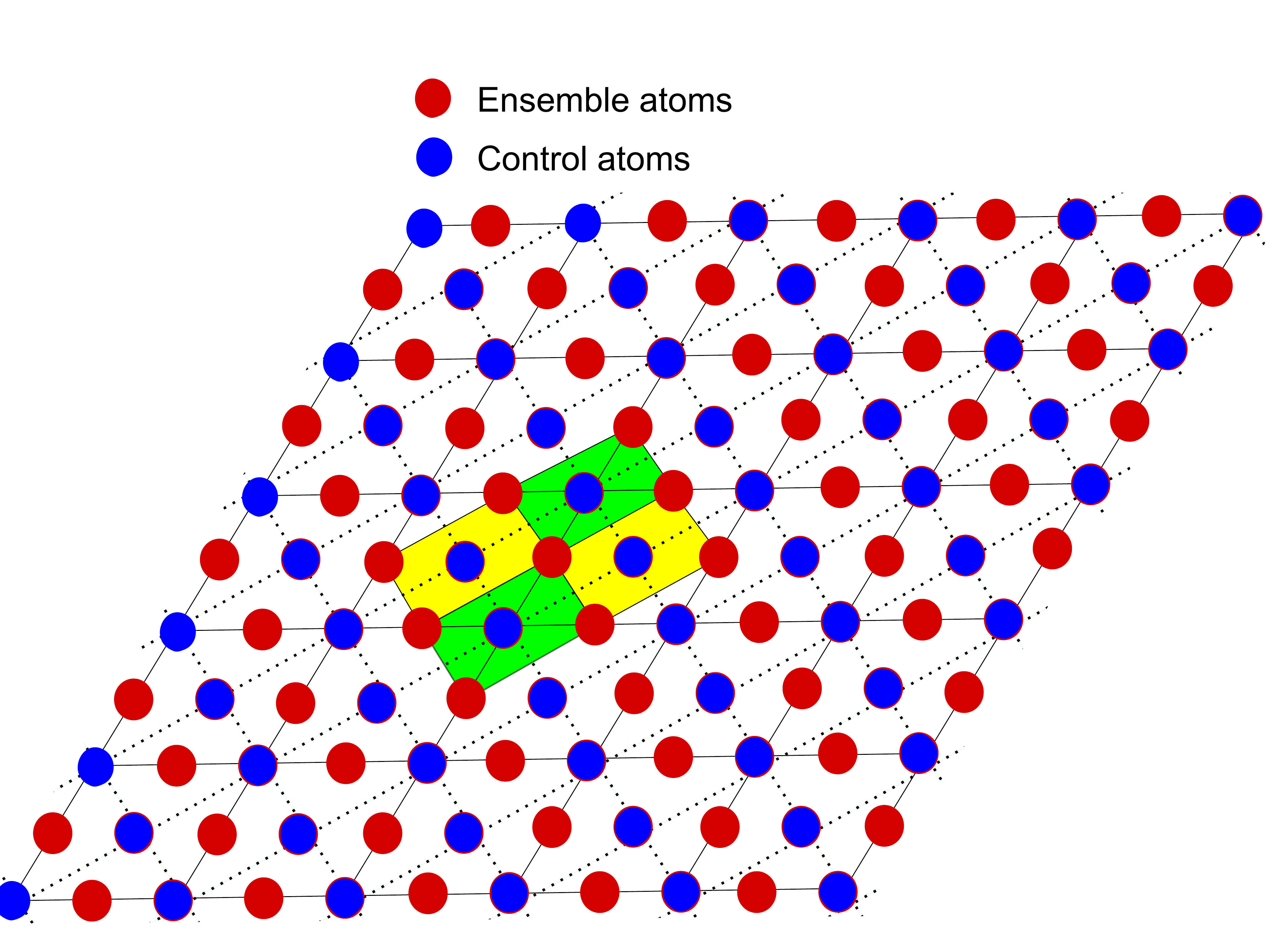}
\caption{Schematic setup for the simulation of the (2+1)-dimensional $U(1)$ LGT proposed in Ref.~\cite{Tagliacozzo2013}.
Control atoms (blue) sit in the vertices and center of plaquettes and ensure Gauss law and plaquette interactions of ensemble atoms (red) on the links.
Source: Ref.~\cite{Tagliacozzo2013}, reprinted with permission by the authors and Elsevier. }
\label{fig:rydberg_tagliacozzo}
\end{figure}

Rydberg atoms are highly excited atomic states, with an electron in a level with high principal number $n$,
and particular properties that can be exploited for quantum information processing \cite{Jones2017}.
Due to their size, Rydberg atoms have a large dipole moment.
This affects the atoms in their neighbourhood, shifting their energy levels, which
prevents the occurrence of a second Rydberg excitation within a certain radius, a phenomenon 
known as Rydberg blockade.
The long-range of dipole forces allows also the interaction of Rydberg atoms at long distances.

A scheme to use Rydberg atoms as a platform for digital quantum simulation was 
first introduced by Weimer et al.~\cite{Weimer2010rydberg}.
The proposal considers atoms trapped in an optical lattice or a magnetic trap.
Some of them represent the degrees of freedom of the quantum many-body problem, and 
additional auxiliary atoms are included to be used as control.
They need to be individually addressed to realize the mesoscopic Rydberg gate 
introduced in Ref.~\cite{Mueller2009rydberg}.
The control atoms
mediate the coherent many-body interactions of their neighbours (ensemble),
and can also be used to read out the ensemble state.
In addition, the optical pumping of the controls results in collective dissipation in the ensemble.
This can be used as cooling mechanism to prepare desired ground states.
In Ref.~\cite{Weimer2010rydberg}, the implementation is described for the particular case of 
Kitaev's toric code (at low energy a $Z_2$ LGT).

Tagliacozzo et al.~\cite{Tagliacozzo2013} put forward a proposal, based on the same setup,
for the digital simulation of an Abelian LGT, more concretely a pure-gauge $U(1)$
quantum link model in 2+1 dimensions. 
In this proposal, atoms are trapped in a two-dimensional lattice, 
with ensemble atoms (representing the gauge degrees of freedom)
sitting on the links, and control atoms occupying the vertices and the center of each plaquette
 (see Fig.~\ref{fig:rydberg_tagliacozzo}).
These controls are respectively used to ensure the Gauss law
and to implement plaquette terms.
In this proposal, the gauge invariance is imposed at the initial step using the Rydberg gates to 
engineer the dissipative preparation of the system in the gauge-invariant subspace.
This is followed by the evolution under a changing Hamiltonian (adiabatically, or using optimal control techniques) 
to obtain the ground state of the Hamiltonian with the desired parameters.
This evolution, discretized in Trotter steps, can also be implemented by
Rydberg gates, while the dissipative processes can be used at the same time to control gauge-violating errors.
The ground state can be used to probe confinement, by including external charges in the initial
state.

The more complex case of non-Abelian models was 
addressed by Tagliacozzo et al.~\cite{Tagliacozzo2013a}.
The simulation of the smallest $SU(2)$ LGT with a similar setup requires 
two atoms per link (each with an effective two-dimensional subspace), 
plus the additional Rydberg controls in vertices and plaquettes.
A simplification is possible in the strong coupling regime, that
uses only one, as in the Abelian case.

Zhang et al.~\cite{Zhang2018polyakov}
proposed the use of Rydberg interactions in an optical lattice with a multiladder geometry
to implement an analog simulation of the (1+1)-dimensional Abelian Higgs model. In this proposal, related to Ref.~\cite{Bazavov2015gaugeinv} 
(described in Sec.~\ref{subsec:ultracold}), the atomic degrees of freedom encode directly the physical subspace of the LGT.
In this simulation setup, the legs of the ladder correspond to different states of the
one-dimensional sites, and the long range interactions of Rydberg atoms would be exploited to implement the 
interaction between them.

Recently, Bernien et al.~\cite{Bernien2017expt}
have realized a quantum many-body simulation with
a chain of 51 individually trapped Rydberg atoms.
Besides demonstrating the potential of the platform for further quantum simulations,
this experiment found anomalously long-lived oscillations after a quantum quench, 
which has triggered the quest for a theoretical explanation \cite{Turner2018scars,Khemani2019pxp,Lin2019scars}.
In a recent work, Surace et al.~\cite{Surace2019rydberg} 
claimed that this experiment constitutes already a quantum simulation of an LGT.
They found an exact mapping between the states of the Rydberg atom chain and 
the electric field configurations of the $U(1)$ quantum link model with spin $1/2$.
The mapping corresponds to a gauge-invariant formulation, after the matter
has been integrated out using the Gauss law, which can be connected to the Rydberg blockade constraint.
The observed slow dynamics is then a signal of the string inversion phenomenon.

Notarnicola et al.~\cite{Notarnicola2019realtime} have proposed another setup to simulate the (1+1)-dimensional $Z_2$ theory,
mapping the gauge invariant subspace (again with integrated-out matter) onto the states of an atomic lattice. 
Each link is represented by three atoms.
Thanks to the Rydberg blockade, and by suitably choosing the encoding of neighbouring links, the available states are matched to the allowed configurations in the 
LGT, and the setup can be used to explore the dynamics of the $Z_3$ chain.

Concerning more complex simulations in more than one spatial dimension, Celi et al.~\cite{Celi2019emerging} suggested a setup of Rydberg 
atoms, arranged in orientable pairs in a suitable spatial configuration. Each pair incarnates a spin-$1/2$ degree of freedom in a dual formulation
 of the Rokhsar-Kivelson model, a discrete $U(1)$ LGT, which can then be simulated with this platform.

% is there cQED??

\subsection{Quantum Computation}
\label{subsec:QComp}

\begin{figure}
\centering
\includegraphics[width=.85\columnwidth]{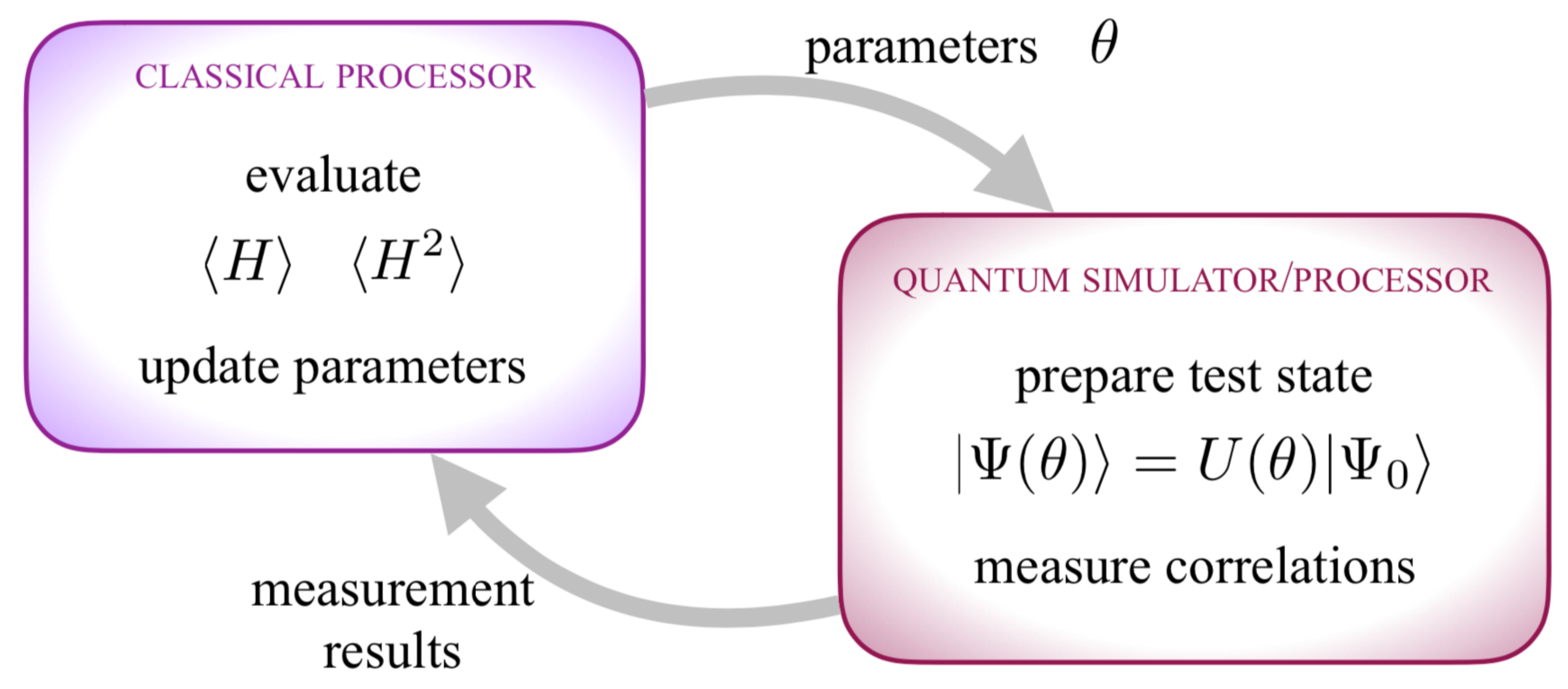}
\caption{Scheme of the feedback loop in hybrid VQE/VQS.}
\label{fig:vqs}
\end{figure}

Compared to the simulation platforms we have just discussed, a quantum computer is 
a more general programmable digital device. It should be able to run an arbitrary
sequence of gates, in a fault-tolerant way, so it needs to implement some form of error correction.
In particular, it could run any digital quantum simulation.

The question of whether a universal quantum computer will be able to 
efficiently simulate an LGT was first addressed by Yamamoto and Byrnes~\cite{Yamamoto2006qc}.
They considered pure-gauge compact QED, $SU(2)$ and $SU(3)$ LGT, in any number of spatial dimensions,
and introduced possible encodings of the respective link Hilbert spaces
in terms of qubits. For these representations, the cost of simulating digitally the evolution with the corresponding
Hamiltonians, including preparation of initial states, and measurement of Wilson loops, was upper bounded by a polynomial in the
number of sites of the lattice. The analysis focused on the minimal resources
for an arbitrary architecture, without considering any possible optimization (such as exploiting symmetries for a more efficient encoding) 
or the effect of errors (e.g.\ gauge-violating).

A significant boost to the investigation of quantum algorithms applied to high energy problems
was provided in 2011 by Jordan, Lee and Preskill~\cite{Jordan2012science,Jordan2011scalar},
who developed a quantum algorithm for the computation of scattering probabilities of a continuum scalar field.
The algorithm is based on a discretization of the space, and a cutoff of the field, such that a representation using qubits
is possible, and applies a Suzuki-Trotter expansion of evolution operators to adiabatically prepare the initial states and switch on and off the interaction,
and to simulate the evolution.
Jordan et al.\ demonstrated that the algorithm is efficient (the cost scales polynomially) in the number of particles involved, their energies, and
the required precision, both in the strong and weak coupling regimes. The analysis included the discretization errors, which
would be fundamental to attain the continuum limit.
In later works by other authors, the discretization errors induced by the digital representation of continuum fields using qubit registers 
was analyzed~\cite{Macridin2018fermionboson,Klco2019scalar},  
and optimized strategies were proposed to find a suitable trade-off between the digitization error and other sources of noise.
Jordan et al.\ extended their result to the fermionic case~\cite{Jordan2014fermions}, a result recently improved
by Moosavian and Jordan~\cite{Moosavian2018ferm}, using the substitution of the adiabatic preparation step by a
direct construction of the MPS ground state for the interacting lattice theory (which would be obtained in a previous 
classical computation).

These results are not directly applicable to a gauge theory, but they address general questions
about the complexity of quantum field theories, and in particular the cost of the continuum limit, 
which will be relevant in other more general theories.
Addtionally,
Jordan et al.~\cite{Jordan2018bqp} recently addressed the complementary question, namely what is the computational power of QFT itself.
They demonstrated that a closely related QFT problem, the computation of a vacuum-to-vacuum transition in the presence of time and space-dependent 
external fields, is BQP-complete, and thus has the computational power of a universal quantum computer.

Formulating similar algorithms for LGT involves specific issues. One of them is the representation of the
gauge-variant states in a finite number of qubit registers, since, besides the potential need to truncate 
the Hilbert space of the links, gauge invariance implies a redundancy in the straightforward description. 
A possible strategy, as mentioned for the TNS and quantum simulations, is to restrict the
representation to the physical subspace. This is explicit in the dual formulation proposed by
Kaplan and Stryker~\cite{Kaplan2018u1} for the pure-gauge $U(1)$ case in two and three dimensions.
Alternatively, one may impose constraints to force the Gauss law. Related to this possibility,
Stryker~\cite{Stryker2019oracles} suggested the use of oracles to check for violations of the constraint,
and constructed them explicitly in terms of quantum gates for the $U(1)$ and $Z_N$ cases in
up to three spatial dimensions.

%%  Lamm et al.~\cite{Lamm2019qsim} give a construction of a Hamiltonian starting from the group representation of the link variables (same scheme as Zohar~\cite{Zohar2015})
%%  They construct a generic Hamiltonian, from a transfer matrix that is built from gauge invariant plaquette and link terms, and then simulate (with a QC classical simulation) in a two plaquette system for $D_4$ and $\mathbb{Z}_2$ with fermionic matter. I do not see the point.

% near scale
\begin{center}
\textbf{Near term QC}
\end{center}
The algorithms and the results mentioned above make it evident that a quantum computer will allow
computations that override the classical simulation methods
in the realm of quantum field theory, especially regarding the most difficult scenarios (e.g.\ dynamics and non-perturbative regimes).
But building a fully-fledged fault-tolerant quantum computer with a sufficiently large number of logical qubits,
 still represents a very serious technical challenge, and will probably only happen in the long term.

Nevertheless, thanks to the fast development of quantum technologies in the last few years,
including industrial involvement in the quantum computing research, 
we can expect to have a generation of imperfect quantum computers
that are available in the near term.
These devices, that Preskill~\cite{Preskill2018nisq} named noisy intermediate scale quantum (NISQ),
will have between 50 and a few hundred physical qubits, and will be able to perform
around 1000 gates before noise becomes dominant.

Because a classical computer cannot exactly simulate 50 qubits, 
these resources may be enough to outperform classical computers in 
specific tasks, and there is an ongoing effort to find problems and algorithms capable 
of showing quantum advantage~\cite{Childs2018speedup}.
In particular, in the realm of LGT, different groups have started exploring the possibilities of what 
is now referred to as NISQ-era devices.

This has led to the development of hybrid algorithms, that combine classical processing 
with the use of available quantum hardware,  to optimize the use of the latter.
It is the case of the variational quantum eigensolver (VQE) \cite{Peruzzo2014vqe,Moll2018var}, originally developed
for quantum chemistry.
The quantum hardware is used to prepare an entangled state, specified by some parameters, 
and to measure the correlators that correspond to the individual Hamiltonian terms.
The results are used by the classical processor that computes the energy expectation value and decides how to 
optimize the parameters of the state. With this feedback, the quantum state is prepared again and the process is iterated. 

Klco et al.~\cite{Klco2018qclas} used this approach to study the ground state and dynamics of the Schwinger model. 
Their algorithm makes use of a very efficient encoding of the physical subspace, exploiting all the symmetries of the problem
and applying a truncation in the electric flux per link and the total electric energy of the state. For two spatial sites 
with periodic boundary conditions, the zero charge sector with zero momentum and even parity could be encoded 
using only two qubits. 
The quantum part of the algorithm was subsequently run on the IBM Q platform.
The encoding was computed using classical processing, which would limit the scalability of the approach, as this step
would have an exponentially large cost for larger systems.

Kokail et al.~\cite{Kokail2019selfverifying} presented a modification of the idea, termed variational 
quantum simulation (VQS), in which the 
quantum hardware is not a universal quantum computer, but an analog quantum simulator (see Fig.~\ref{fig:vqs}).
They implemented this VQS on a programmable analog ion-trap quantum simulator, with up to 20 ions, 
for the Schwinger model. Again, the use of the physical subspace, and the problem symmetries (charge conjugation and approximate translational invariance in the bulk)
was key to reduce the number of variational parameters and the depth of the quantum circuit that prepares the test states.
The results in Ref.~\cite{Kokail2019selfverifying} suggest in fact a polynomial scaling of these resources with the system size.
The VQS can verify the quality of its own results by evaluating also the variance of the test states.

Lu et al.~\cite{Lu2019subatomic} presented an implementation of VQE 
on a quantum frequency processor. 
This is a photonic platform, in which qubits are encoded in frequency bins, and gates are implemented using 
electro-optic phase modulators and pulse shapers, standard telecommunications components.
In Ref.~\cite{Lu2019subatomic}, a mode-entangled single photon was used to 
demonstrate the scheme for the Schwinger model, with and without static charges, with eight fermionic sites.

The research in this area is active, and new algorithm proposals, sometimes with demonstrating 
implementations in the available quantum hardware are being developed as we complete this article, for gauge theories,
but also for other quantum field theory problems. 
For instance, a hybrid algorithm  for the spectrum of the $\lambda \Phi^4$ field theory 
has been recently presented by Yeter-Aydeniz et al.~\cite{YeterAydeniz2019scalar},
and we can only expect exciting new developments in this area in the near future.

%\note{A figure missing: sketch of VQS? Results plot from \protect~\cite{Kokail2019selfverifying} (2a)?}

\section{Summary and perspectives}
\label{sec:summary}

The application of quantum information based techniques to lattice gauge theories
is a field undergoing very fast development in various fronts.
On the one hand, classical simulations using the tensor network framework,
founded in the theory of entanglement, are probing their ability to surmount 
the challenges that LGT pose for standard methods.
On the other, quantum simulations and quantum algorithms run in quantum computers 
could offer the potential to surpass any classical computation.
A few proof-of-principle experiments have already taken place, and 
 a large variety of theoretical proposals to realize different models using existing technology
  is paving the way to future, more demanding, experiments in different platforms.
In this article, we have reviewed the status and progress of these attempts.

In the long run, one of the main forces motivating this programme is 
the possibility to apply these novel techniques to QCD,
the theory of the strong interaction, whose non-perturbative regime has for many years been investigated with lattice techniques.
Despite the amazing successes of the latter, in particular Monte Carlo simulations, there are still research areas that are very 
difficult to access, or are inaccessible altogether, using the traditional approaches.
Thus, novel methods are sought for such cases and we argue that these quantum-based techniques can be among the most prospective ones.

On the classical simulation front,
TNs can serve as a remedy for the notorious sign problem of lattice QCD at non-zero baryon density and to investigate real-time dynamics.
However, the employment of this framework for QCD, or for other gauge theories in several spatial dimensions,
is far from trivial.
We have shown several successful applications to (1+1)-dimensional LGTs, including the Schwinger model and the $SU(2)$ and $SU(3)$ theories, and to other lattice quantum field theories, like the $O(N)$ sigma models, scalar field theory or the Thirring model.
The biggest challenge to be overcome is to make efficient use of TN techniques in higher dimensions.
We have discussed the perspectives for this, arguing there are no fundamental obstacles.
Nevertheless, the task is very difficult and clever algorithms need to be devised.
We remark this is somewhat similar to the situation in traditional lattice QCD in its early days -- Wilson's formulation was theoretically very appealing, but its practical realization was considered as very difficult, if possible at all.
It took the community numerous years to arrive at the current state-of-the-art and it required a combination of quickly growing computing power and proper algorithmic solutions.
One can hypothesize that a similar road will have to be traveled in the quest to study QCD with TNs.

Meanwhile, the recent TN studies of low-dimensional theories serve several purposes.
From the viewpoint of future applications, to QCD or other models, they are proofs of concept that TNs are useful in the studies of quantum field theories, with and without gauge symmetry.
This could not be taken for granted before the recent investigations, since TNs are an advantageous language if the studied theory possesses relatively little entanglement.
The body of evidence collected during the last years
has demonstrated that indeed the entanglement properties in physically relevant states of LGT allow for accurate TNS descriptions.
In addition, the studies of low-dimensional theories revealed their several interesting features, adding to the earlier studies with various perturbative and non-perturbative methods.
In particular, numerous studies of real-time properties were carried out, giving novel insights into their non-equilibrium physics.

In the long run, the quantum simulation, either with dedicated devices (quantum simulators) or with universal quantum computers, 
should be able to outperform classical computations, in particular in problems like non-equilibrium dynamics. 
Designing a competitive quantum simulator is a complex task, 
that requires solving a number of challenges, within the possibilities and limitations of a particular experimental setup.
It is remarkable that TNS will provide a best-suited tool to assist in the design and validation of such proposals.
More technical details on the developments in these two complementary avenues towards LGT can be found in the
recent review~\cite{QTFLAGreview}.

Even if the experimental realization of a quantum simulation of an LGT beyond what is classically possible may not happen in the immediate future,
the theoretical works we review here have demonstrated that 
the necessary technology to surmount each of those challenges is already available (see also~\cite{QTFLAGreview}).
A few experiments have already demonstrated the realization of some of the building blocks, 
such as the pioneering trapped-ion quantum simulation of the Schwinger model,
and we can only expect there will be exciting new developments in the future.

A universal quantum computer would extend the functionality of a quantum simulator. Although such 
fully functional device is not yet available, the appearance of smaller scale programmable digital quantum devices
has motivated the investigation of specific algorithms that use them for LGT calculations, an area that is evolving
extremely fast.

The different avenues discussed in this review offer excellent prospects for further scientific developments.
The ultimate dream of this quest, running TN and quantum simulations of QCD, could uncover some essential features of the strong interaction.
But the road leading there will be itself as exciting and interesting.
Relevant theories will be investigated, which will potentially allow insights never obtained before with other methods.
Complementary, also the classical and quantum methods will undergo maybe unforeseen improvements, as they develop to face the challenges found on the way.

%\noteMC{How to cite QTFLAG review}

\acknowledgments
We are grateful to  Enrique Rico, Luca Tagliacozzo, Savvas Zafeiropoulos and Erez Zohar
for their constructive comments on earlier versions of this work.
M.C.B.\ is partly supported by the Deutsche Forschungsgemeinschaft (DFG, German Research Foundation) under Germany's Excellence Strategy -- EXC-2111-- 390814868,
and by the EU-QUANTERA project QTFLAG (BMBF grant No. 13N14780). She also acknowledges the hospitality of KITP (supported in part by the National Science Foundation under Grant No. NSF PHY-1748958) and ICMAT, where parts of this works were completed.
K.C.\ is partly supported by the National Science Centre (Poland) grant SONATA BIS 2016/22/E/ST2/00013.

%\begin{figure}[h]
%\includegraphics[width=.8\columnwidth]{figures/scales}
%\caption{Scaling of $\log(N/\delta)$ (solid lines) and $\log(\sqrt{N}/\delta)$ (dotted lines) vs $\delta$ compared to $1/\delta$ (solid black line) for the range of system sizes we study.
%It seems that we will not be in the regime where the exponential in \eqref{eq:scaleCos} and \eqref{eq:scaleCheby} becomes dominant, but
%for the variances we reach, the scaling is going to be dominated by the $\log$ part.}
%\label{fig:scalesDelta}
%\end{figure}

\bibliography{TN_for_LGT2,TN_for_LGT3}
%\bibliography{TN_for_LGT3}
\end{document}